\documentclass[12pt]{article}

\usepackage{amsmath}
\usepackage{amssymb}
\usepackage{latexsym}
\usepackage{graphicx}
\usepackage{psfrag,fancyhdr,epsfig}

\newcommand{\bmath}[1]{\mbox{\boldmath{$#1$}}}

\begin{document}

\begin{titlepage}

\vspace*{1cm}
\begin{center}
{\bf \Large Brane Decay of a (4+n)-Dimensional Rotating\\[1mm] Black Hole. II:
spin-1 particles}

\bigskip \bigskip \medskip

{\bf M. Casals}$^1$, {\bf P. Kanti}$^2$ and
{\bf E. Winstanley}$^3$

\bigskip
$^1$ {\it School of Mathematical Sciences, University College
Dublin,\\ Belfield, Dublin 4, Ireland}

$^{2}$ {\it Department of Mathematical Sciences, University of Durham,\\
Science Site, South Road, Durham DH1 3LE, United Kingdom}

$^3$ {\it Department of Applied Mathematics, The University of Sheffield,\\
Hicks Building, Hounsfield Road, Sheffield S3 7RH, United Kingdom}

\bigskip \medskip
{\bf Abstract}
\end{center}
The present works complements and expands a previous one, focused on the emission
of scalar fields by a $(4+n)$-dimensional rotating black hole on the brane, by
studying the emission of gauge fields on the brane from a similar black hole.
A comprehensive analysis of the particle, energy and angular momentum emission
rates is undertaken, for arbitrary angular momentum of the black hole
and dimensionality of spacetime.
Our analysis reveals the existence of a number of distinct features associated
with the emission of spin-1 fields from a rotating black hole on the brane,
such as the behaviour and magnitude of the different emission rates, the angular
distribution of particles and energy, the relative enhancement compared to the
scalar fields, and the magnitude of the superradiance effect. Apart from their
theoretical interest, these features can comprise clear signatures of the
emission of Hawking radiation from a  brane-world black hole during its spin-down
phase upon successful detection of this effect during an experiment.

\end{titlepage}

\section{Introduction}

One of the most exciting consequences of the scenario of Large Extra Dimensions
\cite{ADD} is that it opened the way for probing physics beyond the scale of
quantum gravity by postulating the existence of a low fundamental gravitational
scale $M_*$. If this scale is sufficiently low, trans-planckian particle
collisions could easily be achieved in various environments. The products
of such collisions cannot be ordinary particles any more but rather heavy
or extended objects ($p$-branes, string balls, string states, etc) arising in
the context of the more fundamental theory valid above $M_*$. As this
ultimate fundamental theory should also include gravity, strong gravitational
phenomena could also be observed at the quantum level, and the creation of
tiny black holes as the result of trans-planckian collisions \cite{creation}
is highly likely.

According to the theory with Large Extra Dimensions, all Standard Model fields
are restricted to live on a (3+1)-dimensional brane, while gravity is allowed
to propagate in the $(4+n)$-dimensional bulk (for some early works, see
\cite{early}). The theory predicts the existence of $n$ additional spacelike
compact dimensions, all having -- in the simplest case -- the same size
$L$. The collision of highly energetic particles on the brane may
then produce a tiny black hole that is attached to our brane but, being
a gravitational object, extends off our brane as well. In order to be able
to treat the black hole as a classical object, it will be assumed that the
black hole mass $M_{BH}$ is considerably larger than the fundamental scale
of gravity $M_*$. Under the additional assumption that the horizon radius
$r_h$ of the produced black hole is much smaller than the size of the extra
compact dimensions $L$, this object can be considered to live in a
$(4+n)$-dimensional non-compact, empty\footnote{The brane
self-energy can be naturally assumed to be of the order of the fundamental
Planck scale $M_*$ in order to avoid a hierarchy problem. But since
$M_* \ll M_{BH}$, its effect on the gravitational background is negligible.}
space; its properties differ from a 4-dimensional black hole with
the same mass and strongly depend on the number $n$ of additional dimensions
present \cite{ADMR}.

Small higher-dimensional black holes may be created during trans-planckian
collisions either at ground-based colliders \cite{colliders} or in high
energy cosmic-ray interactions in the atmosphere of the earth \cite{cosmic}
(for an extensive discussion of the phenomenological implications and a more
complete list of references, see the reviews \cite{Kanti, reviews, Harris}).
After a short `balding' phase, during which the black hole will shed all
additional quantum numbers apart from its mass, angular momentum and
charge, the more familiar Kerr-like phase commences, during which the black
hole will lose mainly its angular momentum. After that, a Schwarzschild phase
follows with the, now spherically-symmetric, black hole gradually losing its
actual mass. The energy loss of the black hole takes place through the
emission of Hawking radiation \cite{hawking} (and through superradiance
during its Kerr-like phase) that is, at the same time, the most prominent
signature of black hole creation. This consists of the emission of elementary
particles both in the bulk and on the brane, and it will be characterized
by a very distinct thermal spectrum.

Due to the simplicity of the corresponding gravitational background that
describes it, the Schwarzschild phase of the life of a small higher-dimensional
black hole underwent an intensive and thorough study during recent
years. The emission of Hawking radiation from such a black hole has been
studied both analytically \cite{kmr1, Frolov1, kmr2}
and numerically \cite{HK1}. These works have provided us with both analytical
formulae and exact numerical results for the emission rates, and demonstrated
how the latter strongly depend on the existence of additional spacelike
dimensions in nature. As was shown in detail in \cite{HK1}, the number
of particles and amount of energy emitted by the black hole per unit time
was strongly enhanced by the presence of extra dimensions in the theory.
Remarkably, it was found that the type of particles emitted also
depends on how many dimensions exist in the bulk: scalar particles are
predominantly emitted by a black hole living in a spacetime with a small
number of (or no) extra dimensions, while gauge bosons are the preferred
particles emitted by a black hole living in a spacetime with $n>5$.
The spectrum of Hawking radiation may be a source of valuable information
on other fundamental parameters of the theory, such as the value of the
cosmological constant \cite{BGK} and the string coupling associated with the
presence of higher-derivative gravitational (Gauss-Bonnet) terms
\cite{Barrau}. Other studies have focused on the effect of the mass
of the emitted particles on the spectrum \cite{Jung-mass, Doran} as well
as the charge of the black hole \cite{Jung-charge}.

Although initially almost totally ignored, the progress in studying the
emission of Hawking radiation during the Kerr-like phase in the life of a
small higher-dimensional black hole has been very rapid during the last
year. Two early works \cite{Frolov2,IOP1} focused on the special case of
a 5-dimensional rotating black hole: the first one provided analytical
formulae for the emission rates, and the second derived analytical results for
the energy emission rate in the limit of low energy and low angular momentum
of the black hole. The derivation of exact results valid for arbitrary values
of the energy of the emitted particle and for arbitrary values of the
angular momentum of the black hole required the numerical integration of
both radial and angular parts of the equation of motion of the specific
particle. The same task demands also the determination of the angular
eigenvalue, that links radial and angular equations together, and which
does not exist in closed form for arbitrary energy and angular momentum.
The first exact numerical results for the emission of Hawking radiation
from a $(4+n)$-dimensional rotating black hole, in the form of scalar fields,
as well as for the energy amplification due to the superradiance effect,
were presented in \cite{HK2}. There it was shown that the energy emission
is greatly enhanced by both the dimensionality of spacetime and the angular
momentum of the black hole. In \cite{IOP2}, results on the power (energy)
spectrum, as well as its angular distribution in the 5-dimensional case were
presented, but once more the analysis relied on the assumption of low energy
and low angular momentum. Results on the superradiance effect in the case of
a higher-dimensional black hole were also presented in \cite{IOP-proc} and
\cite{Jung-super}, while the emission in the bulk was addressed in
\cite{Nomura, Jung-bulk}.

Recently, a comprehensive study of the emission of Hawking radiation from a
higher-dimensional black hole in its spin-down phase appeared in the literature
\cite{DHKW}. This work focused on the emission of scalar fields on
the brane, and presented exact results for the fluxes of particles, energy and
angular momentum, for arbitrary values of the dimensionality of spacetime,
angular momentum of the black hole, and energy of the emitted particle.
Also, the angular distribution of the emitted radiation -- a distinctive signature
of emission from a rotating black hole -- was investigated: it was demonstrated
that the uniform distribution of particles in the case of a spherically-symmetric
black hole is now replaced by a clearly oriented one that is transverse to the
rotation axis of the black hole.

As the techniques used in the aforementioned work were strongly spin-dependent,
the analysis for the emission of higher-spin fields was left for subsequent
works. Here, we undertake the effort to fulfil this task for the case of
gauge fields. As these fields are by definition restricted to live on the
brane, only the brane emission channel will be studied. To this end, we will
use semiclassical techniques that were originally developed for the case of
4-dimensional black holes and are here generalized to
cover the case of higher-dimensional black hole line-elements projected onto
a brane. As these techniques were developed particularly for the case of spin-1
particles, we are forced to leave the study of fermions for another work.

The outline of our paper is as follows: in Section 2, we present the theoretical
framework for our analysis that includes the form of the gravitational background,
and the derivation of the radial and angular part of the spin-1 equation of motion
on the brane. In Section 3, we review, and generalize where necessary, the principles
leading to the expressions for the transmission (or absorption) coefficient and of
the various Hawking radiation emission fluxes. The techniques used to numerically
integrate both the angular and radial parts of the spin-1 equation of motion -- to
derive the angular eigenvalue, the spin-weighted spheroidal harmonics, and the
radial function -- are described in detail in Section 4.
The readers who are not interested in numerical techniques can move directly to
Section 5, where our results are presented. Subsection 5.1 presents our results
for the transmission coefficient and its behaviour in terms of the number of extra
dimensions and angular momentum of the black hole. Subsections 5.2, 5.3 and 5.4 are
devoted to the computation of the particle, energy and angular momentum fluxes,
respectively: their dependence on both the dimensionality of spacetime and
angular momentum of the black hole is investigated and, in the case of the first
two types of fluxes, their angular distribution is also studied. Subsection 5.5
presents the total emissivities from a higher-dimensional rotating black hole
on the brane in terms of the angular momentum, number of additional dimensions
and azimuthal angle. The final subsection 5.6 addresses the issue of superradiance,
and the energy amplification is computed in terms of the angular momentum and
dimensionality of spacetime. We finish with a summary of our results and conclusions,
in Section 6.


\section{Gauge Field in a Brane-Induced Rotating Black Hole Background}
\label{qft}

As is well known, the generalization of the 4-dimensional, Kerr black hole background
to a rotating, uncharged black hole living in a $(4+n)$-dimensional spacetime is given
by the Myers-Perry  solution \cite{MP}. Under the assumption that all ordinary matter --
including ourselves -- is restricted to live on a brane embedded in this background,
we are interested in the type of line-element that describes the geometry on our
brane. This follows by projecting the $(4+n)$-dimensional Myers-Perry background
onto the brane, and is found to have the form \cite{Kanti}
\begin{equation}
\begin{split}
ds^2=\left(1-\frac{\mu}{\Sigma\,r^{n-1}}\right)dt^2&+\frac{2 a\mu\sin^2\theta}
{\Sigma\,r^{n-1}}\,dt\,d\varphi-\frac{\Sigma}{\Delta}dr^2 \\[3mm] &\hspace*{-1cm}
-\Sigma\,d\theta^2-\left(r^2+a^2+\frac{a^2\mu\sin^2\theta}{\Sigma\,r^{n-1}}\right)
\sin^2\theta\,d\varphi^2\,,
\end{split} \label{induced}
\end{equation}
where
\begin{equation}
\Delta = r^2 + a^2 -\frac{\mu}{r^{n-1}}\,, \qquad
\Sigma=r^2 +a^2\,\cos^2\theta\,.
\label{master}
\end{equation}
The mass and angular momentum (transverse to the $r \varphi$-plane) of the black hole
are then given by
\begin{equation}
M_{BH}=\frac{(n+2) A_{n+2}}{16 \pi G}\,\mu\,,  \qquad
J=\frac{2}{n+2}\,M_{BH}\,a\,, \label{def}
\end{equation}
with $G$ being the $(4+n)$-dimensional Newton's constant, and $A_{n+2}$
the area of an $(n+2)$-dimensional unit sphere, given by
\begin{equation}
A_{n+2}=\frac{2 \pi^{(n+3)/2}}{\Gamma[(n+3)/2]}\,.
\end{equation}
Above, we have assumed that the ($4+n$)-dimensional black hole was
created by the collision of two highly energetic particles moving on our 3-brane.
Assuming that the thickness of our brane is much smaller than the size $L$ of the
extra dimensions, the two particles will have a non-zero impact parameter only
along our brane; the produced black hole will therefore acquire only one non-zero
angular parameter about an axis (the one transverse to the $r \varphi$-plane) in
our brane.

The black hole horizon is given by solving the equation $\Delta(r)=0$, which, for
$n\geq1$, leads to a unique solution given by
\begin{equation}
r_{h}^{n+1}=\frac{\mu}{1+a_*^2}\,,
\label{horizon}
\end{equation}
where we have defined $a_*=a/r_{h}$. We should note here that while for $n=0$ and
$n=1$, there is a maximum possible value of $a$ that guarantees the existence of
a real solution to the equation $\Delta=0$, for $n>1$ there is no fundamental
upper bound on $a$ and a horizon $r_{h}$ always exists. An upper bound can
nevertheless be imposed on the angular momentum parameter of the black hole by
demanding the creation itself of the black hole from the collision of the two
particles. The maximum value of the impact parameter between the two particles
that can lead to the creation of a black hole was found to be \cite{Harris}
\begin{equation}
b_\text{max}=2 \,\biggl[1+\biggl(\frac{n+2}{2}\biggr)^2\biggr]^{-\frac{1}{(n+1)}}
\mu^{\frac{1}{(n+1)}}\,,
\end{equation}
an analytic expression that is in very good agreement with the numerical results
produced in \cite{creation}c. Then, by writing $J=b M_{BH}/2$ \cite{IOP1}, for the
angular momentum of the black hole, and using Eq. (\ref{horizon}) and the second of
Eq. (\ref{def}), we obtain
\begin{equation}
a^\text{max}_*=\frac{n+2}{2}\,.
\label{amax}
\end{equation}

We now proceed to derive the equation of motion of a gauge field in the
induced-on-the-brane gravitational background (\ref{induced}).
The quantization of an electromagnetic field on a rotating black hole background
is considerably more involved than for a non-rotating black hole. While there is
an extensive literature on the quantization of a scalar field on a Kerr black hole
spacetime, rather less work has been done on spin-1 fields. For example, the first
comprehensive study of the stress-energy tensor on the entire geometry has only been
recently undertaken \cite{Casals:2005kr}. However, the formalism itself has been
sufficiently well developed for our purposes for Kerr black holes
\cite{Casals:2005kr,Teukolsky:1972my,Teukolsky:1973ha,Chandrasekhar:1985kt,
Chrzanowski:1975wv,Chrzanowski:1976jb,Cohen:1974cm,Wald:1978vm,press}
and, here, we shall extend this to the more general background metric (\ref{induced}).
In this section, we will only outline the key features of the theory, referring the
interested reader to \cite{Casals:2005kr} (and references therein) for further details.

We use the Newman-Penrose formalism \cite{Newman:1961qr} to derive the equations for
the electromagnetic field components, written in terms of the Newman-Penrose scalars:
\begin{equation}
\phi_{-1} \equiv F_{{\bmath{l}}{\bmath{m}}}\,;
\qquad
\phi_{0} \equiv \left(F_{{\bmath{l}}{\bmath{n}}}+F_{{\bmath{m}}^*{\bmath{m}}}\right)/2\,;
\qquad
\phi_{+1} \equiv F_{{\bmath{m}}^*{\bmath{n}}}\,,
\label{NPscalars}
\end{equation}
where $\{{\bmath{l}},{\bmath{n}},{\bmath{m}},{\bmath{m}}^*\}$ are the null basis vectors.
Here, we follow the notation of \cite{Casals:2005kr,Carter:1987hk,Price:1971fb}, and
use the subscript $h$ on $\phi$, taking the values (0, $\pm$1), to describe the
helicity of the spin-1 field -- note that both helicities ($h=\pm 1$) will contribute to
the Hawking radiation. We should emphasize that, in our terminology, helicity is in
effect a label for the field modes, and we do not strictly link this to any particular
physical field configuration. This latter point is explored in more detail in
\cite{Casals:2005kr,CasalsPhD}.

The tetrad we use is the Kinnersley tetrad \cite{Chandrasekhar:1985kt}, which readily
generalizes to the metric (\ref{induced}) with $\Delta $ given by Eq. (\ref{master})
rather than the usual Kerr metric form.{\footnote{See \cite{Kanti,IOP1}
for details of the Newman-Penrose coefficients in this modified tetrad.}}
The analysis of Teukolsky \cite{Teukolsky:1972my,Teukolsky:1973ha} showed that the
partial differential equations for $\phi _{-1}$ and $\rho ^{-2}\phi _{+1}$
[where $\rho=-1/(r-ia\cos\theta)$ is one of the spin coefficients] are in this case
separable.
It is straightforward to show that this result also extends to our more general situation.
The Maxwell scalars (\ref{NPscalars}) can be decomposed into a Fourier mode sum
\begin{equation}
\phi _{h} =\int_{-\infty}^{+\infty}
\, d{\omega}\sum_{\ell =|h|}^{+\infty}
\,\sum_{m=-\ell }^{+\ell }\,\sum_{P=\pm 1}
{}_{h}a _{\ell m\omega P} \,
{}_{h}\phi _{\ell m\omega}\,,
\label{phifourier}
\end{equation}
where ${}_{h}a_{\ell m\omega P}$ are the coefficients of the Fourier series.
The parameter $\ell$ labels the eigenvalues of the angular Teukolsky equation
(see Eq. (\ref{angular}) below).
The parameter $P$ is sum\-med in (\ref{phifourier}) over the values $+1$ and $-1$,
corresponding to the two linearly independent polarization states of the electromagnetic
potential.
An electromagnetic potential mode of polarization $P$ is an eigenstate of
the parity operator
\begin{equation}
\mathcal{P}:(\theta,\phi)\to(\pi-\theta,\phi+\pi)\,.
\label{parity}
\end{equation}
with eigenvalue $P$.
The decomposition of the potential into eigenstates of the parity operator $\mathcal{P}$
is the natural choice due to the invariance of the metric (\ref{induced}) under this
operation. The Fourier modes in (\ref{phifourier}) are then separated according to
\begin{equation}
{}_{\pm 1}\phi _{\ell m\omega} = {}_{\pm 1}N_{\ell m \omega}\,\rho^{(1\pm1)}\,
{}_{\mp 1}R_{\ell m\omega}(r)\,{}_{\mp 1}Z_{\ell m\omega}(\theta ,\varphi )\,
e^{-i\omega t}\,,
\label{separation}
\end{equation}
where ${}_{\pm 1}N_{\ell m \omega}$ is a suitable normalization constant,
${}_{\mp 1}R_{\ell m\omega }(r)$ is a radial function
and ${}_{\mp 1}Z_{\ell m\omega}(\theta ,\varphi )$ is the angular function.
The equation for the Maxwell scalar $\phi _{0}$ is not separable; instead,
this scalar can be written in terms of the radial and angular functions for
$h=\pm1$. The answer is given explicitly in
\cite{Casals:2005kr,Chandrasekhar:1985kt,Chrzanowski:1976jb} for Kerr
black holes ($n=0$), but here we do not need it further.
The Fourier modes of the electromagnetic potential
can be reconstructed from the Maxwell scalars
${}_{\pm 1}\phi _{\ell m \omega}$ -- explicit formulae for this are given in
\cite{Casals:2005kr,Chrzanowski:1975wv}, again for Kerr black holes.

The angular function ${}_{h}Z_{\ell m\omega}(\theta, \varphi)$ is defined as
\begin{equation}
{}_{h}Z_{\ell m\omega}(\theta,\varphi)\equiv
\frac{(-1)^{m+1}}{\sqrt{2\pi}}\,{}_{h}S_{\ell m\omega}(\theta)\,e^{+im\varphi}\,,
\end{equation}
where ${}_{h}S_{\ell m\omega}(\theta)$ is the spin-weighted spheroidal harmonic
\cite{Casals:2004zq,press1,staro,fackerell,Seidel:1988ue,breuer,brw}.
Introducing the new variable $x=\cos \theta$, the latter
satisfies the equation
\begin{equation}
\left[
\frac{d}{dx}\left((1-x^{2})\frac{d}{dx}\right)+a^2 \omega^{2}x^{2}-2h
a \omega x-\frac{(m+h x)^{2}}{1-x^{2}}+{}_{h}\mathcal{A}_{\ell m\omega}+h
\right] {}_{h}S_{\ell m\omega}(x)=0\,.
\label{angular}
\end{equation}
In the above, ${}_{h}\mathcal{A}_{\ell m\omega}$ is the separation constant,
usually written as
\begin{equation}
{}_{h}\mathcal{A}_{\ell m\omega} = {}_{h}\lambda_{\ell m\omega}
-a^2 \omega^2+2m a \omega,
\label{angeigen}
\end{equation}
with ${}_{h}\lambda_{\ell m\omega}$ regarded as the eigenvalue of the equation
(\ref{angular}). The spin-weighted spheroidal harmonics satisfy the following
symmetry relations:
\begin{eqnarray}
{}_{h}S_{\ell m\omega}(\theta) & = &
(-1)^{\ell +m}\,{}_{-h}S_{\ell m\omega}(\pi-\theta)
\nonumber \\
{}_{h}S_{\ell m\omega}(\theta) & = &
(-1)^{\ell+h}\,{}_{h}S_{\ell -m-\omega}(\pi-\theta)
\nonumber \\
{}_{h}S_{\ell m\omega}(\theta) & = &
(-1)^{h+m}\,{}_{-h}S_{\ell -m-\omega}(\theta)\,,
\label{Ssymm}
\end{eqnarray}
and are normalized according to
\begin{equation}
\int _{0}^{\pi } d\theta \sin\theta \, {}_{h}S_{\ell m\omega}^2(\theta)=1.
\label{S-normalization}
\end{equation}
The solution of the angular equation (\ref{angular}) is discussed in Section
\ref{numerical}.

The radial function ${}_{h}R_{\ell m\omega }(r)$, on the other hand, is found to satisfy
the following equation
\begin{equation}
\Delta ^{-h}\,
\frac {d}{dr}
\left(\Delta^{h+1}\,
\frac {d{}_{h}R_{\ell m\omega}}{dr}\right)+{}_{h}V\,{}_{h}R_{\ell m\omega}=0\,.
\label{radial}
\end{equation}
For $h=+1$, the potential ${}_{h}V={}_{+1}V$ is given by \cite{Kanti,IOP1}
\begin{equation}
\label{radialpoth+}
{}_{+1}V=\frac{K^2-iK\Delta '}{\Delta}+4i \omega r
+\left( \Delta '' -2 \right)
-{}_{+1}\lambda_{\ell m\omega}\,,
\end{equation}
where we have defined: $K=(r^2+a^2)\,\omega-am$.
Throughout this work, we use a prime $(')$ to denote a derivative with respect to
the radial coordinate $r$. For both Kerr ($n=0$) and 5-dimensional Kerr-like black
holes of the form given in Eq. (\ref{induced}), the term $\Delta ''-2$ vanishes
identically. The $h=+1$ radial equation (\ref{radial}) with the potential
(\ref{radialpoth+}) fits the pattern of the master equation found for various
spins in \cite{Kanti,IOP1}.
However, there is a rather surprising result for helicity $h=-1$, when the
radial function ${}_{-1}R_{\ell m\omega }(r)$ still satisfies (\ref{radial})
but the potential ${}_{-1}V$ has a different form:
\begin{equation}
\label{radialpoth-}
{}_{-1}V=\frac{K^2+iK\Delta '}{\Delta}-4i \omega r
-{}_{-1}\lambda_{\ell m\omega} .
\end{equation}
Note that the $\Delta ''-2$ term is absent in this case. Thus, while the master
equation in \cite{Kanti,IOP1} describes the helicity $h=+1$ radial
functions, it does not describe the helicity $h=-1$ radial functions. We suspect
that something similar may happen with spin-1/2 fields as well. The solution of
the radial equation (\ref{radial}) is also described in detail in Section
\ref{numerical}.

\section{Hawking radiation in the form of gauge fields}
\label{Hawking}
For an evaporating black hole on the brane, the relevant quantum state is the (``past'')
Unruh vacuum $|U^{-}\rangle $. There are some subtleties in the construction of
this state for an electromagnetic field on a rotating black hole background which
are described in detail in \cite{Casals:2005kr}.
The field is expanded in terms of the usual ``in'' and ``up'' modes, for which the
radial functions ${}_{h}R_{\ell m\omega }(r)$ have the asymptotic forms \cite{Casals:2005kr}:
\begin{eqnarray}
{}_{h}R^{\text{in}}_{\ell m\omega} & \sim &
\begin{cases}
\label{Rin}
{}_{h}R^{\text{in,tra}}_{\ell m\omega}\,
\Delta^{-h}\,e^{-i\tilde{\omega}r_*} & (r\rightarrow r_h) \\[2mm]
{}_{h}R^{\text{in,inc}}_{\ell m\omega}\,\frac{\textstyle e^{-i\omega r_*}}
{\textstyle r}
+{}_{h}R^{\text{in,ref}}_{\ell m\omega}\,\frac{\textstyle e^{+i\omega r_*}}
{\textstyle r^{1+2h}} &
(r\rightarrow +\infty)
\end{cases}
\\[3mm]
{}_{h}R^{\text{up}}_{\ell m\omega} & \sim &
\begin{cases}
\label{eq:R_up}
{}_{h}R^{\text{up,inc}}_{\ell m\omega}\,e^{+i\tilde{\omega}r_*}
+{}_{h}R^{\text{up,ref}}_{\ell m\omega}\,\Delta^{-h}\,e^{-i\tilde{\omega}r_*} &
\qquad (r\rightarrow r_h) \\[2mm]
{}_{h}R^{\text{up,tra}}_{\ell m\omega}\,\frac{\textstyle e^{+i\omega r_*}}
{\textstyle r^{1+2h}} &
\qquad (r\rightarrow +\infty)\,.
\end{cases}
\end{eqnarray}
In the above, we have used the abbreviation
\begin{equation}
{\tilde {\omega }} = \omega - m\Omega ,
\end{equation}
with $\Omega $ the horizon angular velocity
\begin{equation}
\Omega=\frac{a_*}{(1+a_*^2)\,r_h}\,,
\label{angvel}
\end{equation}
and the usual tortoise coordinate $r_{*}$ defined through the relation
\begin{equation}
\frac{dr_*}{dr}=\frac{r^2+a^2}{\Delta(r)}\,.
\label{tortoise}
\end{equation}
The coefficients ${}_{h}R^{\text{in,tra}}_{\ell m\omega}$,
${}_{h}R^{\text{in,inc}}_{\ell m\omega}$, ${}_{h}R^{\text{in,ref}}_{\ell m\omega}$,
${}_{h}R^{\text{up,inc}}_{\ell m\omega}$, ${}_{h}R^{\text{up,ref}}_{\ell m\omega}$,
${}_{h}R^{\text{up,tra}}_{\ell m\omega}$ appearing in Eqs. (\ref{Rin}) and
(\ref{eq:R_up}) are arbitrary integration constants.
The correct expression for the unrenormalized expectation value of the stress-energy
tensor in the state $|U^{-}\rangle $ is given in terms of a mode sum as follows
\cite{Casals:2005kr}:
\begin{eqnarray}
\hspace*{-0.7cm}\langle U^{-} | T_{\mu \nu } | U^{-} \rangle & = &
\frac{1}{2}\sum_{\ell mP}
\left(
\int_0^{\infty} d {\tilde{\omega}}\,
\coth\left(\frac{\pi\tilde{\omega}}{\kappa_+}\right) \right.
\nonumber \\ & & \hspace*{-0.5cm} \qquad \qquad \times
\left\{T_{\mu\nu}
\left[{}_{h}\phi_{\ell m\omega}^{\text{up}},
{}_{h}\phi_{\ell m\omega}^{\text{up} *}\right]
+(-1)^{\vartheta}\mathcal{P} \left(T_{\mu\nu}
\left[{}_{h}\phi_{\ell m\omega}^{\text{up}},
{}_{h}\phi_{\ell m\omega}^{\text{up} *}\right]
\right)\right\}
\nonumber \\ & &
\left. +\int_0^{\infty} d{\omega}\,
\left\{T_{\mu\nu}
\left[{}_{h}\phi_{\ell m\omega}^{\text{in}},
{}_{h}\phi_{\ell m\omega}^{\text{in} *}\right]
+(-1)^{\vartheta}\mathcal{P}\left(T_{\mu\nu}
\left[{}_{h}\phi_{\ell m\omega}^{\text{in}},
{}_{h}\phi_{\ell m\omega}^{\text{in} *}\right]
\right)\right\}
\right),
\label{STforU-}
\end{eqnarray}
where the quantity
$T_{\mu\nu} \left[{}_{h}\phi_{\ell m\omega}, {}_{h}\phi_{\ell m\omega}^{*}\right] $
is the classical electromagnetic stress-energy tensor
\begin{eqnarray}
T_{\mu\nu} \left[{}_{h}\phi_{\ell m\omega}, {}_{h}\phi_{\ell m\omega}^{*}\right]
&= &
\left\{
{}_{-1}\phi_{\ell m \omega} \,
{}_{-1}\phi_{\ell m \omega}^*{\bmath {n}}_{\mu}{\bmath {n}}_{\nu}
+2 \, {}_{0}\phi_{\ell m \omega} \, {}_{0}\phi_{\ell m \omega}^*
\left[{\bmath {l}}_{(\mu} {\bmath {n}}_{\nu)}+
{\bmath {m}}_{(\mu}{\bmath {m}}_{\nu)}^*\right]
\right. \nonumber \\ & &
+_{+1}\phi_{\ell m \omega}\, {}_{+1}\phi_{\ell m \omega}^*{\bmath {l}}_{\mu}{\bmath {l}}_{\nu}
-4 \, {}_{0}\phi_{\ell m \omega} \, {}_{-1}\phi_{\ell m \omega}^*
{\bmath {n}}_{(\mu}{\bmath {m}}_{\nu)}
\nonumber \\  & & \left.
-4\, {}_{+1}\phi_{\ell m \omega} \, {}_{0}\phi_{\ell m \omega}^*
{\bmath {l}}_{(\mu}{\bmath {m}}_{\nu)}
+2 \, {}_{+1}\phi_{\ell m \omega} \,{}_{-1}\phi_{\ell m \omega}^*
{\bmath {m}}_{\mu}{\bmath {m}}_{\nu}\right\}
\nonumber \\ & &
+{\mbox {complex conjugate}},
\end{eqnarray}
given in terms of the tetrad basis vectors.
In (\ref{STforU-}), the variable $\vartheta $ has been defined so that $(-1)^{\vartheta }$
is equal to $-1$ if one index of the component of the stress tensor is $\theta $
and the other one is not, and it is equal to $+1$ otherwise.
The key feature of the unrenormalized $\langle U^{-} | T_{\mu \nu } | U^{-} \rangle $
(\ref{STforU-}) is the term involving the parity operator ${\mathcal {P}}$ (\ref{parity}).
This ensures that the stress-energy tensor as a whole is invariant under the
parity operator.

Our interest in this paper is in the outgoing fluxes of particles $N$, energy $E$ and
angular momentum $J$, which means that we do not need to compute all the components
of the renormalized stress-energy tensor.
We require only the components $T_{tr}$ and $T_{r\varphi }$, and our analysis is further
simplified by the fact that these components do not require renormalization.
This latter fact is proved for a quantum scalar field in \cite{Frolov:1989jh}.
Their argument is, however, purely geometric, and can readily be extended to the
case of an electromagnetic field, thereby showing that the counterterms in
Ref. \cite{Christensen:1978yd} vanish for these components of the stress-energy tensor.
In addition, a similar argument can be used to show that the linearly divergent term
\cite{Jensen:1988sf} also vanishes for these components.
We may then use the results \cite{Casals:2005kr,Frolov:1989jh}
\begin{eqnarray}
\frac {dE}{dt} & = &
\int _{S_{\infty }} d\theta \, d\varphi \, r^{2} \sin \theta \,
\langle U^{-} | T^{tr} | U^{-} \rangle ,
\nonumber \\
\frac {dJ}{dt} & = &
\int _{S_{\infty }} d\theta \, d\varphi \, r^{2} \sin \theta \,
\langle U^{-} | T^{r}_{\varphi } | U^{-} \rangle ,
\end{eqnarray}
where the integration is taken over the sphere at infinity.
The contribution to $T_{tr}$ for an ``up'' mode takes the form:
\begin{eqnarray}
\label{Ttrup}
& &
\Delta \left(T_{tr}{}_{\ell m \omega }^{\text{up}}
+\mathcal{P}T_{tr}{}_{\ell m \omega }^{\text{up}}\right) =
\nonumber \\ & & \qquad \qquad
\frac{\mathbb{T}_{\ell m\omega}}{4\pi^2\Sigma}
\Big\{ -\omega\Sigma\left({}_{-1}S_{\ell m\omega}^2+{}_{+1}S_{\ell m\omega}^2\right)
+\frac{a^3\cos\theta\sin^2\theta}{\Sigma}\left({}_{-1}S_{\ell m\omega}^2
-{}_{+1}S_{\ell m\omega}^2\right)
 \nonumber \\ & &  \qquad \qquad
+a\sin\theta\left({}_{-1}S_{\ell m\omega}\partial_{\theta}{}_{-1}S_{\ell m\omega}
-{}_{+1}S_{\ell m\omega}\partial_{\theta}{}_{+1}S_{\ell m\omega}\right)\Big\} .
\end{eqnarray}
In the above equation, ${\mathbb {T}}_{\ell m \omega }$ is the transmission coefficient,
defined as a ratio of energy fluxes:
\begin{equation}
\mathbb{T}_{\ell m\omega} =
\frac{d E^{\text{(tra)}}_{\ell m\omega}/dt}{d E^{\text{(inc)}}_{\ell m\omega}/dt} ,
\end{equation}
with $E^{\text{(inc)}}$ and $E^{\text {(tra)}}$ denoting, respectively, the incident
energy on the black hole and the energy transmitted down the event horizon.
The quantity ${\mathbb {T}}_{\ell m \omega }$ is the analogue of the ``absorption
probability'' defined in \cite{DHKW}.
The transmission coefficient ${\mathbb {T}}_{\ell m \omega }$ can be written in terms
of $W[{}_{+1}R,{}_{-1}R^{*}]^{\text{in}}_{\ell m\omega}$, the Wronskian of the ``in'' mode
radial functions, as follows \cite{Casals:2005kr}:
\begin{equation}
\mathbb{T}_{\ell m\omega}
=\frac{-i}{2^4\omega^3}W[{}_{+1}R,{}_{-1}R^{*}]^{\text{in}}_{\ell m\omega} ,
\label{greybody}
\end{equation}
where we have chosen the normalization
$ {}_{-1}R^{\text{in,inc}}_{\ell m\omega}={}_{-1}R^{\text{up,inc}}_{\ell m\omega}=1$.
The contribution to $T_{tr}$ for an ``in'' mode is the same as for an ``up'' mode
but with a change of sign. There are also similar expressions describing the
contributions to $T_{r\varphi}$ for both the ``in'' and ``up'' modes.

For the fluxes in which we are interested, only the behaviour of the stress tensor at
infinity is needed. As $r\rightarrow \infty $, the expression (\ref{Ttrup})
simplifies to:
\begin{equation}
\left(T_{tr}{}_{\ell m \omega }^{\text{up}}
+\mathcal{P}T_{tr}{}_{\ell m \omega }^{\text{up}}\right) \sim
-\frac{\mathbb{T}_{\ell m\omega}}{4\pi^2r^{2}}
 \omega \left({}_{-1}S_{\ell m\omega}^2+{}_{+1}S_{\ell m\omega}^2\right)\,.
 \label{Ttrinfang}
\end{equation}
It is important to note that both the positive and negative helicity angular functions
contribute to this expression.
In \cite{IOP1}, only one helicity was included for each mode, which
resulted in the angular distributions in that paper being highly asymmetric about
the equator. By using the correct expression given above, we will find later that
the angular distributions are in fact symmetric about the equator, as would be
expected from a physical point of view.

We now use (\ref{Ttrinfang}) and the corresponding result for $T_{r\varphi }$ to
find the following formulae for the
differential emission rates of particles, energy and angular momentum, integrated over
all angles $\theta $, as a function of the mode energy $\omega $:
\begin{eqnarray}
\frac {d^{2}N}{dt  d\omega}  & = &
\frac {1}{2\pi} \sum _{\ell =1}^{\infty }
\sum _{m=-\ell }^{\ell } \sum _{P=\pm 1}
\frac {1}{\exp\left(\tilde{\omega}/T_\text{H}\right) - 1}
{\mathbb {T}}_{\ell m \omega }\,,
\label{flux} \\
\frac {d^{2}E}{dt  d\omega }  &  = &
\frac {1}{2\pi} \sum _{\ell =1}^{\infty }
\sum _{m=-\ell }^{\ell } \sum _{P=\pm 1}
\frac {\omega }{\exp\left(\tilde{\omega}/T_\text{H}\right) - 1}
{\mathbb {T}}_{\ell m \omega }\,,
\label{power} \\
\frac {d^{2}J}{dt  d\omega}  & = &
\frac {1}{2\pi} \sum _{\ell =1}^{\infty }
\sum _{m=-\ell }^{\ell } \sum _{P=\pm 1}
\frac {m}{\exp\left(\tilde{\omega}/T_\text{H}\right) - 1}
{\mathbb {T}}_{\ell m \omega }\,,
\label{ang-mom}
\end{eqnarray}
as well as the angular distributions of particles and energy:
\begin{eqnarray}
\frac {d^{3}N}{d(\cos\theta)dt  d\omega}  & = &
\frac {1}{4\pi} \sum _{\ell =1}^{\infty }
\sum _{m=-\ell }^{\ell } \sum _{P=\pm 1}
\frac {1}{\exp\left(\tilde{\omega}/T_\text{H}\right) - 1}
{\mathbb {T}}_{\ell m \omega }
 \left({}_{-1}S_{\ell m\omega}^2+{}_{+1}S_{\ell m\omega}^2\right)\,,
\label{fluxang} \\
\frac {d^{3}E}{d(\cos\theta)dt  d\omega }  &  = &
\frac {1}{4\pi} \sum _{\ell =1}^{\infty }
\sum _{m=-\ell }^{\ell } \sum _{P=\pm 1}
\frac {\omega }{\exp\left(\tilde{\omega}/T_\text{H}\right) - 1}
{\mathbb {T}}_{\ell m \omega }
\left({}_{-1}S_{\ell m\omega}^2+{}_{+1}S_{\ell m\omega}^2\right)\,,
\label{powerang}
\end{eqnarray}
where $T_\text{H}$ is the Hawking temperature of the $(4+n)$-dimensional,
rotating black hole, given by
\begin{equation}
T_\text{H}=\frac{(n+1)+(n-1)\,a_*^2}{4\pi\,(1+a_*^2)\,r_{h}}\,.
\label{temp}
\end{equation}
An important part of the above expressions is the sum over the polarization $P$,
which is in agreement with \cite{Page:1976ki}.
The transmission coefficient ${\mathbb {T}}_{\ell m \omega }$ is independent
of $P$, so in effect we simply multiply by a factor of 2.
As in \cite{DHKW}, we do not study the angular behaviour of $T_{r\varphi }$ (corresponding
to an angular distribution of the flux of angular momentum) as it will not be directly
observable. However, for the interested reader, $T_{r\varphi }$ is studied for the
4-dimensional Kerr black hole in \cite{Casals:2005kr}.

Let us finally note here that the emission of Hawking radiation will be assumed to
take place through the emission of gauge particles with energy much smaller than the
mass of the black hole. This will guarantee that there is no significant back reaction
to the gravitational background due to the change in the black hole mass after the
emission of a particle. Since the black hole radiation spectrum is centered around
the temperature of the black hole, we must demand that $T_\text{H} \ll M_{BH}$.
This eventually translates to $M_{BH} \gg M_*$, a condition that has already been
imposed to guarantee the absence of any quantum corrections to our analysis.


\section{Numerical Analysis}
\label{numerical}

The greybody factor, or transmission coefficient ${\mathbb {T}}_{\ell m \omega }$, given
by Eq. (\ref{greybody}), can be calculated from the solutions of the radial Teukolsky
equation (\ref{radial}) for propagation on the brane black hole background; however,
we still need to solve the angular equation (\ref{angular}) since its eigenvalue appears
in the radial equation. We will also require the spin-weighted spheroidal harmonics
themselves for the angular distributions of the particle and energy fluxes (\ref{fluxang},
\ref{powerang}). In this section, we briefly outline the numerical methods used to
solve both the angular and radial equations of a spin-1 field in the aforementioned
black hole background.

In order to solve the angular equation, we use the shooting method as described in
\cite{Casals:2004zq}. This method is based on the one applied in \cite{DHKW, Num.Rec.}
to solve the spin-0 spheroidal harmonic equation, and here we adapt it to the non-zero
spin case. We start by using the Frobenius method, and re-write the angular function
as follows:
\begin{equation} \label{eq:asympt. S for x->+/-1}
{}_{h}S_{\ell m\omega}(x)=(1-x)^{\alpha_+}(1+x)^{\alpha_-}{}_{h}y_{\ell m\omega}(x)\,,
\end{equation}
where
\begin{equation}
\begin{aligned}
\alpha_+ &=\frac{|m+h |}{2}\,, & \qquad \qquad
\alpha_-&=\frac{|m-h |}{2}\,.
\end{aligned}
\end{equation}
The function ${}_{h}y_{\ell m\omega}(x)$ then satisfies the differential equation
\begin{equation} \label{eq:ang. teuk. eq. for y}
\begin{aligned}
& \Bigg\{
 (1-x^{2})\frac {d^2}{dx^2}-2\left[\alpha_+-\alpha_-+(\alpha_++\alpha_-+1)x\right]
 \frac {d}{dx}+
{}_{h}\lambda_{\ell m\omega}+h(h+1)-a^2 \omega^2
\\ &
 \qquad \qquad \qquad \quad
+2m a \omega-(\alpha_++\alpha_-)(\alpha_++\alpha_-+1)+a^{2} \omega ^2 x^{2}-2h
a \omega x \Bigg\} {}_{h}y_{\ell m\omega}(x)=0\,.
\end{aligned}
\end{equation}
From Eq. (\ref{eq:ang. teuk. eq. for y}), it is clear that ${}_{h}y''_{\ell m\omega}$
cannot be calculated at exactly the end-points $x=\pm 1$ since these are (regular) singular
points. We will, therefore, integrate this equation from the point $x_1\equiv -1+\epsilon$
up to the point $x_2\equiv +1-\epsilon$, where $\epsilon\ll 1$.

The shooting method for solving Eq. (\ref{eq:ang. teuk. eq. for y}) requires an ansatz,
${}_{h}\hat{\lambda}_{\ell m\omega}$, for the eigenvalue ${}_{h}\lambda_{\ell m\omega}$
of the angular equation. Imposing regularity of the solution at $x=-1$ determines
the values of ${}_{h}y_{\ell m\omega}$ and ${}_{h}y'_{\ell m\omega}$ at $x_1$, up to an
arbitrary overall normalization. The equation is then integrated from $x_1$ to $x_2$,
as an initial value problem. If the solution obtained at $x_2$ does not comply with the
boundary condition of regularity at $x=+1$, the value of
${}_{h}\hat{\lambda}_{\ell m\omega}$ is improved. The differential equation is then
integrated again with the new value of ${}_{h}\hat{\lambda}_{\ell m\omega}$ until regularity at $x=+1$ is met within
the desired accuracy; the derived solution for the spin-weighted spheroidal harmonics
is then normalized so that Eq. (\ref{S-normalization}) is satisfied.

\begin{figure}[t]
\begin{center}
\mbox{
\includegraphics[height=5.5cm,clip]{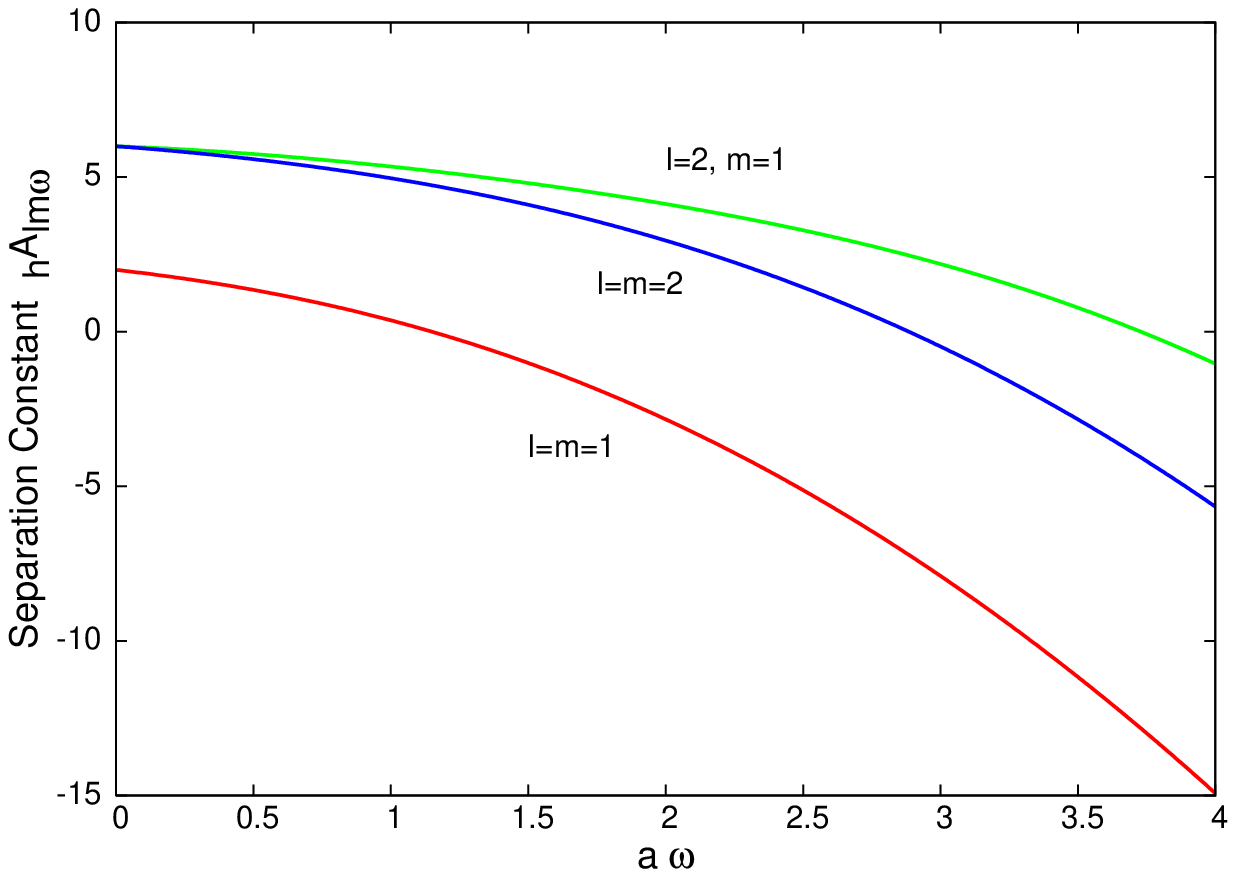}}
{\includegraphics[height=5.5cm,clip]{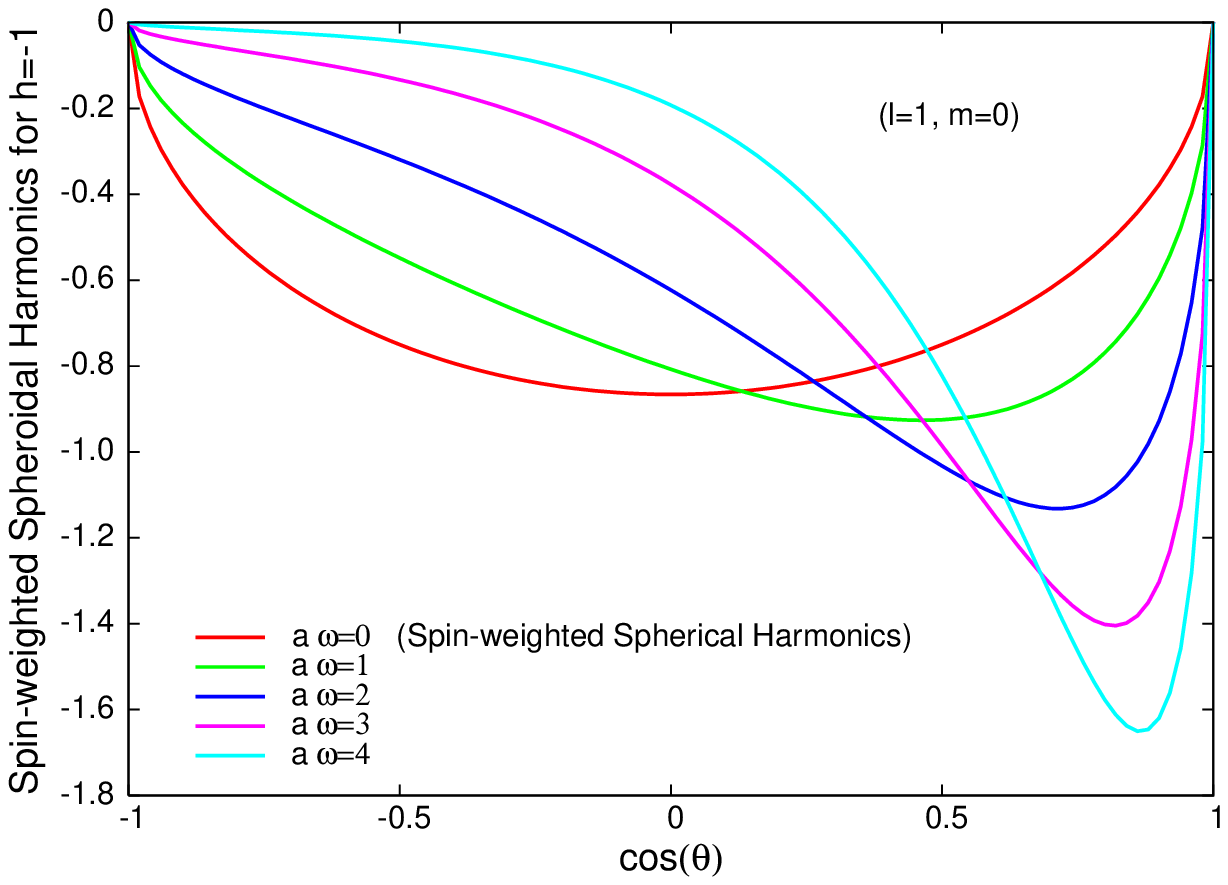}}
\caption{{\bf (a)} The separation constant ${}_{h}\mathcal{A}_{\ell m\omega}(a \omega)$,
for the modes $\ell=m=1,2$ and $\ell=2, m=1$, and helicity $h=-1$; {\bf (b)} Spin-1
weighted spheroidal harmonics for a gauge field with $h=-1$, $\ell=1$, $m=0$, and for
$a \omega=(0,1,2,3)$. }
\label{spheroidal}
\end{center}
\end{figure}

For an ansatz ${}_{h}\hat{\lambda}_{\ell m\omega}$ for the angular
eigenvalue, the numerical solution ${}_{h}y^{\text{num}}_{\ell m\omega}$ of
Eq. (\ref{eq:ang. teuk. eq. for y}) is a combination of both a regular and
an irregular solution, i.e.
\begin{equation}
{}_{h}y^{\text{num}}_{\ell m\omega}=A({}_{h}\hat{\lambda}_{\ell m\omega})\,
{}_{h}y_{\ell m\omega}
+B({}_{h}\hat{\lambda}_{\ell m\omega})\,{}_{h}y^{\text{irreg}}_{\ell m\omega}\,,
\end{equation}
where ${}_{h}y_{\ell m\omega}$ and ${}_{h}y^{\text{irreg}}_{\ell m\omega}$ are the
regular and irregular solutions at $x=\pm 1$, respectively, and $A$ and $B$ unknown
functions of ${}_{h}\hat{\lambda}_{\ell m\omega}$. The value of
${}_{h}\hat{\lambda}_{\ell m\omega}$ must be modified so that only the regular
term $A\,{}_{h}y_{\ell m\omega}$ is retained. For general spin\footnote{In the scalar case,
regularity at $x=+1$ may be imposed by requiring that ${}_{0}\hat{\lambda}_{\ell m\omega}$
is a zero of the function $g({}_{0}\hat{\lambda}_{\ell m\omega})\equiv
{}_{0}y^{' \text{num}}_{\ell m\omega}(x_2)-{}_{0}y'_{\ell m\omega}(x_2)$; this function
clearly tends to zero as ${}_{0}\hat{\lambda}_{\ell m\omega}$ approaches the correct
eigenvalue while it tends to infinity when ${}_{0}\hat{\lambda}_{\ell m\omega}$ is far from
it due to the behaviour of the irregular solution.
This method (see also a variation in \cite{DHKW}) can be used since, for
scalar fields, the analytic value ${}_{h}y'_{\ell m\omega}(x_2)$ is known because
${}_{0}y'_{\ell m\omega}(x)\propto {}_{0}y'_{\ell m\omega}(-x)$.
However, for a non-zero spin field, we have
${}_{h}y'_{\ell m\omega}(x)\propto {}_{-h}y'_{\ell m\omega}(-x)$, and therefore we do
not know the analytic value ${}_{h}y'_{\ell m\omega}(x_2)$ for a particular value
$h\neq 0$ of the helicity; therefore, an alternative method must be used.},
this is achieved by requiring that ${}_{h}\hat{\lambda}_{\ell m\omega}$ is a zero
of the function
\begin{equation}
g({}_{h}\hat{\lambda}_{\ell m\omega})\equiv
{}_{h}y^{' \text{num}}_{\ell m\omega}(x_2)-{}_{h}y^{' \text{approx}}_{\ell m\omega}(x_2)\,,
\end{equation}
where ${}_{h}y^{' \text{approx}}_{\ell m\omega}(x_2)$ is an approximation to the (unknown)
regular value:
\begin{equation}
{}_{h}y_{\ell m\omega}^{' \text{approx}}(x_2)\simeq
\frac{{}_{h}y^{\text{num}}_{\ell m\omega}(x_2)}{{}_{h}y_{\ell m\omega}(x_2)}
\,{}_{h}y'_{\ell m\omega}(x_2)\,.
\end{equation}
The fraction ${}_{h}y'_{\ell m\omega}(x_2)/{}_{h}y_{\ell m\omega}(x_2)$ can be found
analytically by calculating a series expan\-sion of ${}_{h}y_{\ell m\omega}$ around $x=+1$.
This series and further details can be found in \cite{CasalsPhD}.
As an illustrative result, in Fig. \ref{spheroidal}, we depict the value of the
separation constant
${}_{h}\mathcal{A}_{\ell m\omega}$~=~${}_{h}\lambda_{\ell m\omega}+\\2ma\omega-a^2\omega^2$,
and of the spin-1 weighted spheroidal harmonics for various field modes.

Once the eigenvalues of the angular equation (\ref{angular}) are found, we can proceed
to solve the radial equation (\ref{radial}).
The radial Teukolsky equation in four dimensions contains a long-range potential
in the spin-1 case, which cannot readily be integrated numerically.
We use the following definition of long/short-rangeness~\cite{ar:Sasa&Naka'82}:
a second-order differential equation
\begin{equation}
\frac {d^2Y(x)}{dx^2}+{\mathcal {A}}(x)\,
\frac {dY(x)}{dx}+{\mathcal {B}}(x)\,Y(x)=0
\end{equation}
is said to be short-range if, and only if, ${\mathcal {A}}(x)=O(x^{-d})$ and
${\mathcal {B}}(x)=b_{\pm}^2+O(x^{-d})$, where $d\geq 2$ and $b_{\pm}$ are constants,
when $x\to \pm \infty$.
If this condition is guaranteed, then the asymptotic form of the solution is
\begin{equation} \label{eq:asympt. form of short-range sln.}
Y(x) \sim
\begin{cases}
e^{\pm ib_+x}   & (x\to +\infty) \\
e^{\pm ib_-x}   & (x\to -\infty)\,.
\end{cases}
\end{equation}
Detweiler showed \cite{ar:Detw'76,Detw77} that a particular transformation from the
radial function ${}_{-1}R_{\ell m\omega}$ to a new radial function $X_{\ell m\omega}$
leads to a differential equation which is real and contains a short-range potential
in terms of the tortoise co-ordinate $r_*$ (\ref{tortoise}). This new differential
equation, being short-range (in the homogeneous case), is suitable for numerical
integration. This method is followed and detailed in \cite{Casals:2005kr} for the
four dimensional, Kerr-Newman case.

In this work, we have followed Detweiler's method and generalized his differential
equation valid for the Kerr black hole to any 4-dimensional line-element of
the form (\ref{induced}), including the case of a higher-dimensional black hole
background induced on the brane. It can be easily checked that the
Teukolsky-Starobinski\u{\i} identities \cite{Chandrasekhar:1985kt,press}
are valid for a general metric function $\Delta(r)$; in particular,
\begin{equation} \label{eq:Teuk-Starob. id.}
{}_{+1}R_{\ell m\omega}=
2\left( \frac {d}{dr}-\frac{iK}{\Delta}\right)
\left( \frac {d}{dr}-\frac{iK}{\Delta}\right){}_{-1}R_{\ell m\omega}\,.
\end{equation}
It can also be easily checked that the same is true for all the formulae
in \cite{ar:Detw'76} until we reach the point where the potential is written
out explicitly, i.e. Eq. (61) in \cite{ar:Detw'76}. For a general function $\Delta$,
we define a new radial function $X_{\ell m \omega}$ according to
\begin{equation} \label{eq:def. of X}
X_{\ell m\omega}\equiv(r^2+a^2)^{1/2}\Delta^{-1/2}\chi_{\ell m\omega}\,,
\end{equation}
where
\begin{equation} \label{eq:chi as func of Rs and Rs'}
\chi_{\ell m\omega}=\alpha\,{}_{-1}R_{\ell m\omega}+\beta\,
\frac {d{}_{-1}R_{\ell m\omega}}{dr}\,.
\end{equation}
The functions $\alpha $ and $\beta $ appearing in the definition of
$\chi _{\ell m \omega }$ are quite complex. To define them, first write
$\alpha $ and $\beta $ in terms of two new quantities $p$ and $q$ as follows:
\begin{equation}
\alpha=p+a_Dq\,, \qquad \qquad
\beta=b_Dq\,,
\end{equation}
where
\begin{eqnarray}
a_D&=&-\frac{2}{\Delta}\left[2K^2+\Delta(iK'-{}_{-1}\lambda_{\ell m\omega})\right]\,,
\nonumber  \\
b_D&=&\frac{4iK}{\Delta}\,.
\end{eqnarray}
The $p$ and $q$ functions are then themselves defined as
\begin{eqnarray}
p&=&{}_1B_{\ell m\omega}\left(\frac{2K^2}{\Delta}
-{}_{-1}\lambda_{\ell m\omega}+{}_1B_{\ell m\omega}\right)^{-1/2}\,,
\nonumber \\
q&=&\frac{p\Delta}{2{}_1B_{\ell m\omega}}\,,
\end{eqnarray}
where, finally,
\begin{equation}
{}_1B_{lm\omega}^2={}_{-1}\lambda_{lm\omega}^2+4ma\omega-4a^2\omega^2\,.
\end{equation}
The function $X_{\ell m \omega }$ now satisfies the differential equation
\begin{equation} \label{eq:diff. eq. for X,s=-1}
\frac {d^2X_{\ell m\omega}}{dr_*^2}
-{}_{-1}\mathcal{U}X_{\ell m\omega}=0\,,
\end{equation}
with
\begin{equation} \label{eq:potential -1mathcalU}
\begin{aligned}
{}_{-1}\mathcal{U}&=\frac{\Delta{}_{-1}\lambda_{\ell m\omega}-K^2}{(r^2+a^2)^2}
+\frac{\Delta^2(Kp')'}{(r^2+a^2)^2Kp}- \left(\frac{\Delta}{r^2+a^2}\right)^{3/2}
\left[\frac{\Delta^{1/2}}{(r^2+a^2)^{1/2}}\right]''\,.
\end{aligned}
\end{equation}
Note that the signs of $a_D$ and $b_D$ above differ from those in \cite{ar:Detw'76} so
that the sign in the Teukolsky-Starobinski\u{\i} identity (\ref{eq:Teuk-Starob. id.})
agrees with that in \cite{press}.

The explicit form of the potential (\ref{eq:potential -1mathcalU}) given by Detweiler
uses $\Delta=r^2-2Mr+a^2$, and therefore it only applies to the four dimensional,
Kerr background. Here, we have calculated it for a general function $\Delta(r)$,
and the result is:
\begin{equation}
\begin{aligned}
{}_{-1}\mathcal{U}&=-\frac{\left[\omega (r^2+a^2)-am\right]^2}{(r^2+a^2)^2}+
\frac{\Delta{}_{-1}\lambda_{\ell m\omega}}{(r^2+a^2)^2}
\\ &-
\frac{\Delta\left\{ 2(2r^2-a^2)\Delta+(r^2+a^2)\left[\Delta''(r^2+a^2)-
2r\Delta'\right]\right\}}{2(r^2+a^2)^4}\\
&-\frac{\Delta\left\{2(5r^2+\nu^2)\Delta-(r^2+\nu^2)\left[5r\Delta'+
\frac{\Delta''}{2}(r^2+\nu^2)\right]\right\}}{(r^2+a^2)^2\left[(r^2+\nu^2)^2+
\eta\Delta\right]} \\
&+\frac{6 r(r^2+\nu^2)^2\Delta\left[2r\Delta-(r^2+\nu^2)\Delta'\right]}
{(r^2+a^2)^2\left[(r^2+\nu^2)^2+\eta\Delta\right]^2}
-\frac{\eta\Delta \left(\Delta'\right)^2 \left[2(r^2+\nu^2)^2-\eta\Delta \right]}
{4(r^2+a^2)^2\left[(r^2+\nu^2)^2+\eta\Delta\right]^2}\,, \label{pot-U}
\end{aligned}
\end{equation}
where
\begin{equation}
\begin{aligned}
\nu^2 &\equiv a^2-am/\omega\,, \\[1mm]
\eta&\equiv \frac{{}_1B_{\ell m\omega}-{}_{-1}\lambda_{\ell m\omega}}{2\omega^2}\,.
\end{aligned}
\end{equation}
When $\Delta(r)$ is given by Eq. (\ref{master}), the potential ${}_{-1}\mathcal{U}$
is found to be real and short-range, for any number of extra dimensions $n\geq 0$.
That is, the differential equation (\ref{eq:diff. eq. for X,s=-1}) is short-range,
in terms of $r_*$, in both limits $r\to +\infty$ and $r\to r_h$ for any $n\ge 0$.
Furthermore, it can be checked that this differential equation
is short-range, in terms of the independent variable
$r$, in the limit $r\to +\infty$, for $n\geq 1$.
Note, however, that it is \textit{not} short-range in terms of $r$ in the limit
$r\to r_h$ for any $n\geq 0$.

The relation $r_*=r_*(r)$ can be easily found in closed form in the cases $n=0\to 3$
but not in the cases $n=4\to 7$. We, therefore, found it useful to define a new
variable $r_{\Delta}$ such that $r_{\Delta }$ behaves like $r_*$, for $r\to r_h$,
and like $r$, for $r\to +\infty$ (to ensure short-rangeness), and such that we can
find the relation $r_{\Delta}=r_{\Delta}(r)$ in closed form $\forall n\geq 1$.
The differential equation will be short-range (for $n\geq 1$) with respect to this new radial variable,
both for $r\to r_h$ and $r\to +\infty$. A radial variable $r_{\Delta}$ that
satisfies these properties can be defined via
\begin{equation}
\frac {dr_{\Delta}}{dr}\equiv \frac{r}{r-r_h}\,.
\end{equation}
The relation $r_{\Delta}=r_{\Delta}(r)$ in closed form is then given by
\begin{equation}
r_{\Delta}=r+r_h\log\left(\frac{r-r_h}{r_h}\right)\,.
\end{equation}
This relation can be inverted with the use of Lambert's W function~\cite{lambert} to yield
\begin{equation}
r=r_h\left[W\left(\frac{1}{r_h}\,e^{(r_{\Delta}-r_h)/r_h}\right)+1\right]\,.
\end{equation}
Note that $r_{\Delta}\sim r_*$, as $r\to r_h,+\infty$.
If Eq. (\ref{eq:diff. eq. for X,s=-1}) is re-expressed in terms of $r_{\Delta}$, the
resulting differential equation is
\begin{equation} \label{eq:diff. eq. for X,r_d,s=-1}
\frac {d^2X_{\ell m\omega}}{dr_{\Delta}^2}+
\frac{(r-r_h)\left[rF_{\Delta}'(r^2+a^2)+F_{\Delta}(a^2-r^2)\right]}
{r^2F_{\Delta}(r^2+a^2)}\,\frac {dX_{\ell m\omega}}{dr_{\Delta}}
-\left(\frac{r^2+a^2}{rF_{\Delta}}\right)^2{}_{-1}\mathcal{U}(r)X_{\ell m\omega}=0\,,
\end{equation}
where
\begin{equation}
F_{\Delta}\equiv \frac{\Delta}{r-r_h}
\end{equation}
is finite at the horizon. In this work, we numerically solved the differential equation
(\ref{eq:diff. eq. for X,s=-1}) in the cases $n=0\to 3$, and the differential equation
(\ref{eq:diff. eq. for X,r_d,s=-1}) in the cases $n=4\to 7$.

As in \cite{DHKW}, we solved the radial equation (either (\ref{eq:diff. eq. for X,s=-1})
or (\ref{eq:diff. eq. for X,r_d,s=-1}) as applicable), to find the ``in'' modes, for
which the function $X_{\ell m \omega }$ has the asymptotic forms
\begin{equation} \label{eq:X_in}
\begin{aligned}
X^{\text{in}}_{\ell m\omega} & \sim
\begin{cases}
B^{\text{in}}_{\ell m\omega}\,e^{-i\tilde{\omega}r_*} & (r\rightarrow r_h) \\[2mm]
e^{-i\omega r_*}+A^{\text{in}}_{\ell m\omega}\,e^{+i\omega r_*} & (r\rightarrow +\infty)\,.
\end{cases}
\end{aligned}
\end{equation}
Our ultimate aim is to calculate the greybody factor $\mathbb{T}_{\ell m\omega}$.
To find an expression for this in terms of the function $X_{\ell m \omega }$, we first
note that the inverse of relation (\ref{eq:chi as func of Rs and Rs'}) is given
in \cite{ar:Detw'76}, and it is equally valid for a general function $\Delta(r)$:
\begin{equation} \label{eq:Rs,Rs' as funcs. of chi and chi'}
{}_{-1}R_{\ell m\omega}=
\left[\alpha\left(\alpha+\beta'\right)-\beta\left(\alpha'+\frac{\beta}{\Delta}\,
{}_{-1}V\right)\right]^{-1} \left[\left(\alpha+\beta'\right)\chi_{\ell m\omega}-
\beta\,\frac {d\chi_{\ell m\omega}}{dr}\right]\,.
\end{equation}
Using this relation, we can obtain the coefficients of the radial Teukolsky function
${}_{-1}R^{\text{in}}_{\ell m\omega}$ at $r\to +\infty$, which correspond to the
normalization chosen in Eq. (\ref{eq:X_in}):
\begin{equation} \label{eq:R_1 coeffs from X's}
\begin{aligned}
\frac{{}_{-1}R^{\text{in,ref}}_{\ell m\omega}}
{{}_{-1}R^{\text{in,inc}}_{\ell m\omega}A^{\text{in}}_{\ell m\omega}}&=
\frac{4\omega^2}{{}_1B_{\ell m\omega}}\,, &\qquad\quad
{}_{-1}R^{\text{in,inc}}_{\ell m\omega}&=\frac{1}{2^{3/2}|\omega|}\,.
\end{aligned}
\end{equation}
These expressions, together with Eq. (\ref{greybody}), allow us to obtain the following
simple expression for the greybody factor:
\begin{equation} \label{T-final}
\mathbb{T}_{\ell m\omega}
=\frac{\tilde{\omega}}{\omega}\,|B^{\text{in}}_{\ell m\omega}|^2\,.
\end{equation}
By using the above result, we may then proceed to compute the fluxes of particles,
energy and angular momentum, given in Eqs. (\ref{flux}--\ref{powerang}),
by a brane-induced black hole.


\section{Numerical Results}

We now turn to the presentation of the exact numerical results, first, for the
transmission coefficient and, subsequently, for the flux, power and
angular momentum spectra for the emission of gauge fields on the brane from a
higher-dimensional, rotating black hole. As in the case of the emission of scalar fields
\cite{DHKW}, we wish to study the dependence of the various emission rates on the
dimensionality of spacetime $n$ and the angular momentum parameter $a_*$. Also, the
angular distribution of the emitted number of particles and energy will be derived
-- as we will see, the particular shape of the angular distribution comprises a clear
signature not only of the emission during the spin-down phase in the life of the black
hole but also of the type of emitted particles. Finally, the total emissivities of flux,
power and angular momentum, as well as the superradiance effect, will be discussed,
and more particularly their dependence on the two fundamental parameters $n$ and $a_*$.

\begin{figure}[t]
\begin{center}
\begin{tabular}{c c c}
\includegraphics[height=5.5cm,width=5.5cm,clip]{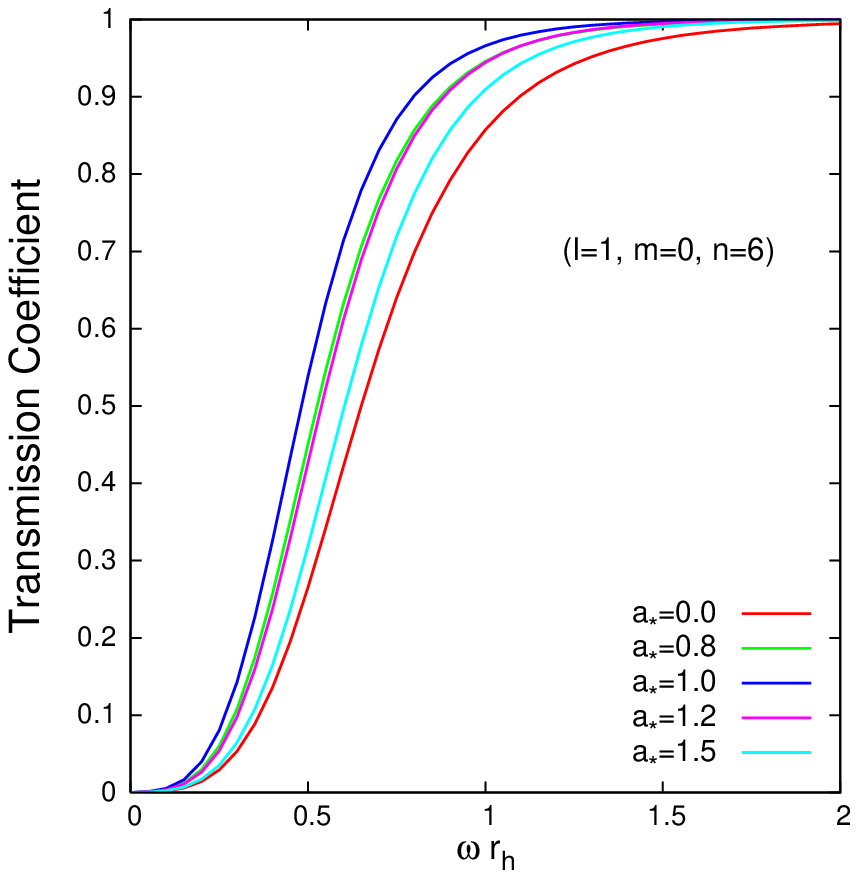} & \hspace*{-1cm}
\includegraphics[height=5.5cm,clip]{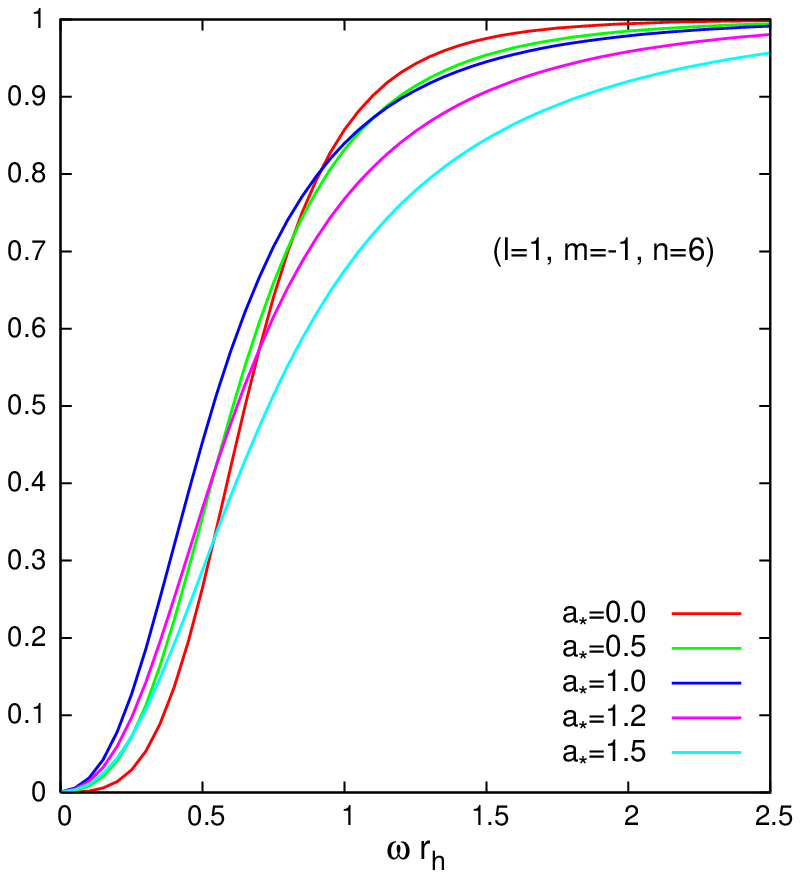} & \hspace*{-1cm}
\includegraphics[height=5.5cm,clip]{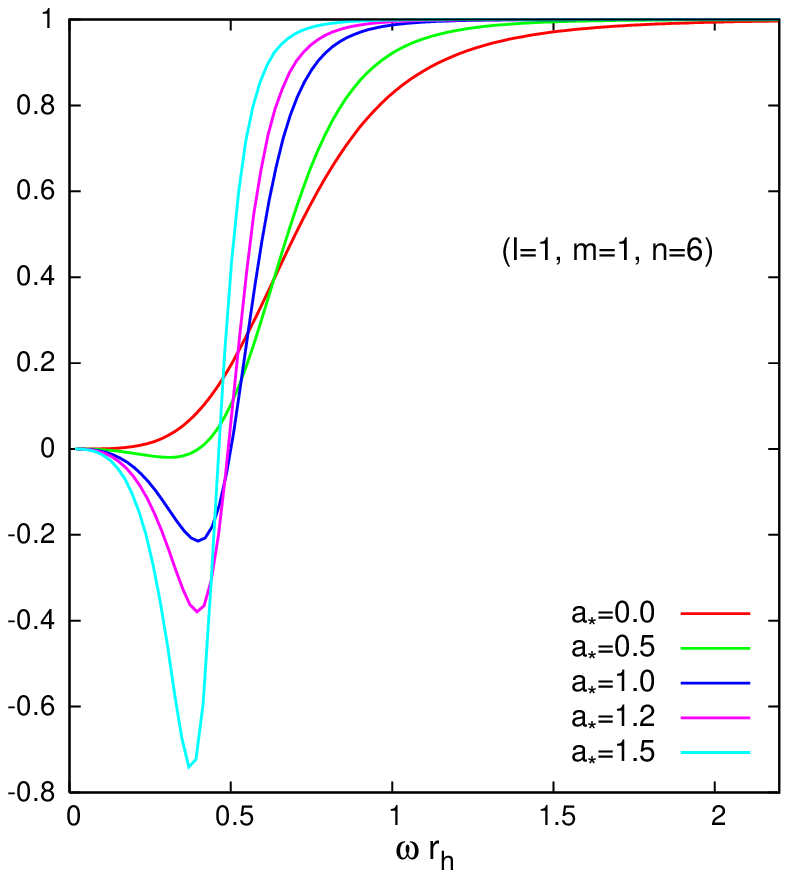} \end{tabular}
\caption{Transmission coefficients for spin-1 emission on the brane from a
10-dimensional black hole with variable $a_*$, for the modes (a) $\ell=1, m=0$, (b)
$\ell=1, m=-1$ and (c) $\ell=1, m=1$.}
\label{transm-a}
\end{center}
\end{figure}

\subsection{Transmission Coefficient}

We start the presentation of our results with the ones for the transmission
coefficient $\mathbb{T}_{\ell m \omega}$, as these follow from the numerical
solution of the radial equations (\ref{eq:diff. eq. for X,s=-1}) and
(\ref{eq:diff. eq. for X,r_d,s=-1}) by using as input the exact eigenvalue
$_{-1}\lambda_{\ell m \omega}$. As we will see, the transmission coefficient
bears a strong dependence on both the angular momentum parameter $a_*$ of
the black hole and the number of additional spacelike dimensions $n$. In
addition, its behaviour depends on the particular mode studied as well as
on the energy regime that is investigated.

Figure \ref{transm-a} depicts the dependence of $\mathbb{T}_{\ell m \omega}$
on the angular momentum parameter $a_*$ -- for simplicity, the number of extra
dimensions has been fixed to $n=6$. The three panels show the modes ($\ell=1,m=0$),
($\ell=1,m=-1$) and ($\ell=1,m=1$), and reveal the non-monotonic behaviour
of $\mathbb{T}_{\ell m \omega}$ as $a_*$ varies. For the ($\ell=1,m=0$) mode, the
transmission coefficient first increases and then decreases, as $a_*$ increases.
This behaviour is justified by, and reflected in, the behaviour of the gravitational
potential (\ref{pot-U}) seen by the propagating gauge field: as $a_*$ starts increasing,
the barrier lowers, thus giving a boost to $\mathbb{T}_{\ell m \omega}$; as $a_*$
increases further, however, the one-peak potential changes to a double-peak form
with the height of the peak located closer to the horizon increasing with $a_*$,
which leads to a suppression of the transmission coefficient.

For the ($l=1, m=-1$) mode, the behaviour of $\mathbb{T}_{\ell m \omega}$ is found to
be $\omega$-depen\-dent, too. For low values of $\omega$, as $a_*$ starts increasing,
the one-peak potential tends to become significantly more localized, leading to
the observed enhancement of the transmission coefficient. For large values of
$\omega$, the gravitational barrier changes to a well whose depth increases with
$a_*$, thus suppressing the value of $\mathbb{T}_{\ell m \omega}$. The dependence
on $\omega$ is even more striking for the third depicted mode ($\ell=1,m=1$).
This mode is a `superradiant' mode, one that is characterized by a negative
value of the transmission coefficient over the energy regime
$\tilde\omega=\omega- m \Omega <0$, according to Eq. (\ref{T-final}), and
thus by a value of the reflection coefficient larger than unity (see also
section 5.6). For fixed dimensionality, Fig. \ref{transm-a}(c) reveals that
$\mathbb{T}_{\ell m \omega}$ is greatly suppressed with $a_*$ over the
superradiance regime but enhanced everywhere else. The behaviour of the
potential again supports the observed behaviour: in the low-energy regime,
an increase in $a_*$ either raises the barrier or deepens the well that
characterizes the gravitational potential, thus suppressing the transmission
coefficient to even smaller values; in the high-energy regime, as $a_*$
increases, the potential is characterized by the simultaneous presence of
a barrier and a well, whose height and depth, respectively, decrease as
$a_*$ increases, thus justifying the observed enhancement.

\begin{figure}[t]
\begin{center}
\thicklines
\includegraphics[height=5.7cm,clip]{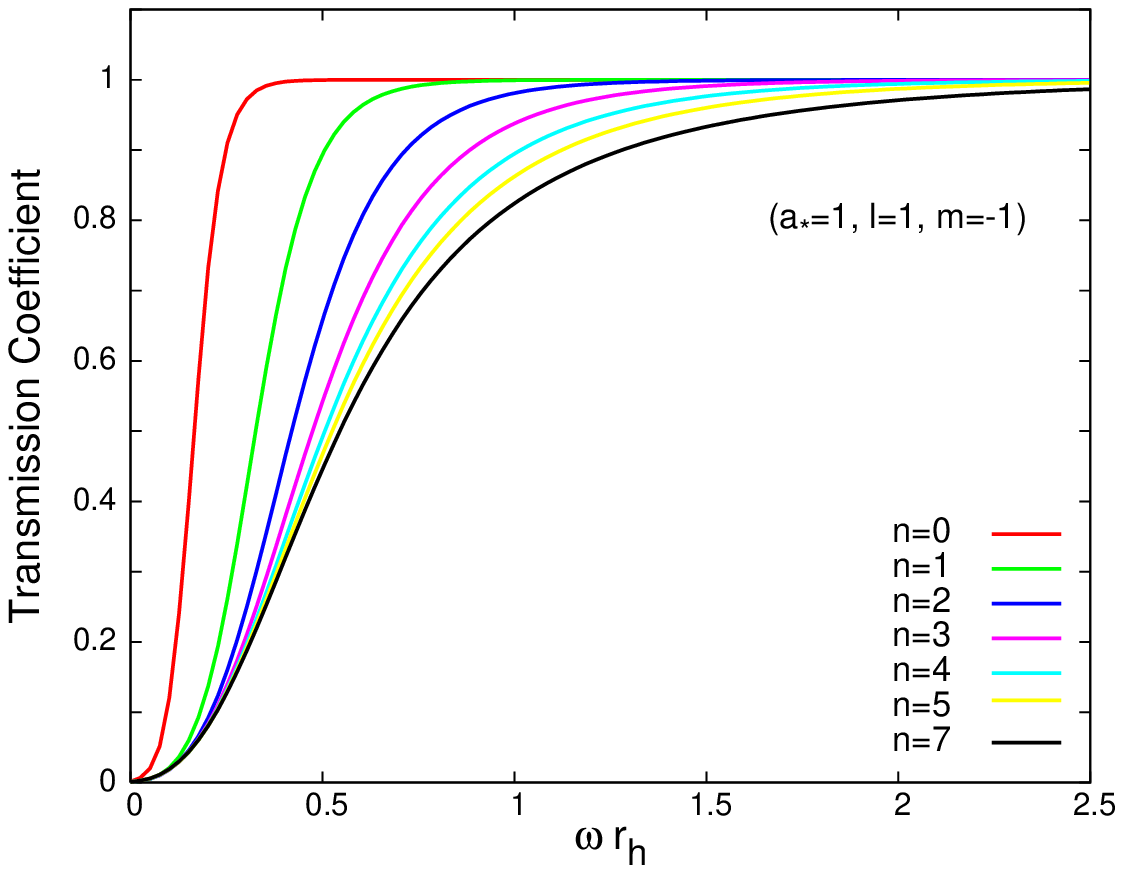}\hspace*{0.3cm}
{\psfrag{x}[][][0.8]{$\omega\,r_\text{h}$}
\psfrag{y}[][][0.8]{$r_\text{h}^2\,d^2N/dt\,d\omega$}
\includegraphics[height=5.7cm,clip]{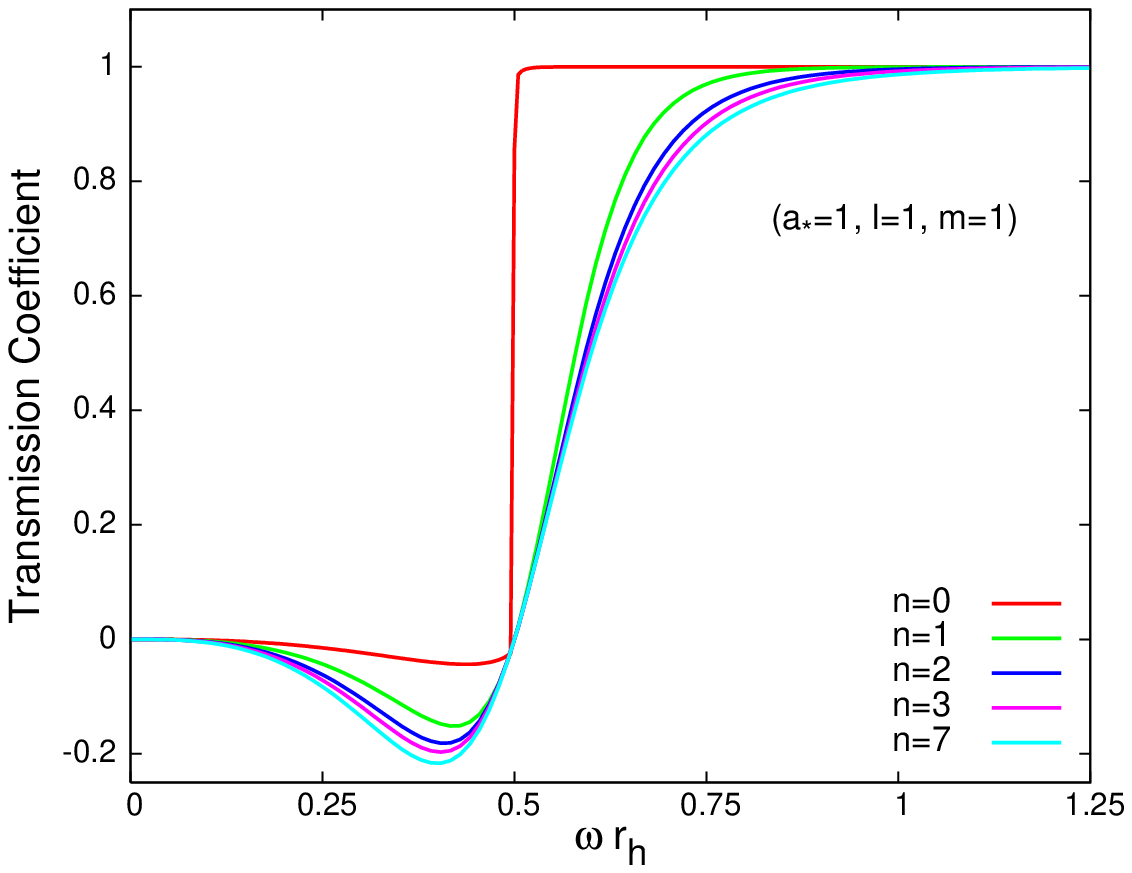}}
\caption{Transmission coefficients for spin-1 emission on the brane from a black
hole with $a_*=1$ and variable $n$, for the modes (a) $\ell=1, m=-1$, and (b)
$\ell=1, m=1$.}
\label{transm-n}
\end{center}
\end{figure}

The dependence of the transmission coefficient on the dimensionality of
spacetime is depicted in Fig. \ref{transm-n}, for a fixed value of the angular
momentum parameter, i.e. $a_*=1$. The left panel depicts the behaviour of a
non-superradiant mode -- in this case the $\ell=1$, $m=-1$ mode -- and reveals the
suppression of $\mathbb{T}_{\ell m \omega}$, as $n$ increases. The above
monotonic behaviour holds for all non-superradiant modes, with $m\le0$, and in
all energy regimes. The right panel of Fig. \ref{transm-n} shows the behaviour
of the corresponding superradiant mode ($l=1$, $m=1$), in which we observe
that the supression of $\mathbb{T}_{\ell m \omega}$ with the number of extra
spacelike dimensions holds also for this type of modes. The suppression is
present also in the superradiant regime, where an increase in the value of
$n$ leads to a further suppression (larger negative value) of the transmission
coefficient. In terms of the gravitational potential, one may easily see that,
for all modes, the height of the barrier -- or, the depth of the well, depending
on the values of $a_*$ and $\omega$ -- increases as $n$ increases, thus
justifying the behaviour depicted in Fig. \ref{transm-n}.

The various emission spectra -- particle, energy and angular momentum -- as
well as their dependence on the parameters $a_*$ and $n$, will be determined
by the behaviour of both the transmission coefficient $\mathbb{T}_{\ell m \omega}$
and the thermal factor ($e^{\tilde\omega/T_\text{H}}-1)^{-1}$,
appearing in Eqs. (\ref{flux})-(\ref{ang-mom}). Having discussed the behaviour
of the transmission coefficient, let us now comment on the dependence of the
thermal factor on $n$ and $a_*$. The behaviour of the latter on the number
of additional dimensions is encoded in the expression of the temperature of
the black hole, given in Eq. (\ref{temp}). A simple study reveals that, for
$a_*$ and $r_h$ fixed, $\partial T_\text{H}/\partial n>0$, and the temperature
of the black hole always increases with the value of $n$, thus giving a boost
to the thermal factor and, therefore, to all emission spectra. As the thermal
factor behaves in exactly the opposite way from the transmission coefficient
in terms of $n$, it remains to be seen which of the two is the prevailing
factor that determines the behaviour of the final emission spectra.

The dependence of the thermal factor on $a_*$ is more elaborate:
in this case we find that, for $n$ and $r_h$ fixed,
\begin{equation}
\frac{\partial }{\partial a_*}\left(\frac{\omega-m \Omega}{T_\text{H}}\right)=
4\pi r_h\,\frac{ma_*^2\,(n-1)+4 a_* \omega -m\,(n+1)}{[\,(n+1) +(n-1)\,a_*^2\,]^2}\,,
\label{derivative}
\end{equation}
according to which the behaviour of the thermal factor, as $a_*$ varies, depends
on the particular mode, the energy and angular-momentum regime, and the dimensionality
of spacetime. For modes with $m=0$, the behaviour of the $(\omega-m \Omega)/T_\text{H}$
is monotonic, and increases with $a_*$ for all values of $n$ and $\omega$, leading
to the suppression of the thermal factor and thus of the various emission spectra.
Combining this result with the behaviour
of the transmission coefficient depicted in Fig. \ref{transm-a}(a), we may conclude
that the $m=0$ modes will significantly contribute to the various spectra only for
black holes with an angular momentum in the low $a_*$-regime, where the suppression
of the thermal factor is small and the transmission coefficient is enhanced.
Turning to the modes with $m<0$, we find that, for $n \leq 1$, the r.h.s of
Eq. (\ref{derivative}) is always positive, and the various spectra are suppressed
for all $a_*$ and $\omega$; for $n>1$, the thermal factor is found to be again
suppressed in the low $a_*$-regime but significantly enhanced in the high $a_*$-regime
and low $\omega$-regime. According to Fig. \ref{transm-a}(b), the transmission
coefficient will also be enhanced for these values of $a_*$ and $\omega$, therefore,
modes with $m<0$ will be mainly emitted in the low-energy regime from black holes
with a large angular momentum. Finally, for modes with $m>0$, we need to distinguish
between the superradiant and non-superradiant energy regimes. In the latter one,
we find that the thermal factor is actually suppressed with $a_*$, while, according
to Fig. \ref{transm-a}(c), the transmission coefficient is enhanced; it is, therefore,
the `battle' between these two quantities that will define the final spectrum for
modes with $m>0$ in the non-superradiant regime. For modes with $m>0$ and energy
in the superradiant regime $\omega -m \Omega <0$, the situation changes
radically: the thermal factor is found to be suppressed with $a_*$, however, its
value is now negative. Since this is the energy regime over which the transmission
coefficient also acquires a negative value, the Hawking radiation flux remains
positive and it is actually significantly enhanced with $a_*$ as both the thermal
factor and $\mathbb{T}_{\ell m \omega}$ are driven towards larger negative values,
as $a_*$ increases.

In the next three subsections, we present the various emission spectra for
spin-1 fields by using the results derived above. Their final expressions
(\ref{flux})-(\ref{powerang}) involve a sum over both $m\in [-\ell, \ell]$
and $\ell\in [1, \infty)$, thus rendering them mode-independent. Although this
makes the contribution of individual modes impossible to discern, the derived
spectra will be the combined effect of all radiated modes and contributing factors.

\subsection{Flux emission spectra}

\begin{figure}[t]
\begin{center}
\thicklines
\mbox{\psfrag{x}[][][0.8]{$\omega\,r_\text{h}$}
\psfrag{y}[][][0.8]{${r_\text{h}^2\,d^2N/dt\,d\omega}$}
\includegraphics[height=5.6cm,clip]{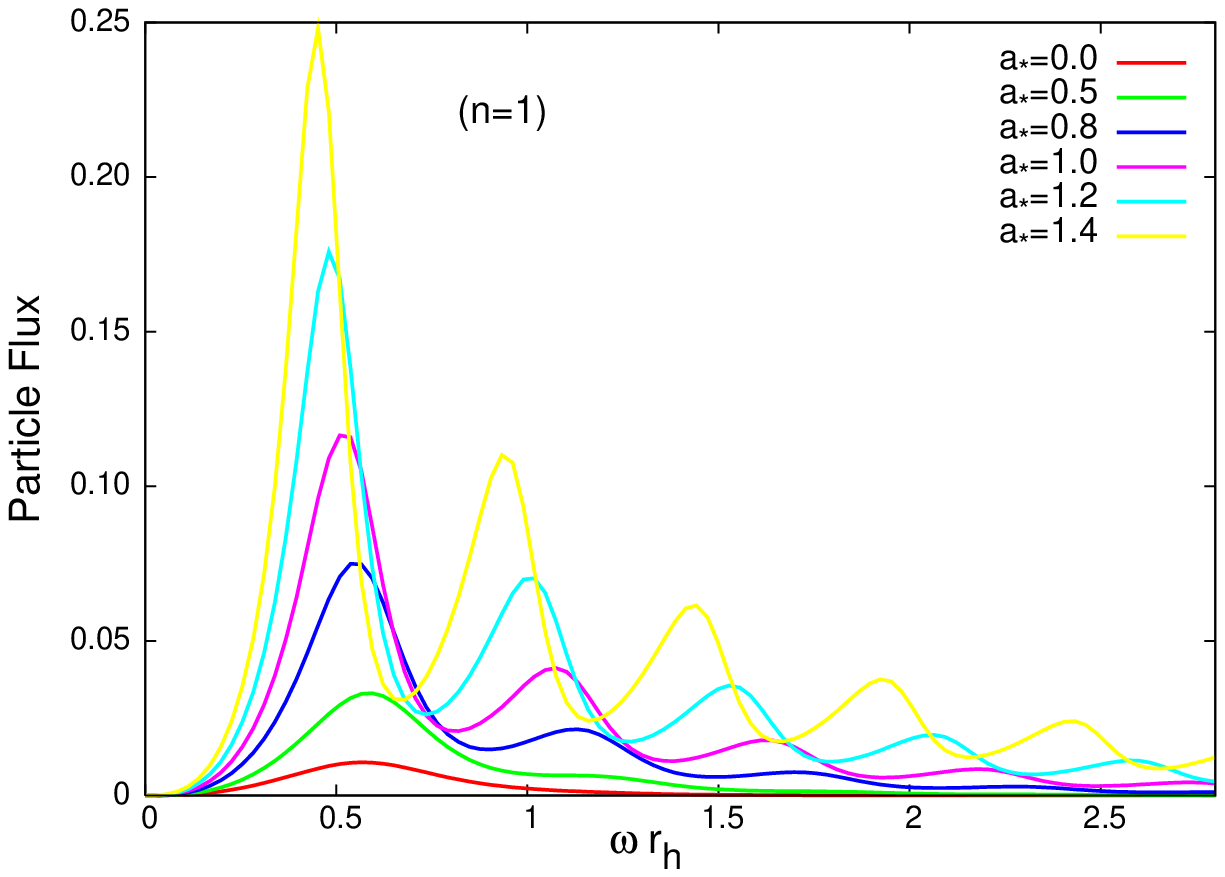}}
{\psfrag{x}[][][0.8]{$\omega\,r_\text{h}$}
\psfrag{y}[][][0.8]{$r_\text{h}^2\,d^2N/dt\,d\omega$}
\includegraphics[height=5.6cm,clip]{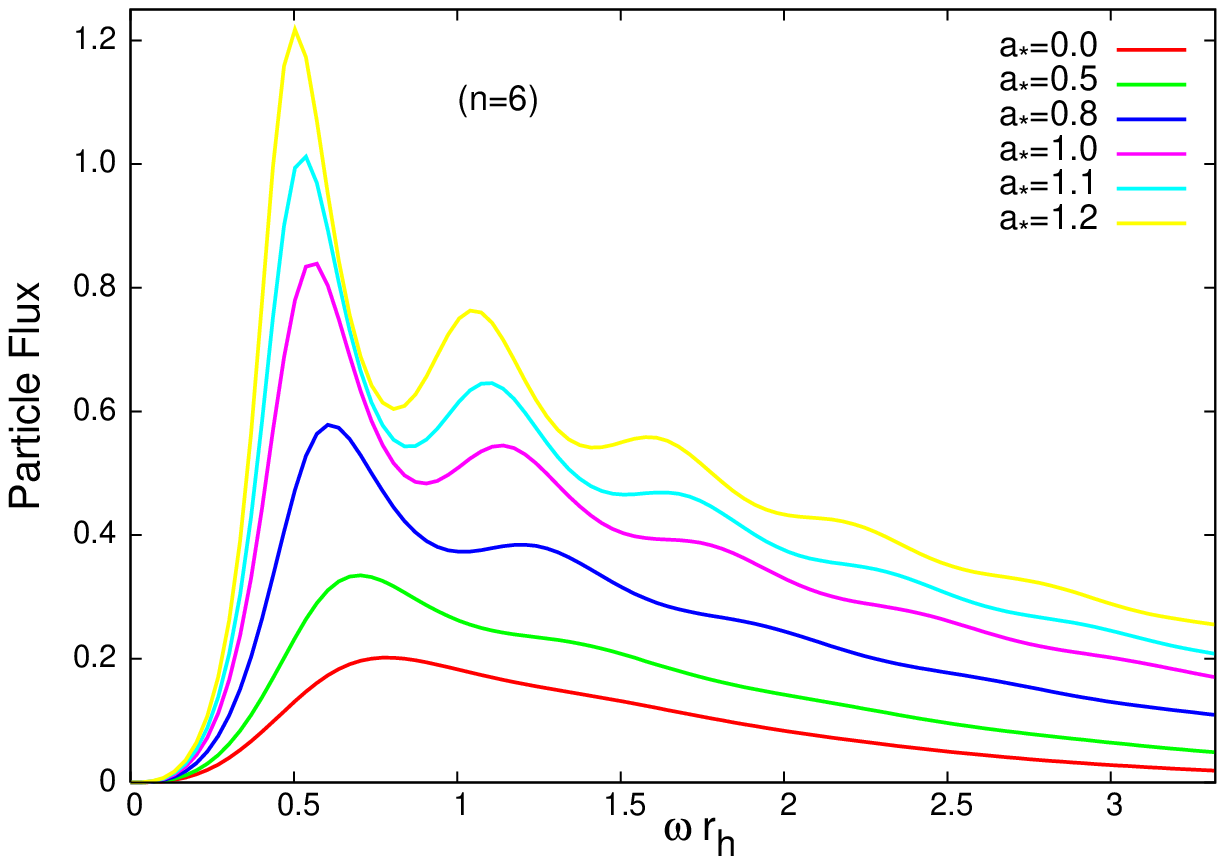}}
\caption{Flux emission spectra for spin-1 particles on the brane from a rotating black hole,
for (a) $n=1$, and (b) $n=6$, and various values of $a_*$.\hspace*{1.5cm}}
\label{s1n1-6flux}
\end{center}
\end{figure}

\begin{figure}[t]
\begin{center}
\psfrag{x}[][][0.8]{$\omega\,r_\text{h}$}
\psfrag{y}[][][0.8]{$r_\text{h}^2\,d^2N/dt\,d\omega$}
\psfig{file=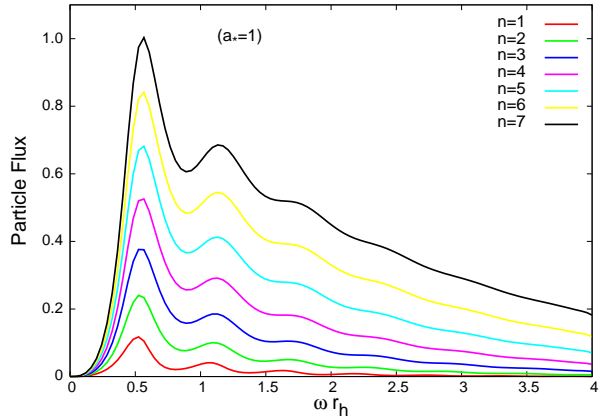, height=5.6cm}
\caption{Flux emission spectra for spin-1 particles on the brane from a rotating black hole,
with $a_*=1$, for various values of $n$.\hspace*{1.5cm}}
\label{s1a1flux}
\end{center}
\end{figure}

The number of gauge bosons emitted by the black hole on the brane per unit time and
unit frequency was computed first, by using Eq. (\ref{flux}). Our results are presented
in Figs. \ref{s1n1-6flux} and \ref{s1a1flux}, in terms of the
dimensionless\,\footnote{Throughout our analysis, we will be assuming, for convenience,
that the black hole horizon radius remains fixed, as the parameters $n$ and $a$ vary.}
energy parameter $\omega r_h$, and for fixed dimensionality and angular momentum
parameter of the black hole, respectively.

%

Figures \ref{s1n1-6flux}(a,b) depict the particle emission rate for a 5-dimensional
($n=1$) and a 10-dimensional ($n=6$) black hole, respectively, and for various values
of $a_*$. Both plots reveal the clear enhancement in the number of particles emitted
per unit time and frequency by the black hole, as the angular momentum of the black
hole increases. In both cases, this enhancement is observed over the whole energy band,
but the exact form of the spectrum depends strongly on the dimensionality of spacetime:
for low values of $n$, there is a clear preference for the emission of low-energy gauge
bosons, although the emission of particles with higher energy becomes increasingly more
likely, as the angular momentum increases; for high values of $n$, the emission of
low-energy quanta is still favoured but the probability for the emission of high-energy
ones has become substantially more important than the one in the low-dimensional case.
The spectra in both cases are characterized by oscillations caused by the higher
partial waves coming gradually into dominance as the energy increases.

\begin{figure}[t]
\begin{center}
\begin{tabular}{c} \hspace*{-0.4cm}
\epsfig{file=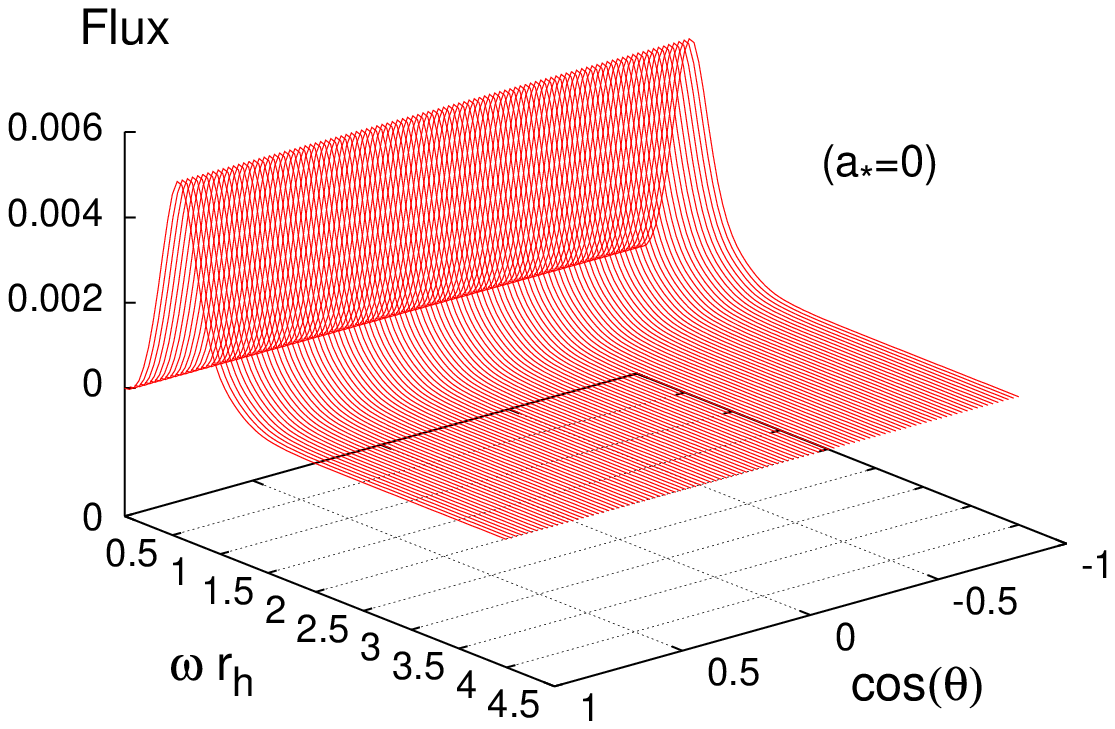, height=4.17cm}\hspace*{-1.15cm}
\epsfig{file=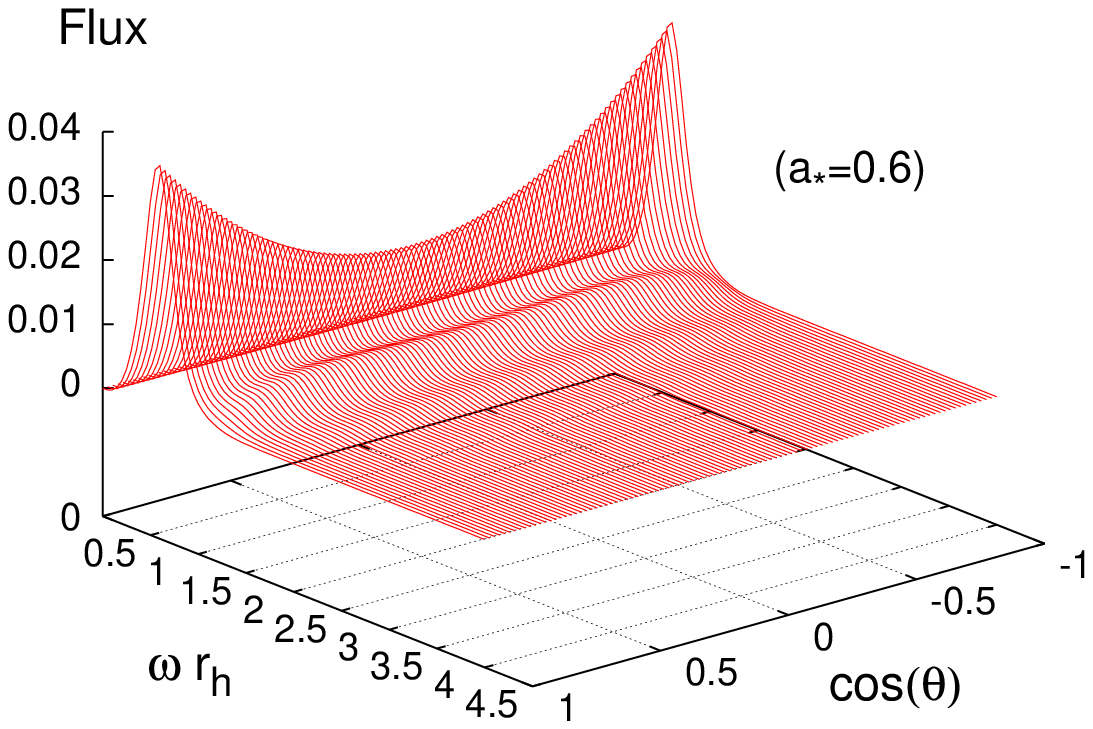, height=4.17cm}\hspace*{-1.15cm}{
\epsfig{file=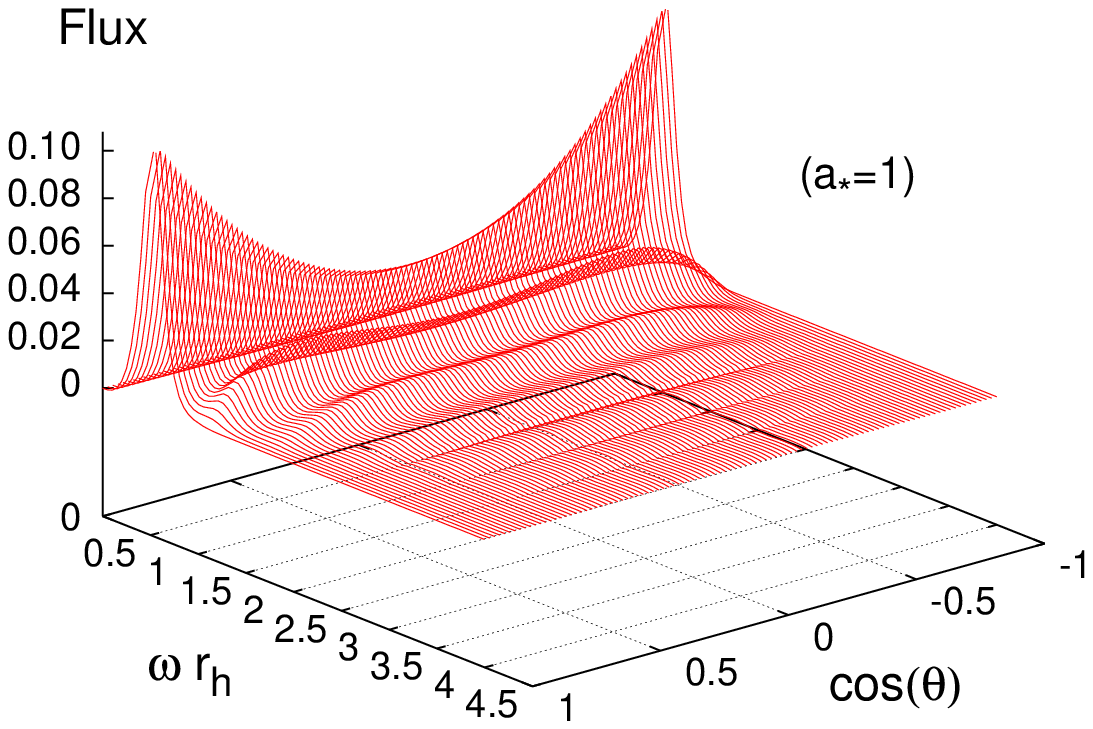, height=4.17cm}}\end{tabular}
\caption{Angular distribution of the flux spectra for gauge boson emission on the brane
from a rotating black hole, for $n=1$ and $a_*=(0,0.6,1)$.}
\label{s1n1f-ang}
\end{center}
\end{figure}
\begin{figure}[t]
\begin{center}
\begin{tabular}{c} \hspace*{-0.4cm}
\epsfig{file=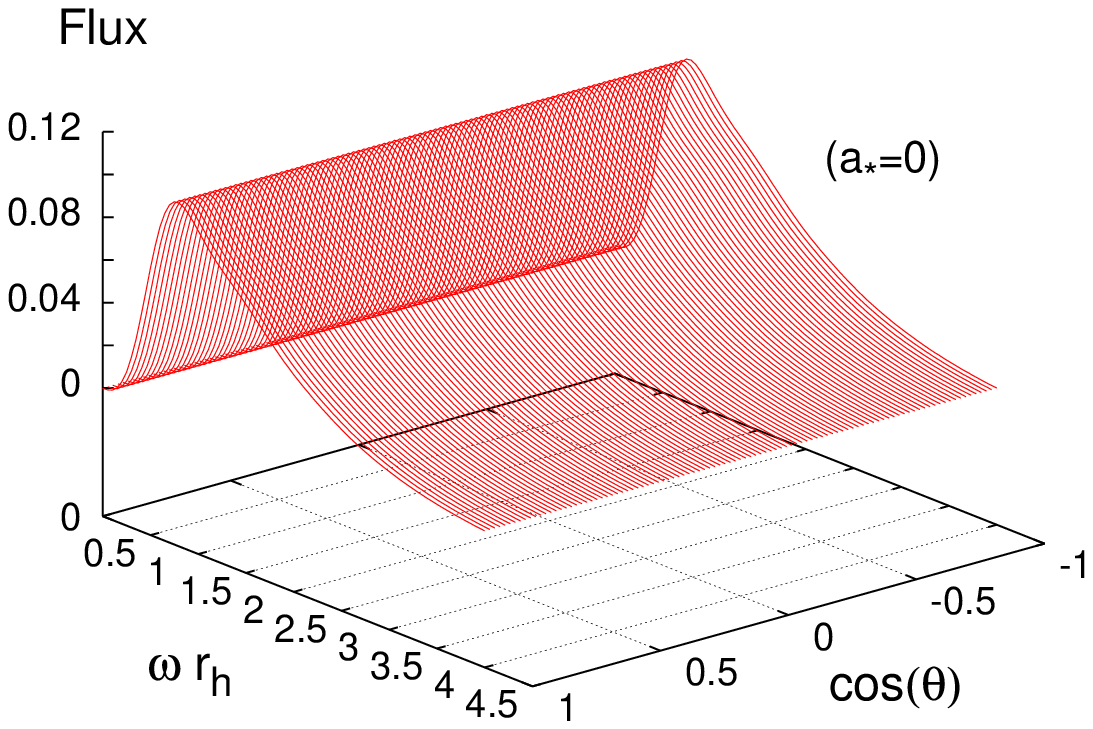, height=4.15cm}\hspace*{-1.15cm}
\epsfig{file=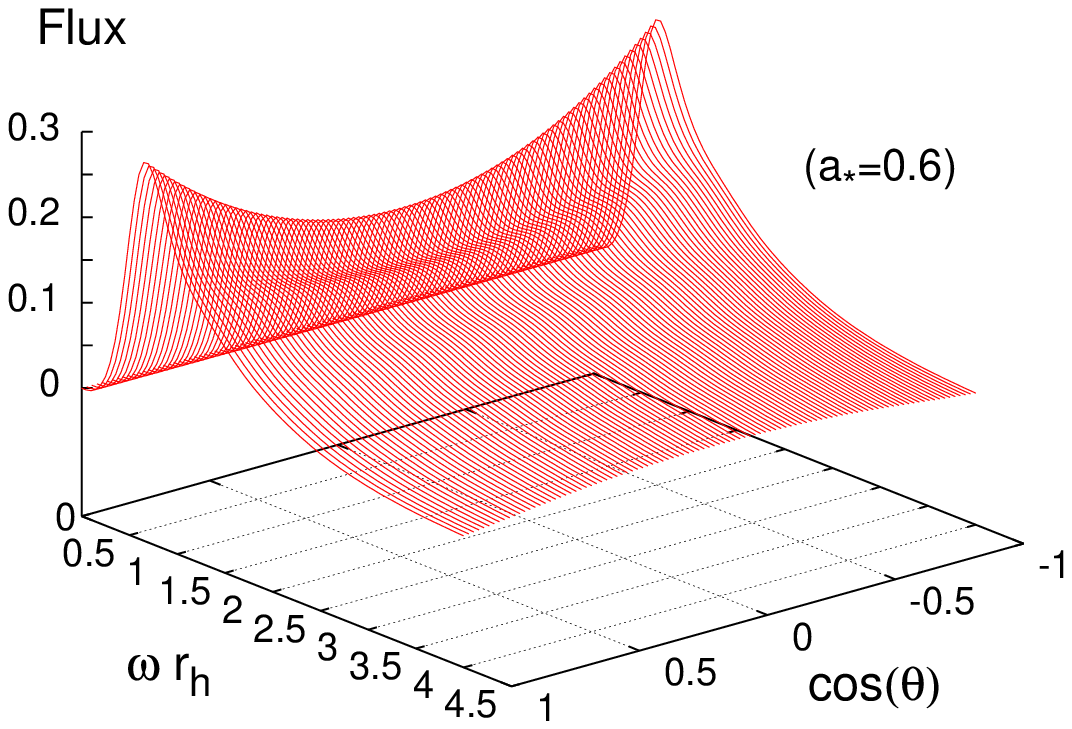, height=4.15cm}\hspace*{-1.15cm}{
\epsfig{file=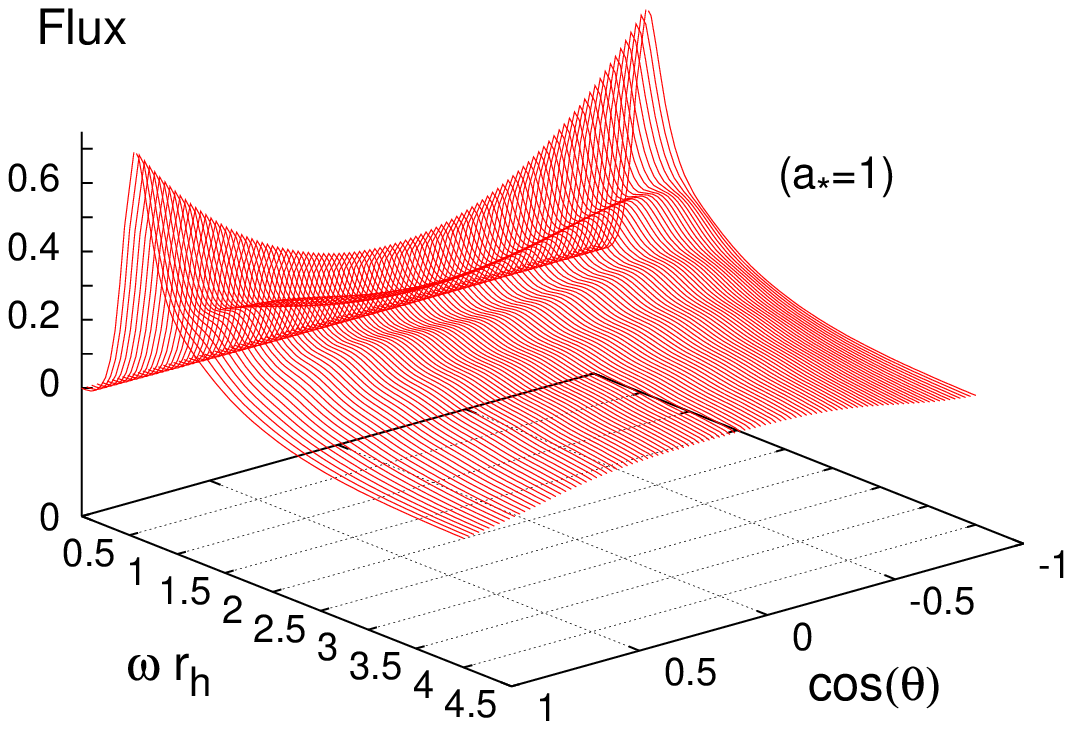, height=4.15cm}}\end{tabular}
\caption{Angular distribution of the flux spectra for gauge boson emission on the brane
from a rotating black hole, for $n=6$ and $a_*=(0,0.6,1)$.}
\label{s1n4f-ang}
\end{center}
\end{figure}

As is well known from previous studies, an increase in the dimensionality of
spacetime causes a significant enhancement in the number of particles emitted per unit
time and frequency by the black hole; this effect was observed in the case of a
higher-dimensional Schwarzschild \cite{kmr1, kmr2, HK1}, Schwarzschild-de-Sitter
\cite{BGK}, and Schwarzschild-Gauss-Bonnet \cite{Barrau} black hole, as well as
in the case of scalar emission from a higher-dimensional Kerr-like black hole
\cite{HK2, DHKW}. The same enhancement is anticipated in the present case
of the emission of gauge bosons, and
this behaviour can be readily seen by comparing the vertical axes of Figs.
\ref{s1n1-6flux}(a) and (b). A more complete picture can be drawn from
Fig. \ref{s1a1flux}, where the flux spectrum is given for fixed angular momentum
($a_*=1$) and variable $n$. As we see, the particle emission curves become
significantly taller and broader, as $n$ increases, which signals the enhancement
of the total emissivity of the black hole by virtually orders of magnitude.

We now turn to the angular distribution of the flux spectra: we study again the two
indicative cases of $n=1$ and $n=6$, in an attempt to detect potential changes in the
angular distribution of particles as the dimensionality of spacetime changes. In each
case, we present spectra for three different values of the angular momentum parameter,
i.e. $a_*=(0, 0.6, 1)$; this will help us see more clearly the effect of the black hole
rotation on the angular distribution pattern. These spectra are shown in Figs.
\ref{s1n1f-ang} and \ref{s1n4f-ang}. For both values of $n$, the spectra show no
angular variation for $a_*=0$, as expected. However, as $a_*$ starts increasing, a
much more interesting pattern emerges: low-energy gauge particles tend to be emitted
along the rotation axis ($\theta=\pm \pi$), but this tendency soon dies out as
the energy of the particles increases. The concentration of the emitted gauge
particles with low energy close to the rotation axis is caused by the non-trivial
coupling between the spin of the particles and the angular momentum of the black hole,
a coupling that was absent in the case of emission of scalar particles \cite{DHKW}.
For higher-energy particles, this effect gradually becomes sub-dominant
compared to the centrifugal force generated by the rotating black hole; this forces
the emitted particles to concentrate around the equatorial region ($\theta=\pi/2$)
instead, an effect that seems to be enhanced as both the dimensionality of spacetime
and angular momentum of the black hole increase.

\smallskip
\begin{figure}[t]
\begin{center}
\mbox{\psfrag{x}[][][0.8]{$\omega\,r_\text{h}$}
\psfrag{y}[][][0.8]{$r_\text{h}\,d^2E/dt\,d\omega$}
\includegraphics[height=5.6cm,clip]{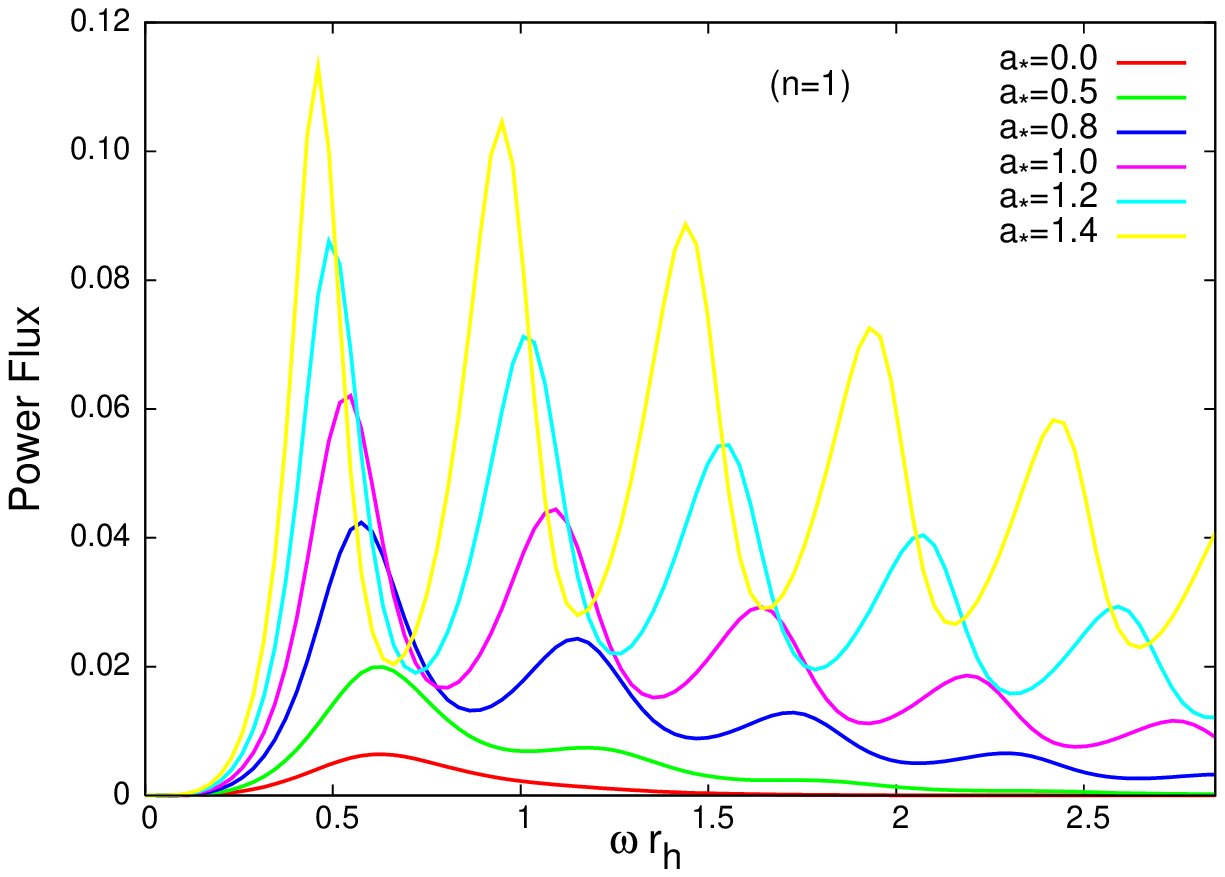}}
{\psfrag{x}[][][0.8]{$\omega\,r_\text{h}$}
\psfrag{y}[][][0.8]{$r_\text{h}\,d^2E/dt\,d\omega$}
\includegraphics[height=5.6cm,clip]{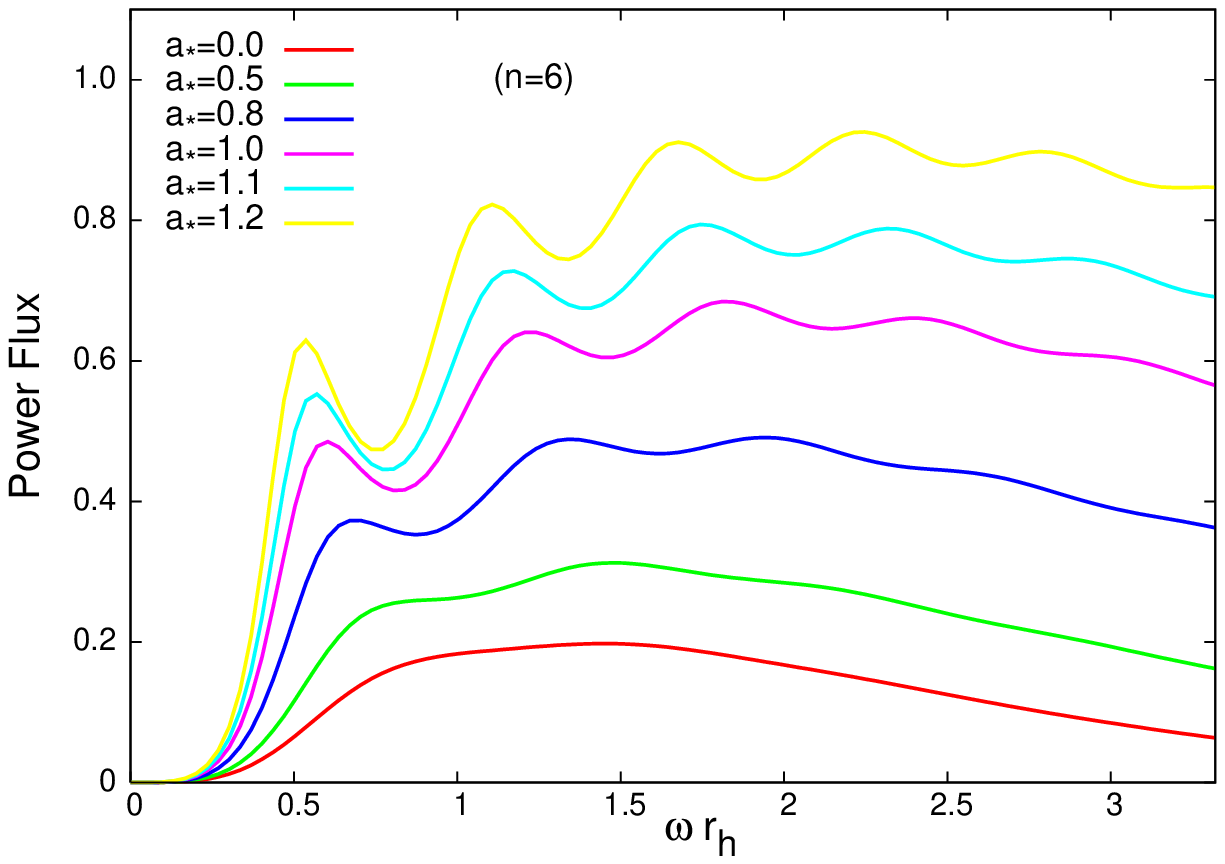}}
\caption{Power emission spectra for spin-1 particles on the brane from a rotating black hole,
for (a) $n=1$, and (b) $n=6$, and various values of $a_*$.\hspace*{1.5cm}}
\label{s1n1power}
\end{center}
\end{figure}
\begin{figure}
\begin{center}
\psfrag{x}[][][0.8]{$\omega\,r_\text{h}$}
\psfrag{y}[][][0.8]{$r_\text{h}\,d^2E/dt\,d\omega$}
\includegraphics[height=5.6cm,clip]{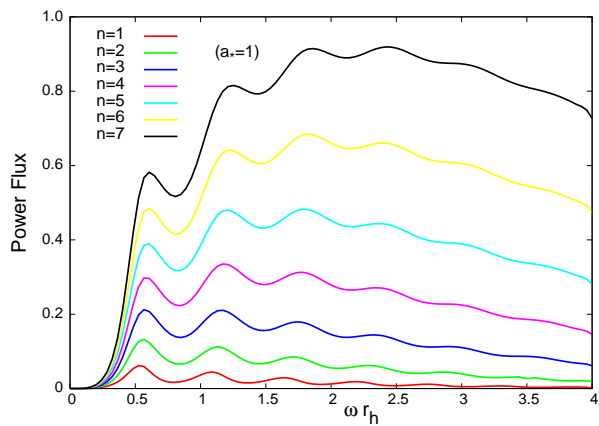}
\caption{Power emission spectra for spin-1 particles on the brane from a rotating black hole,
with $a_*=1$, for various values of $n$.\hspace*{1.5cm}}
\label{s1a1fp}
\end{center}
\end{figure}

\subsection{Power emission spectra}

We now proceed to the power spectrum, i.e. the energy emitted by the black hole in the
form of gauge bosons per unit time and unit frequency. This is given by Eq. (\ref{power}),
and it is displayed in Figs. \ref{s1n1power} and \ref{s1a1fp}, for fixed dimensionality
and angular momentum parameter, respectively, as in the previous subsection.

Figures \ref{s1n1power}(a) and (b) depict the energy emission spectra for a 5-dimensional
and a 10-dimensional black hole, respectively, and for variable $a_*$. In both cases,
the enhancement of the emission rate, with the value of the angular momentum parameter,
is obvious over all energy regimes\footnote{The same enhancement in the emission rate
of gauge particles was found in the low-energy regime in \cite{IOP1} by following an
analytical approach valid in the low-energy and low-angular-momentum limit. The
suppression shown in Figs. 6 and 7 of \cite{IOP1} over the intermediate and high-energy
regimes are an artificial result only, caused by the breakdown of the approximations
used, and therefore should not be trusted.}.  In the $n=1$ case,  the localized -- around the
low-energy regime-- curve, for $a_*=0$, gives its place to a broad, oscillatory curve,
for higher values of $a_*$. Although the emission rate is gradually decreasing, as the
frequency of the particle increases, it retains more than 50\% of its value when the
frequency has increased by a factor of 5. In the $n=6$ case, the situation has changed
even more radically: a peak around the low-energy regime can still be seen in the spectrum
-- remnant of the dominant peak appearing in the corresponding flux spectrum; however,
it is the emission of gauge particles with much higher energy that is the most effective
emission channel now.

The additional enhancement, caused by the increase in the dimensionality of spacetime,
can be easily seen again by comparing the axes of Figs. \ref{s1n1power}(a) and (b).
This enhancement is shown in detail in Fig. \ref{s1a1fp}, for fixed $a_*$ and
$n=1, ..., 7$. The increase in both the height and width of the emission curves,
as the number of additional spacelike dimensions increases, can be clearly seen. From
the above, we anticipate a strong enhancement of the total emissivity, in the
form of gauge particles, of a higher-dimensional rotating black hole with
parameter $n$, a point to be studied in more detail in subsection 5.5.

\begin{figure}[t]
\begin{center}
\begin{tabular}{c} \hspace*{-0.4cm}
\epsfig{file=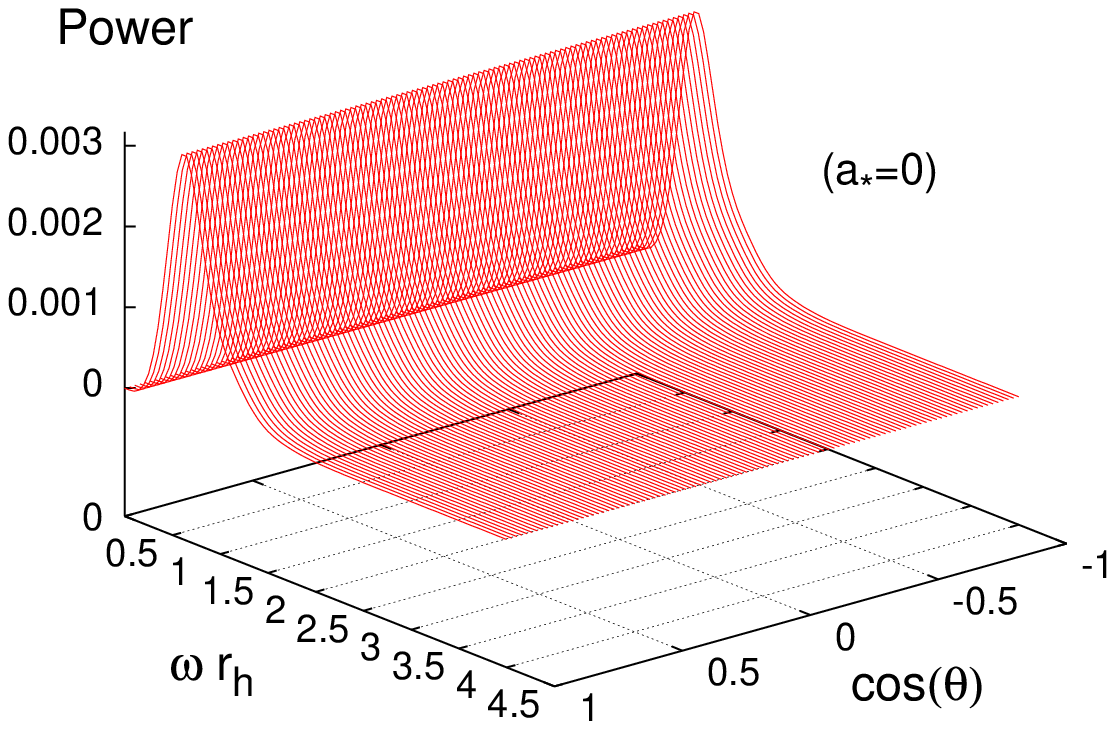, height=4.15cm}\hspace*{-1.15cm}
\epsfig{file=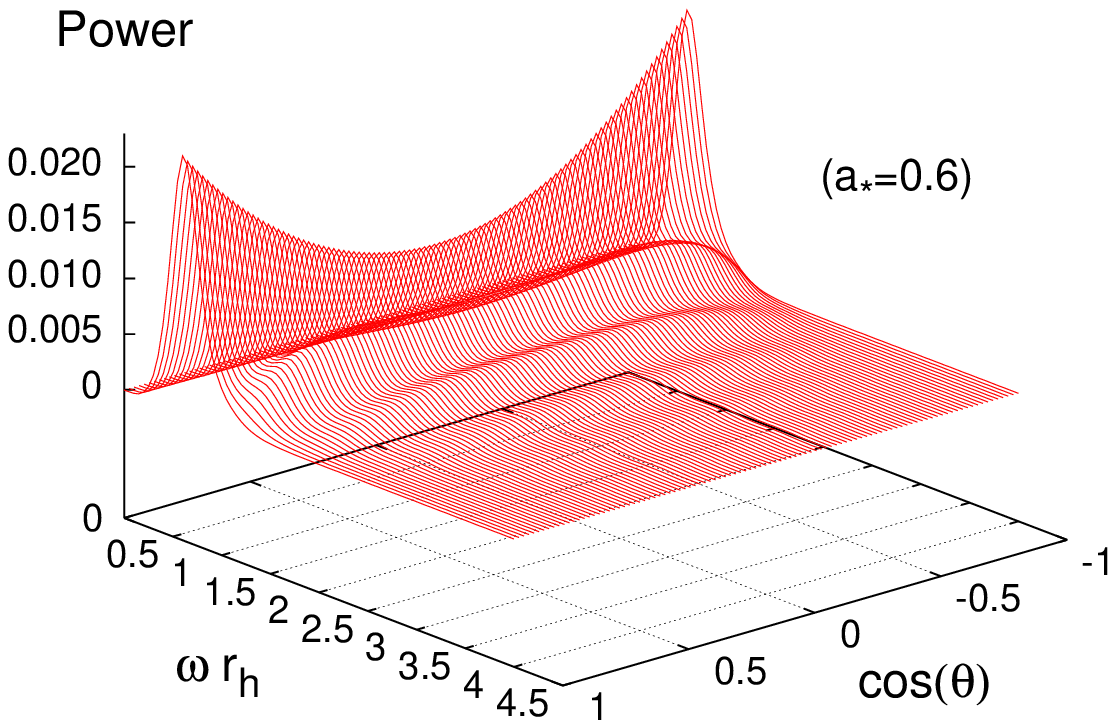, height=4.15cm}\hspace*{-1.15cm}{
\epsfig{file=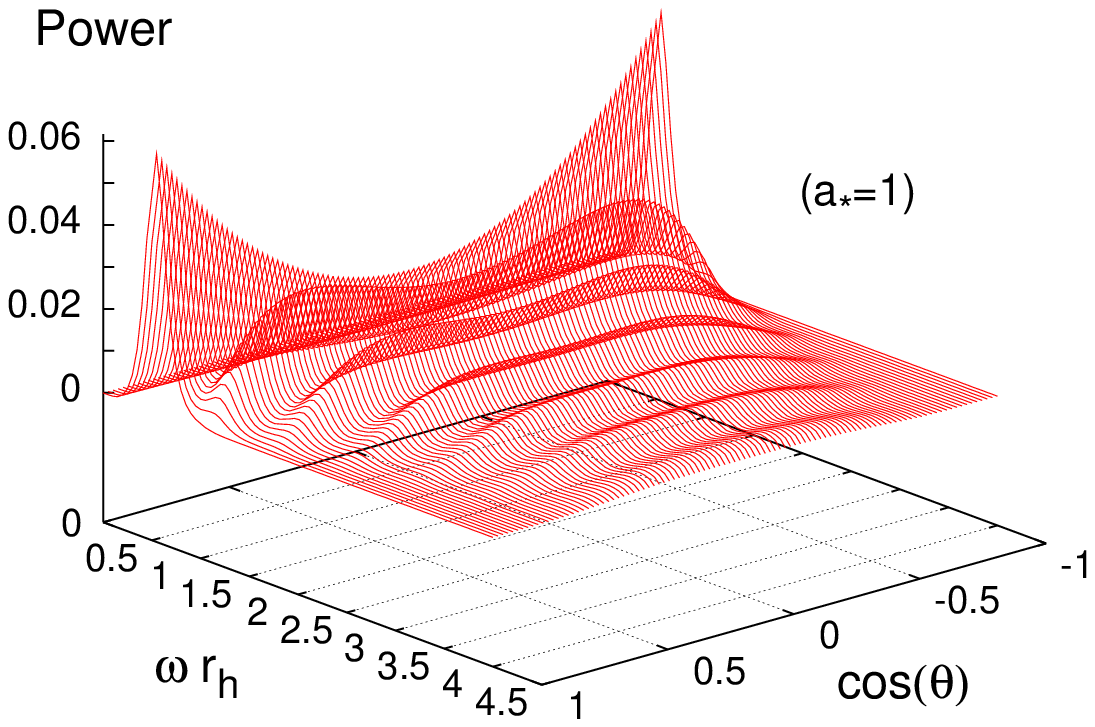, height=4.15cm}}\end{tabular}
\caption{Angular distribution of the power spectra for gauge boson emission from rotating
black holes, for $n=1$ and $a_*=(0,0.6,1)$.}
\label{s1n1p-ang}
\end{center}
\end{figure}
\begin{figure}[t]
\begin{center}
\begin{tabular}{c} \hspace*{-0.4cm}
\epsfig{file=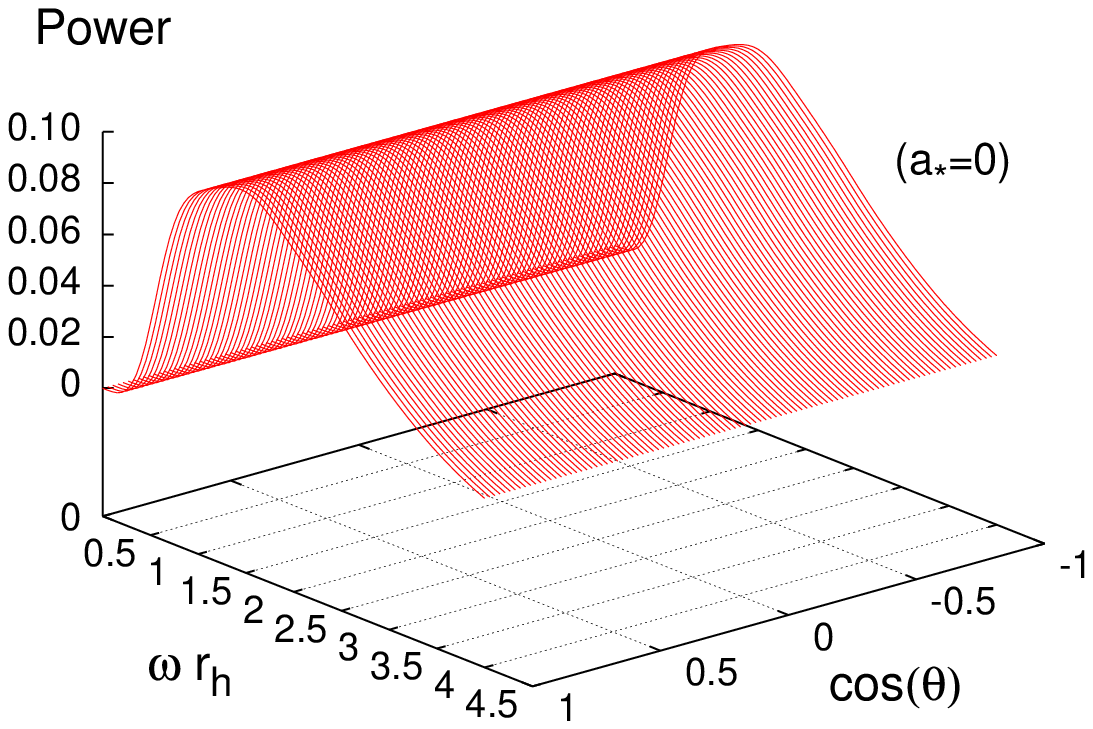, height=4.15cm}\hspace*{-1.15cm}
\epsfig{file=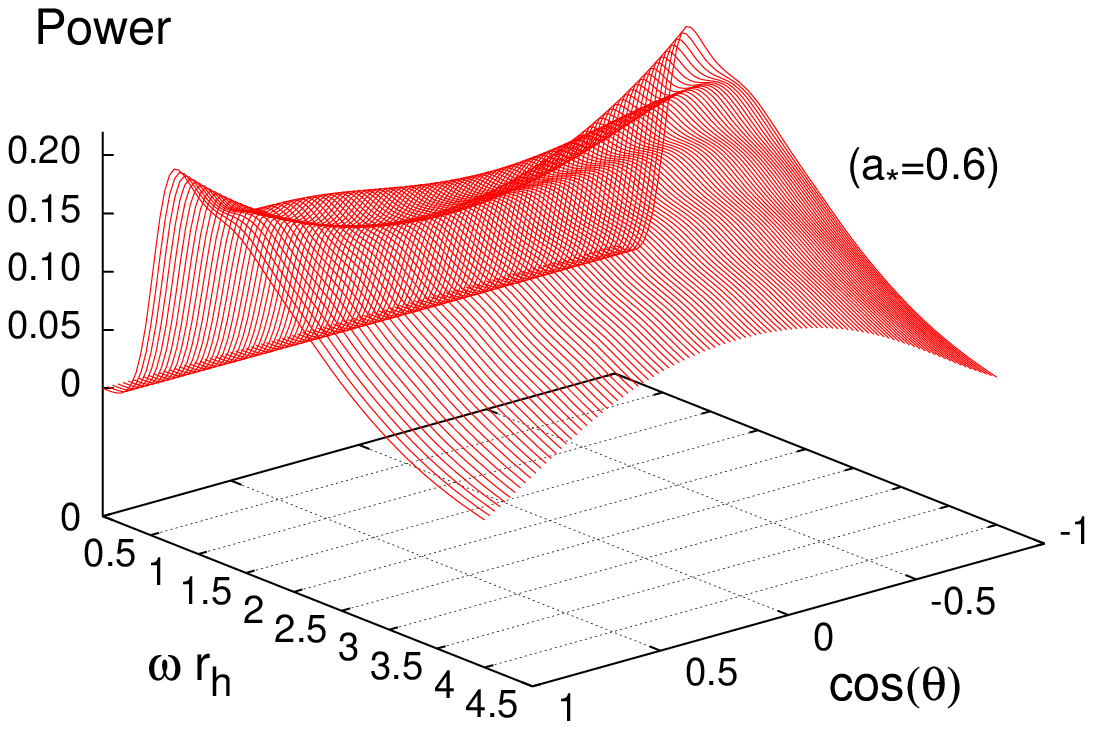, height=4.15cm}\hspace*{-1.15cm}{
\epsfig{file=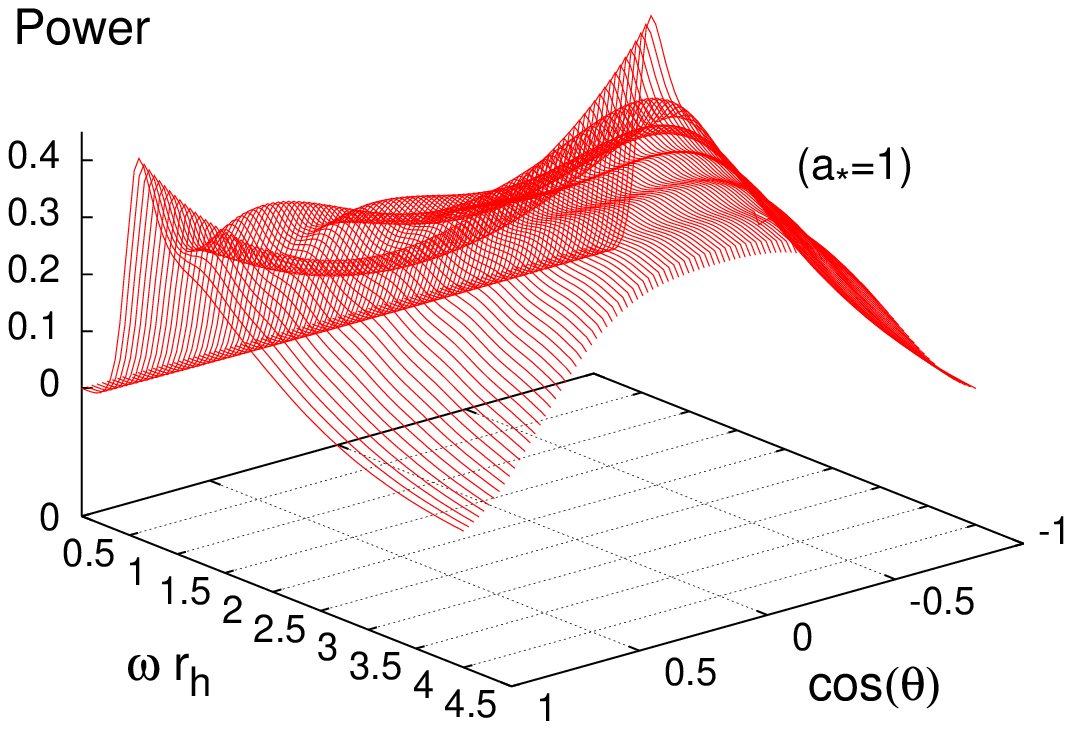, height=4.15cm}}\end{tabular}
\caption{Angular distribution of the power spectra for gauge boson emission from rotating
black holes, for $n=6$ and $a_*=(0,0.6,1)$.}
\label{s1n4p-ang}
\end{center}
\end{figure}

The angular distribution of the power spectra for gauge boson emission on the brane
is shown in Figs. \ref{s1n1p-ang} and \ref{s1n4p-ang}, for the cases of $n=1$ and $n=6$,
respectively, and for $a_*=(0,0.6,1)$. The same features exhibited by the flux spectra
appear also here: the angular variation -- absent for $a_*=0$ -- makes its appearance
in the spectrum as soon as the angular momentum parameter acquires non-zero values.
Again, the low-energy emission channel runs parallel to the rotation axis, while,
for higher-energies, the emission concentrates in the equatorial region. The amount
of energy emitted per unit time and frequency along the equatorial plane increases
also with the angular momentum parameter and the dimensionality of spacetime.

We should note here that our spectra are symmetric under the change $\theta \rightarrow
\pi-\theta$, contrary to what was found in \cite{IOP1}; the difference is due to the
fact that, in our analysis, both polarisations of the gauge field have been taken
into account. Although, the two polarisations may have a different coupling with the
black hole rotation, these couplings are both reversed when moving from one hemisphere
to the other, thus resulting in a symmetric spectrum \cite{Casals:2005kr}.

\smallskip
\begin{figure}[t]
\begin{center}
\mbox{\psfrag{x}[][][0.8]{$\omega\,r_\text{h}$}
\psfrag{y}[][][0.8]{$r_\text{h}\,d^2J/dt\,d\omega$}
\includegraphics[height=5.0cm,clip]{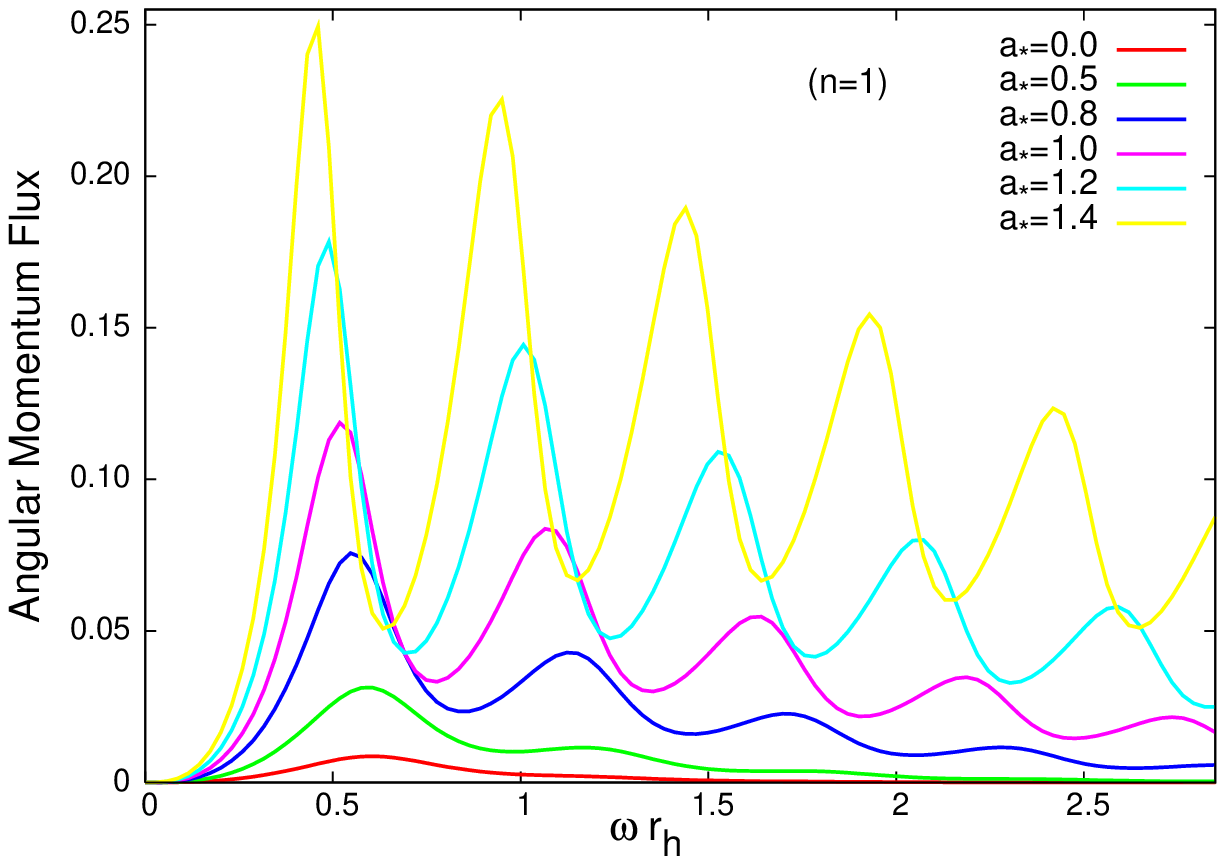}}\hspace*{0.5cm}{
\psfrag{x}[][][0.8]{$\omega\,r_\text{h}$}
\psfrag{y}[][][0.8]{$r_\text{h}\,d^2J/dt\,d\omega$}
\includegraphics[height=5.0cm,clip]{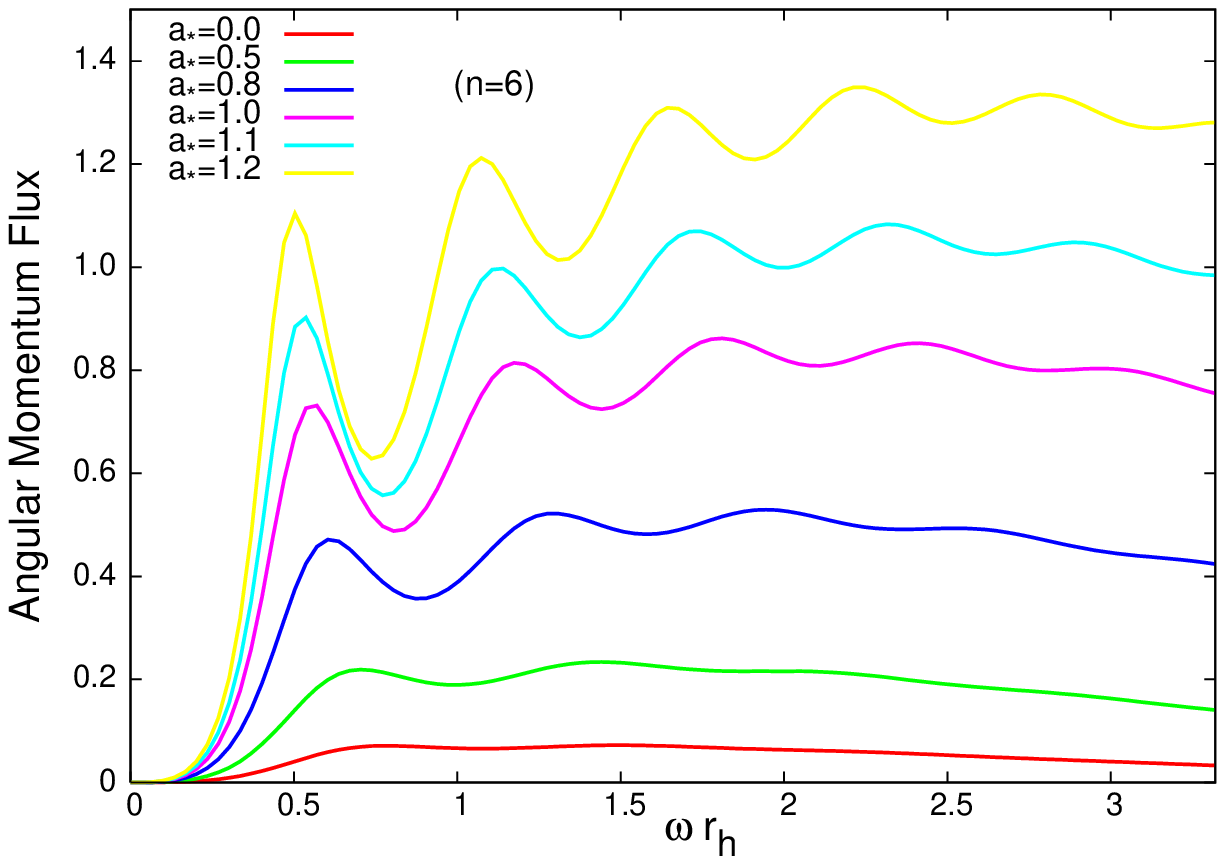}}
\caption{Angular momentum spectra for spin-1 particles from a rotating black hole,
for (a) $n=1$, and (b) $n=6$, and various values of $a_*$.\hspace*{1.5cm}}
\label{s1n1am}
\end{center}
\end{figure}

\subsection{Angular momentum spectra}

The final emission rate that we would like to study is the one of the angular
momentum of the black hole. This differential rate, per unit time and unit
frequency, is first shown in Figs. \ref{s1n1am}(a,b) for $n=1$ and
$n=6$, respectively, and for variable $a_*$. Its behaviour is very similar to
the one exhibited by the power emission spectrum. For $n=1$, the process of the
angular momentum loss takes place predominantly through the emission of low-energy
gauge particles, although the emission of particles with higher energy is still
substantial. For $n=6$, on the other hand, the black hole loses its angular
momentum through a more democratic process of emission, with a slight preference
towards the emission of high-energy particles.

\begin{figure}[t]
\begin{center}
\psfrag{x}[][][0.8]{$\omega\,r_\text{h}$}
\psfrag{y}[][][0.8]{$r_\text{h}^2\,d^2J/dt\,d\omega$}
\includegraphics[height=5.5cm,clip]{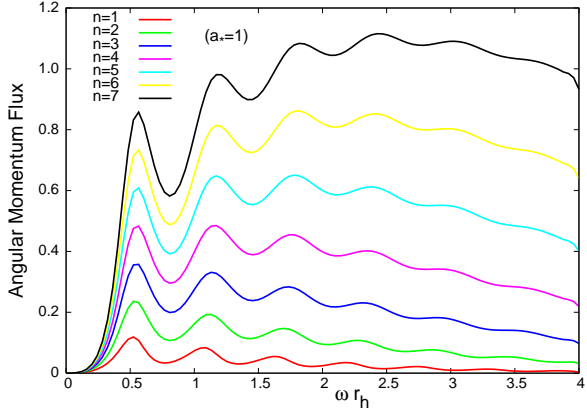}
\caption{Angular momentum spectra for spin-1 particles from a rotating black hole,
for $a_*=1$ and various values of $n$.\hspace*{1.5cm}}
\label{s1a1ang}
\end{center}
\end{figure}

Figure \ref{s1a1ang} depicts the dependence of the angular momentum loss rate on
the dimensionality of spacetime, while fixing the angular momentum parameter at
$a_*=1$. As in the previous two cases of flux and power spectra, this rate is
greatly enhanced, as the dimensionality of spacetime increases. The exact amount
of the enhancement of the total angular momentum emissivity -- which, as we will
see, can reach more than an order of magnitude -- will be discussed in the next
subsection.

\subsection{Total emissivities}

The total emissivities follow by integrating the various differential emission rates
over the frequency regime, and give the total emission rate by the black hole per
unit time. In principle, the integration over the frequency must cover all regimes
in which the black hole radiates. However,
deriving the whole Gaussian emission curve, as the angular momentum parameter and
dimensionality of spacetime increase, soon becomes an unrealistic task. By looking, for
instance, at Figs. \ref{s1a1flux}, \ref{s1a1fp} and \ref{s1a1ang}, one may easily
see that while, for low values of $n$, all emission curves have been  completed by
the time the value of the energy parameter $\omega r_h=4$ is reached, for high values
of $n$, the emission curves are still very close to their peak value.  As both $a_*$
and $n$ take larger values, more and more high-energy particles are emitted by the
black hole, and the tail of the curve extends to values of $\omega r_h$ much larger
than the cutoff value of 4, up to which the radial part of the equation of motion of
the gauge field has been numerically solved. Extending our energy regime towards
much larger values has proved to be an extremely time-consuming process, therefore, in
what follows, we will derive the  ``total emissivities'' by integrating the emission
rates up to our chosen cutoff value. As  is obvious from the results presented
in the previous subsections, an increase in the two fundamental parameters ($a_*$
and $n$) leads to a significant enhancement of the emission rates over the whole
energy regime; the high-energy regime left out of our computation of the total
emissivities, for large values of $a_*$ and $n$, would only increase further this
enhancement. Therefore, the ``total emissivities'' derived, and presented in this
subsection, can be considered as robust lower bounds to the exact ones.

\begin{figure}[t]
\begin{center}
\mbox{\includegraphics[height=5.5cm,clip]{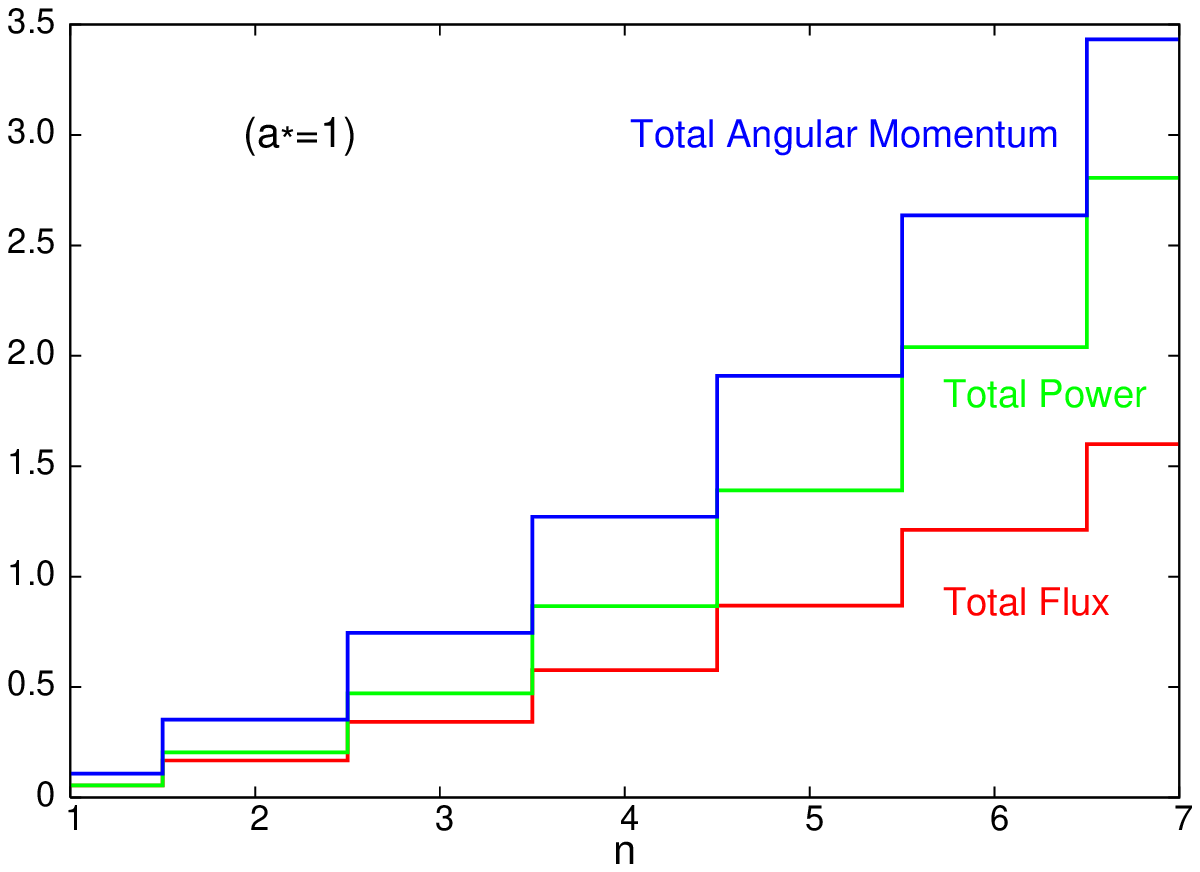}}
{\includegraphics[height=5.5cm,clip]{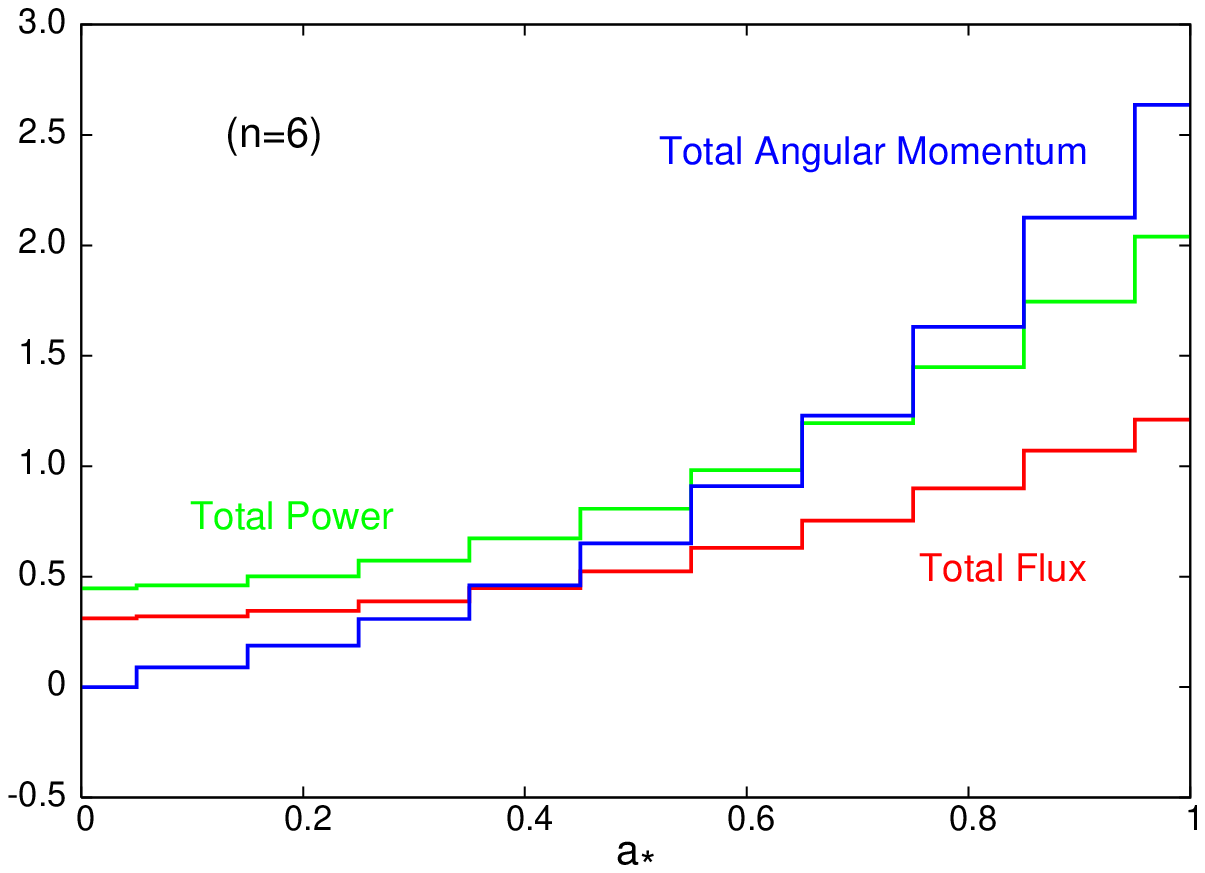}}
\caption{Total emissivities for spin-1 particle emission on the brane from a rotating
black hole as a function of (a) $n$, for $a_*=1$, and (b) $a_*$, for $n=6$.}
\label{totalna}
\end{center}
\end{figure}

By following the above method, we have computed the total emissivities for the
flux, power and angular momentum emitted by a higher-dimensional, rotating black
hole on the brane in the form of gauge fields. The results, for fixed angular
momentum parameter ($a_*=1$) and dimensionality ($n=6$), are depicted in the two
histograms in Figs. \ref{totalna}(a) and (b), respectively. The enhancement of
all three emissivities, that was foreseen while commenting on the results derived
in the previous subsections, can be clearly observed. In the first histogram, the
value of the number of additional spacelike dimensions $n$ varies from 1 to 7.
A simple calculation shows that the number of particles emitted per unit time by
a 11-dimensional, rotating (with $a_*=1$) black hole is 30 times larger than for
a 5-dimensional
one. Similarly, the total amount of energy emitted per unit time and the angular
momentum loss rate for a 11-dimensional black hole are more than 50 and 30 times,
respectively, larger than for a 5-dimensional black hole with the same angular
velocity. For the reason explained in the beginning of this subsection, we expect
these numbers to increase considerably when the high-energy part of the spectrum,
for $n=7$, is included in the calculation.

The second histogram in Fig. \ref{totalna} depicts the enhancement of the three
emissivities when the dimensionality of spacetime is kept fixed (at $n=6$) and
the angular momentum parameter varies. The data, used to produce this plot, reveal
that the number of particles and amount of energy emitted by a 10-dimensional
black hole with $a_*=1$ is 4 and 5 times, respectively, larger than the ones
for a 10-dimensional, non-rotating black hole. As, in the case of $n=6$, a
substantial part of the spectrum has been left out of the computation of the
total emissivities, the aforementioned numbers should be treated as lower
bounds only, with the exact numbers expected to reach values greater than these
by even orders of magnitude. This can be deduced from the fact that, in the case
of $n=1$, where all parts of the spectrum have been successfully derived and thus
included in the computation of the exact total emissivities, the number of particles
and amount of energy emitted per unit time by the 5-dimensional black hole with
$a_*=1$ is already 10 and 15 times, respectively, larger than for its non-rotating,
5-dimensional analogue.

From the first histogram in Fig. \ref{totalna}, we observe that the emission rate
of angular momentum remains always larger than the energy emission rate, with
the former becoming increasingly larger than the latter as $n$ increases. This
seems to support the argument that, during its decay, the angular momentum of
a higher-dimensional black hole will be radiated away faster than its mass, thus
signalling the existence of a  subsequent non-rotating phase in the life of the
black hole. Nevertheless, the second histogram in Fig. \ref{totalna} shows that,
for fixed dimensionality, the angular momentum emission rate strongly depends on
the angular momentum of the black hole: while this rate is again significantly
larger than the energy emission rate for rapidly rotating black holes, it is
significantly smaller for slowly rotating ones. We might, thus, conclude that the
evolution of rotating black holes goes through two different back-to-back stages:
an initial one, where the black hole loses predominantly rotational kinetic energy,
and a second one, where the -- now slowly -- rotating black hole loses mainly
its mass. The exact duration of the two phases, as well as of the subsequent
non-rotating one -- if existent, will strongly depend on the initial angular
momentum of the black hole and the dimensionality of spacetime.

\begin{figure}[t]
\begin{center}
\mbox{\includegraphics[height=5.5cm,clip]{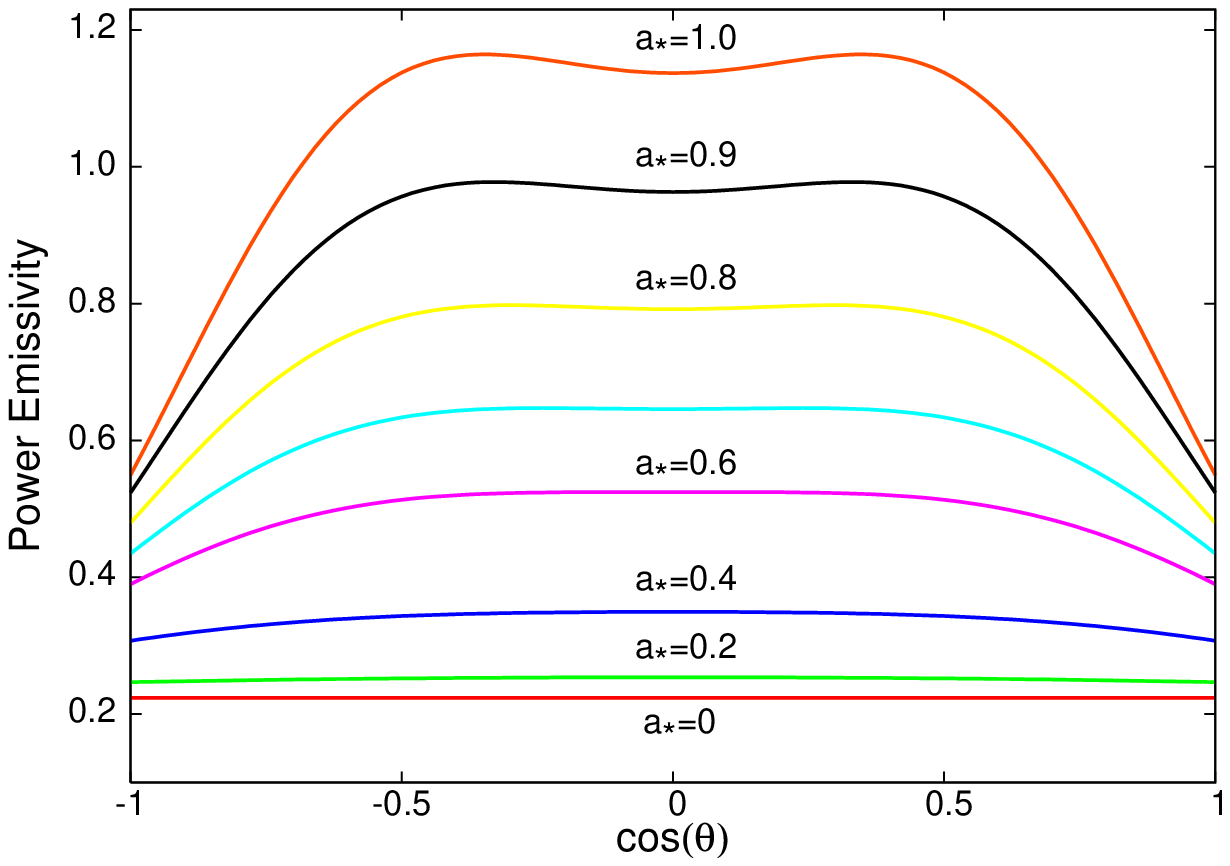}}
{\includegraphics[height=5.5cm,clip]{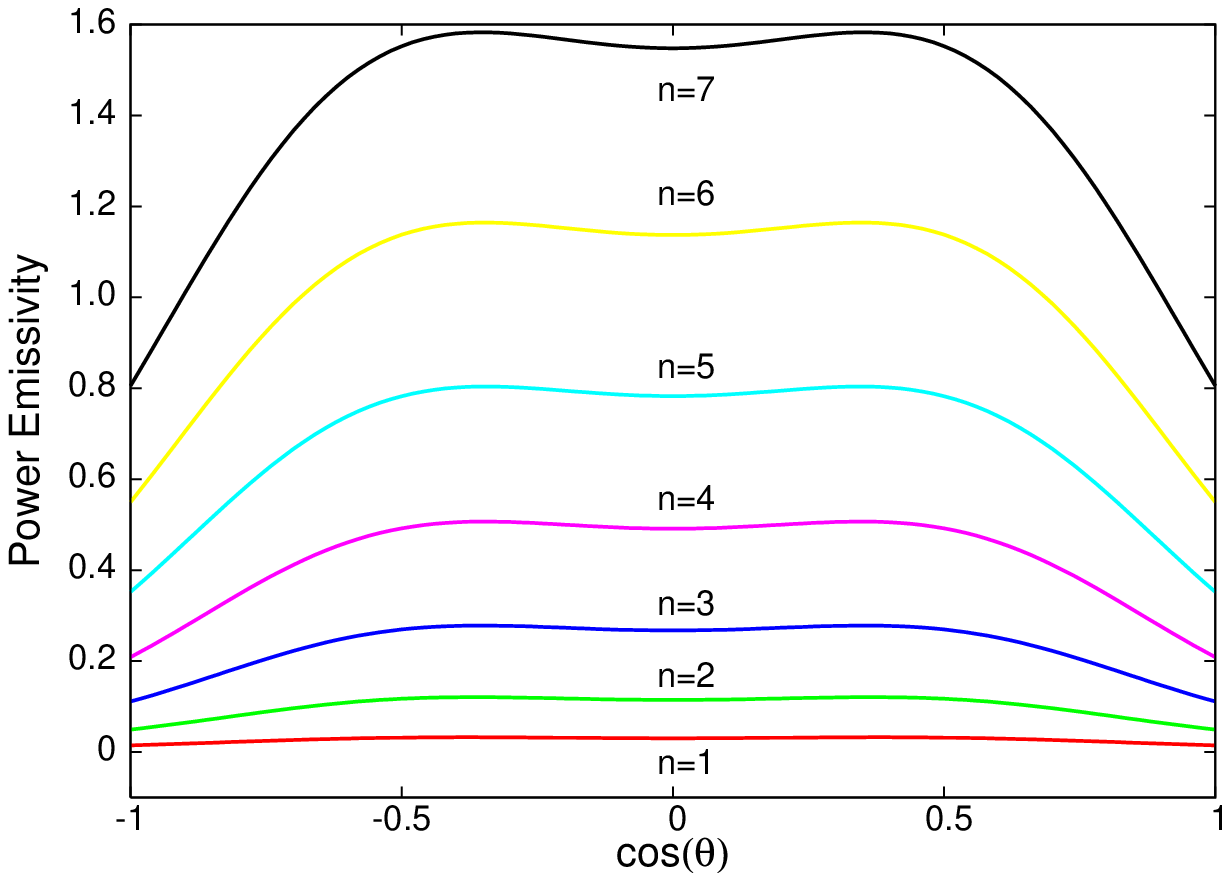}}
\caption{Power emissivity for spin-1 emission on the brane from a rotating black hole
as a function of $\cos \theta$, (a) for $n=6$ and variable $a_*$, and
(b) for $a_*=1$ and variable $n$.}
\label{totalna-theta}
\end{center}
\end{figure}

We finally present the integral over the frequency of the angular distribution of the
power spectrum (\ref{powerang}) to derive the emissivity solely as a function of the
azimuthal angle $\theta $. This is depicted in Figs. \ref{totalna-theta}(a,b),
as a function of $\cos \theta$ for fixed $n$ and variable $a_*$, and vice versa.
As noted before, the shape of the angular distribution pattern is formed under
the effect of the spin-rotation coupling and the centrifugal force, with the former
dominating for low $a_*$ and low $n$, and the latter dominating for high $a_*$ and
high $n$. The two symmetric peaks, due to the spin-rotation coupling, around
$\cos\theta=\pm 1$ cannot be seen in Figs. \ref{totalna-theta} due to their very
small magnitude  -- they barely survive the integration over the frequency due to
the very small energies to which they correspond. Nevertheless, the aforementioned
balance between the two competing forces leads to the type of distributions shown
in Figs. \ref{totalna-theta}, where a twin-peak pattern is formed. The derived shape
of the integral over the frequency of the angular distribution of the power spectrum
agrees with the corresponding one derived in the case of a 4-dimensional, rotating
black hole \cite{Casals:2005kr}.

\subsection{Superradiance}

In this final subsection, we address the issue of superradiance \cite{super},
i.e. the amplification of the amplitude of an incident bosonic wave by a rotating
black hole. This takes place when the transmission coefficient becomes
negative over a particular energy regime, i.e. when $\omega -m \Omega <0$, thus
allowing for values of the reflection coefficient, defined as
$\mathbb{R}_{\ell m \omega}= 1- \mathbb{T}_{\ell m \omega}$, larger than unity.
As noted in section 5.1, this behaviour coincides with
the denominator in the expressions for the various fluxes
(\ref{flux})-(\ref{ang-mom}) becoming also negative, therefore the Hawking
radiation emission rates for these modes remain positive. The superradiance
effect is known to take place in the pure 4-dimensional
case for scalars, gauge bosons and gravitons \cite{press,super, Press:1972}.
In the case of a higher-dimensional black hole, the superradiance effect
associated with either bulk \cite{Frolov2} or brane \cite{HK2, IOP2, Jung-super}
scalar fields has also been investigated. In \cite{HK2}, this effect was
studied in detail in terms of both the dimensionality of spacetime and the
angular momentum of the black hole. It was found that the energy amplification
of 0.3\% for scalars, for a 4-dimensional maximally rotating black hole with
$a_* \simeq 1$ \cite{Press:1972}, increased by more than an order of magnitude for
a 6-dimensional black hole, with the same angular momentum, emitting scalar
fields on the brane. The amplification reached its highest value for the
maximum values of both $n$ and $a_*$ considered in that work, namely $n=6$
and $a_*=4$, and it was found to be around 9\%.

\begin{figure}[t]
\begin{center}
\mbox{
\includegraphics[height=7.3cm]{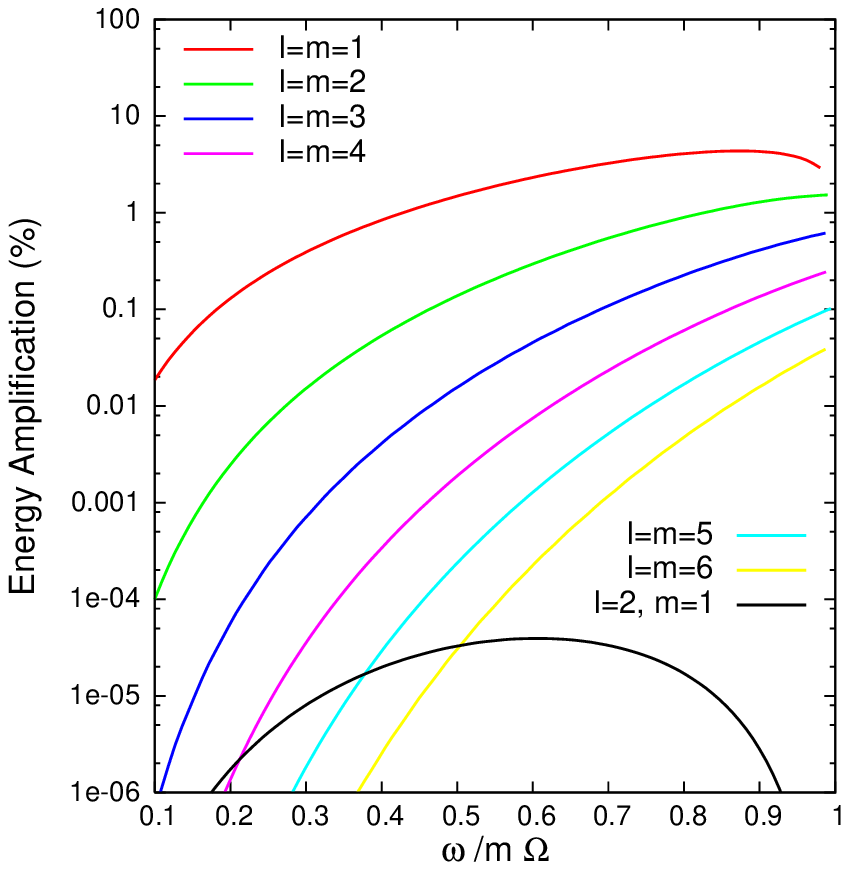}} \hspace*{0.5cm}
{\includegraphics[height=7.3cm]{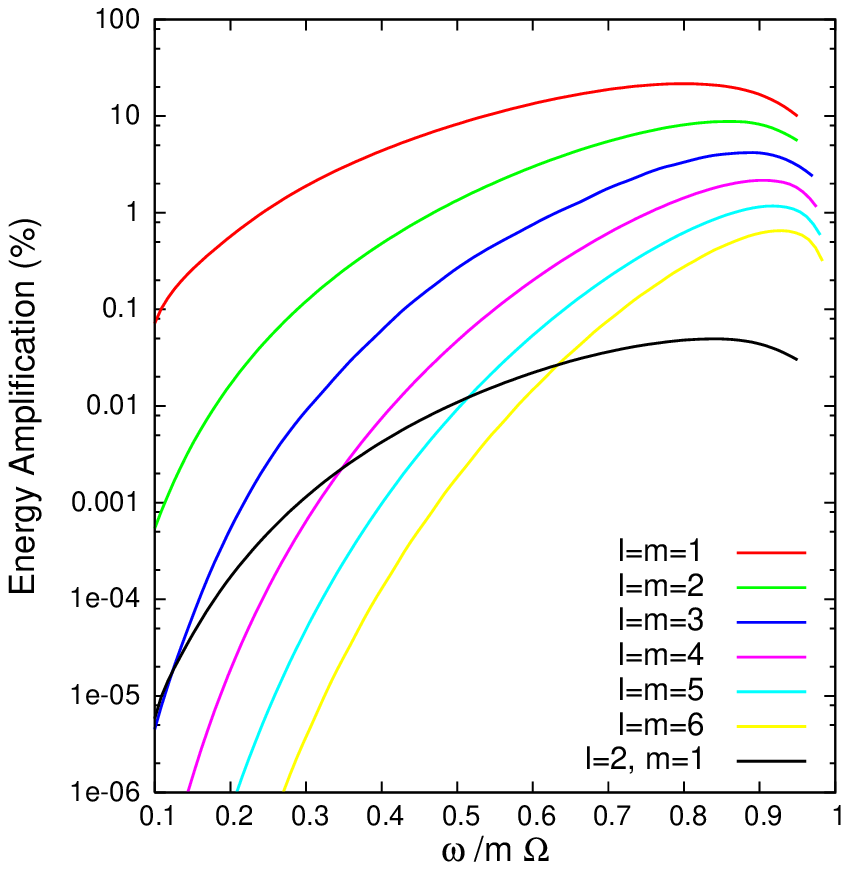}}
\caption{{\bf (a)} Superradiant scattering of gauge bosons by a maximally
rotating ($a=0.99999 M_{BH}$) 4-dimensional black hole; {\bf (b)} Super-radiant
scattering of gauge bosons on the brane by an 11-dimensional black hole
with $a_*=1$.
\hspace*{2.0cm}\label{super4D-11D}}
\end{center}
\end{figure}

By following the conventions of Ref. \cite{press},
in Figs. \ref{super4D-11D}(a) and (b), we depict the energy amplification of
spin-1 fields due to the superradiance effect, for the cases of $n=0$ and
$a_*=0.9955$, and $n=7$ and $a_*=1$, respectively\footnote{The value $a_*=1$
(equivalent to $a=M_{BH}$) corresponds to a maximally rotating
black hole in the 4-dimensional case. This extremal case is highly unstable
from the point of view of numerical analysis, therefore a close enough value,
$a=0.99999 M_{BH}$, equivalent to $a_*=0.9955$, has been used in our analysis,
as  was also done in
\cite{press}.}. The vertical axis in these plots is minus the transmission
coefficient, $-\mathbb{T}_{\ell m \omega}$, expressed as a percentage.
The horizontal axis is the frequency interval over which
$\mathbb{T}_{\ell m \omega}$ is negative, divided by $m \Omega$ to make
the superradiant regime (0,1) for all modes. Figure~\ref{super4D-11D}(a)
shows excellent agreement with the results produced for $n=0$ and
$a=0.99999 M_{BH}$ in \cite{press}: the maximum amplification occurs for
$\ell=m=1$, at $\omega/\Omega = 0.88$, and it has a magnitude of 4.4\%.
However, as the  dimensionality of spacetime increases, even while the
angular momentum remains the same, these numbers change considerably:
according to the data plotted in Fig.~\ref{super4D-11D}(b), the maximum
amplification occurs again for the mode $\ell=m=1$ but this is now at
$\omega/\Omega \simeq 0.8$, and it reaches the figure of 21.7\%. The
behaviour found in section 5.1 -- and depicted in Fig. \ref{transm-n}(b) --
can easily explain the enhancement with $n$ in the superradiance
amplification: for $a_*$ fixed, the transmission coefficient
$\mathbb{T}_{\ell m \omega}$ was found to take increasingly larger negative
values in the superradiance energy regime, as $n$ increased, due to the
increase in the height of the gravitational barrier.

\begin{figure}[t]
\begin{center}
\begin{tabular}{c c c}
\includegraphics[height=5.5cm,width=5.5cm,clip]{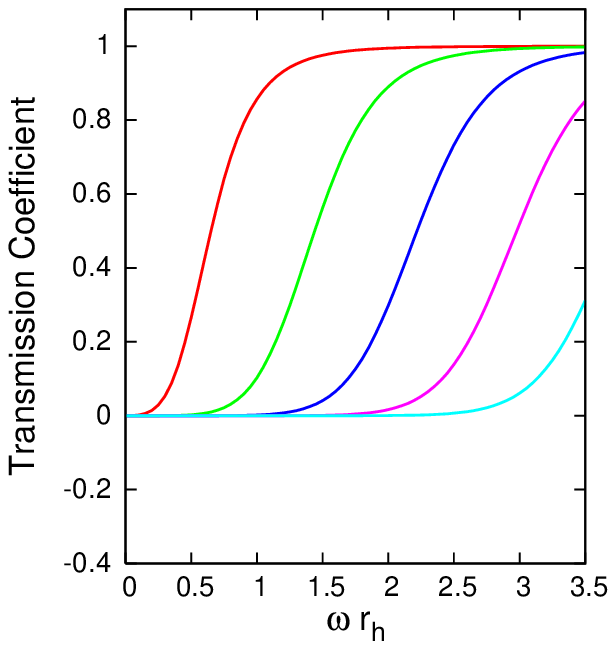} & \hspace*{-0.8cm}
\includegraphics[height=5.5cm,clip]{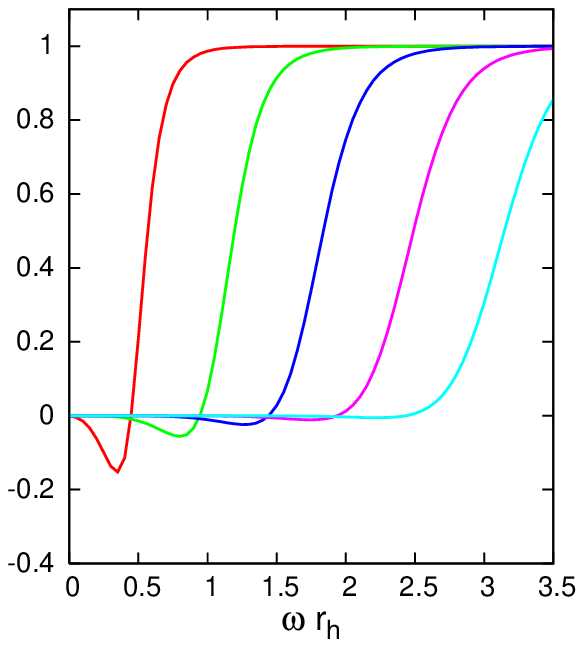} & \hspace*{-0.8cm}
\includegraphics[height=5.5cm,clip]{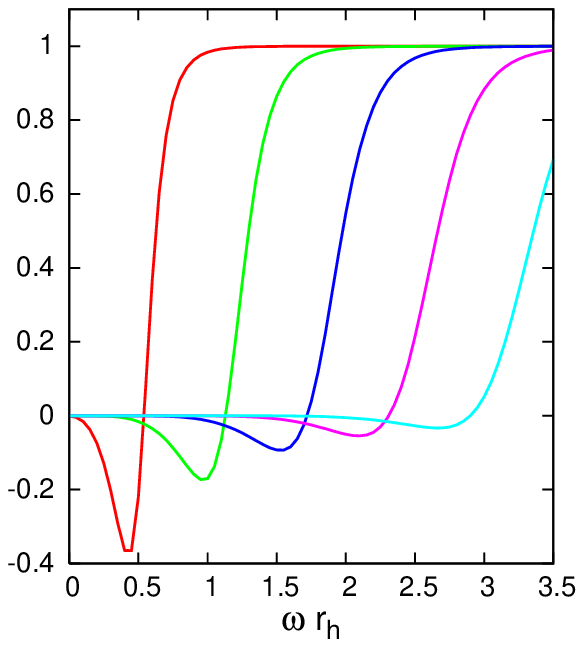} \end{tabular}
\caption{Transmission coefficients for spin-1 emission on the brane, for $n=6$,
from a black hole with (a) $a_*=0$, (b) $a_*=0.9$, and (c) $a_*=1.2$. The curves,
from left to right, correspond to the modes $\ell=m=1, 2, 3, 4$, and 5.}
\label{grey-n6a}
\end{center}
\end{figure}

As the superradiance effect is clearly associated with the rotation velocity of the
black hole, we expect its magnitude to increase with the angular momentum
parameter. Our findings in section 5.1 support this expectation as they clearly
revealed the supression, with $a_*$, of the transmission coefficient,
leading to increasingly larger negative values (see Fig. \ref{transm-a}).
Figure \ref{grey-n6a} depicts the behaviour of the transmission coefficient
$\mathbb{T}_{\ell m \omega}$ for a whole set of spin-1 modes, for a 10-dimensional
black hole ($n=6$) and various values of $a_*$. For the case of $a_*=0$,
shown in Fig. \ref{grey-n6a}(a), the transmission coefficient takes only
positive values, and no superradiance occurs as expected. For $a_*=0.9$,
shown in Fig. \ref{grey-n6a}(b), the transmission coefficient does acquire
negative values over a particular energy regime before it crosses the zero
value and passes into the positive values regime: in this case, as it can
be seen from the different curves, the maximum amplification takes place
for $\ell=m=1$, and it amounts to 15.2\%. As the angular
momentum parameter increases further, the largest negative value of the
transmission coefficient increases, too: for example, for the case of
$a_*=1.2$, shown in Fig. \ref{grey-n6a}(c), the superradiance amplification
reaches the figure of 36.5\%.

As $a_*$ and/or $n$ increase further, we expect the superradiance effect
to become even more important. In an attempt to determine the maximum value
of the energy amplification due to the superradiance, the extreme case
of $n=7$ and $a_*=a^\text{max}_*=4.5$, as dictated by Eq. (\ref{amax}),
was studied. Contrary to the case of the energy amplification of scalar
fields \cite{HK2}, it was found that even for large values of $n$ and
$a_*$, the lowest superradiant mode $\ell=m=1$ is still the dominant
one. For this particular mode, it was found that the energy amplification
reaches the amazing figure of 4200\%, at $\omega/\Omega=0.84$.


\section{Conclusions}

The potential detection of Hawking radiation from a higher-dimensional
black hole formed in the context of a brane-world model (either with
Large Extra Dimensions \cite{ADD}, or warped extra dimensions \cite{RS}
in the limit of $r_h \ll \ell_{AdS}$, with $\ell_{AdS}$ the AdS radius)
demands the derivation of exact results for the various emission rates
during both the spin-down and Schwarzschild phases in the life of the
black hole. After the Schwarzschild phase was exhaustively studied, the
spin-down phase eventually came into the foreground, with a few early and
recent papers offering partial studies of the emission of Hawking
radiation from a rotating brane-world black hole. The first complete study
of the emission of scalar fields on the brane by such a black hole was
performed only recently in \cite{DHKW}. The present paper complements
and expands the latter one by focusing on the emission of gauge bosons
from a $(4+n)$-dimensional rotating black hole on the brane -- where
the gauge bosons are restricted to live by assumption. A comprehensive
analysis of the different emission rates, in terms of the values of the
fundamental parameters of the model, was offered, and additional important
aspects of physical processes associated with a rotating black hole,
such as the angular distribution of the emitted particles and the
superradiance effect, were also studied.

By generalizing techniques that were originally developed for the case of
4-dimen\-sional black holes, to apply in the cases of higher-dimensional black
hole line-elements projected onto a brane, we first derived the equation
of motion for spin-1 particles propagating in the induced-on-the-brane
gravitational background, and decoupled the radial and angular part (apart
from the presence of the angular eigenvalue that appears in both parts
thus linking them).  We then proceeded to review, and generalize where
necessary, the principles leading to the expressions of the transmission
coefficient and of the various Hawking radiation emission fluxes. The
calculation of the transmission coefficient demanded the numerical integration
of the radial part of the equation of motion, while the angular distribution
of the emitted particles demanded the numerical integration of the angular
part - before, however, these two tasks could be performed, the exact value
of the angular eigenvalue had to be computed, again numerically. Our
numerical techniques were discussed in detail in Section 4.

In Section 5, we presented exact numerical results, first, for the transmission
coefficient and, subsequently, for the particle,
energy and angular momentum fluxes. The differential emission rates
per unit time and frequency, integrated over the azimuthal angle $\theta$,
were computed first, and their dependence on the dimensionality of
spacetime, the angular momentum of the black hole and the energy of the
emitted particle was investigated. It was demonstrated that all three
emission rates were enhanced over the whole energy regime as the number
of additional spacelike dimensions and the angular momentum of the black
hole increase. In the case of the particle flux spectra, it was found
that particles with a low energy are more likely to be emitted than
particles with intermediate and high energy, however the emission of the
latter ones becomes increasingly more significant as either $n$ or $a_*$
increase. In the case of power and angular-momentum emission spectra,
for low values of $n$, both of these quantities are predominantly
reduced through the emission of low-energy particles; as the dimensionality
of spacetime takes larger values however, it is the intermediate and
high-energy particles that carry away most of the energy and angular
momentum of the black hole.

\begin{figure}[t]
\begin{center}
\mbox{
\includegraphics[height=5.0cm]{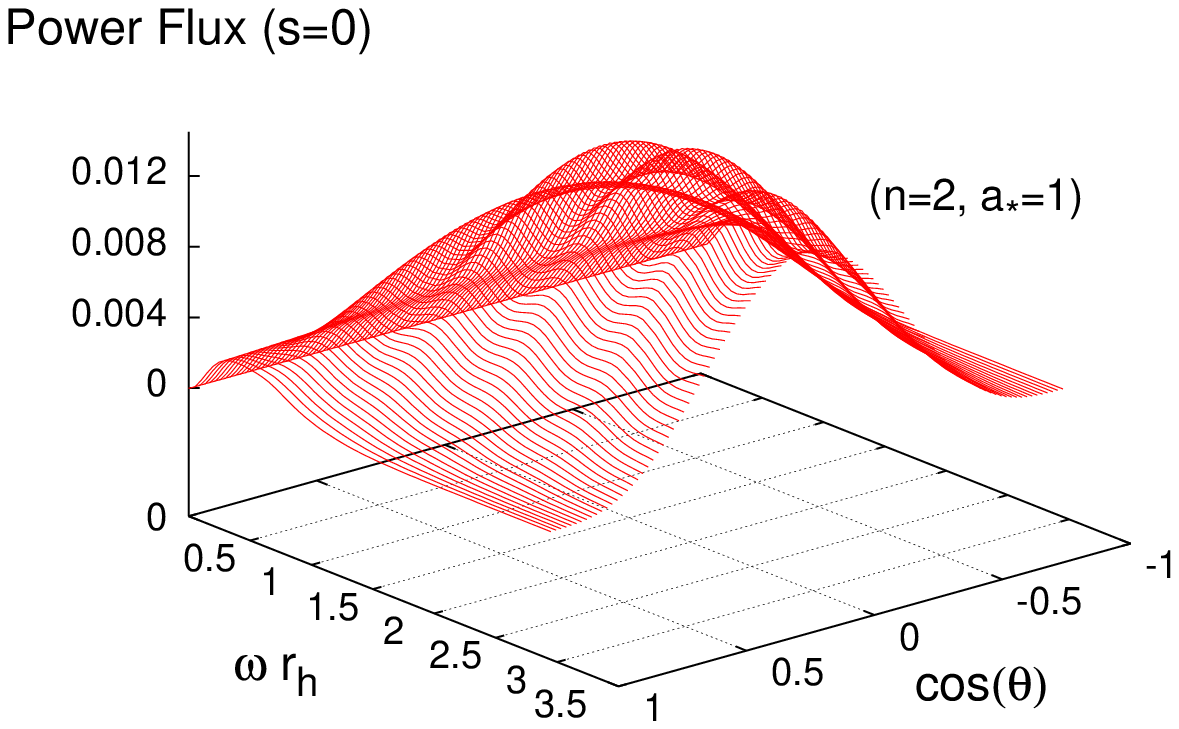}} \hspace*{0.5cm}
{\includegraphics[height=5.0cm]{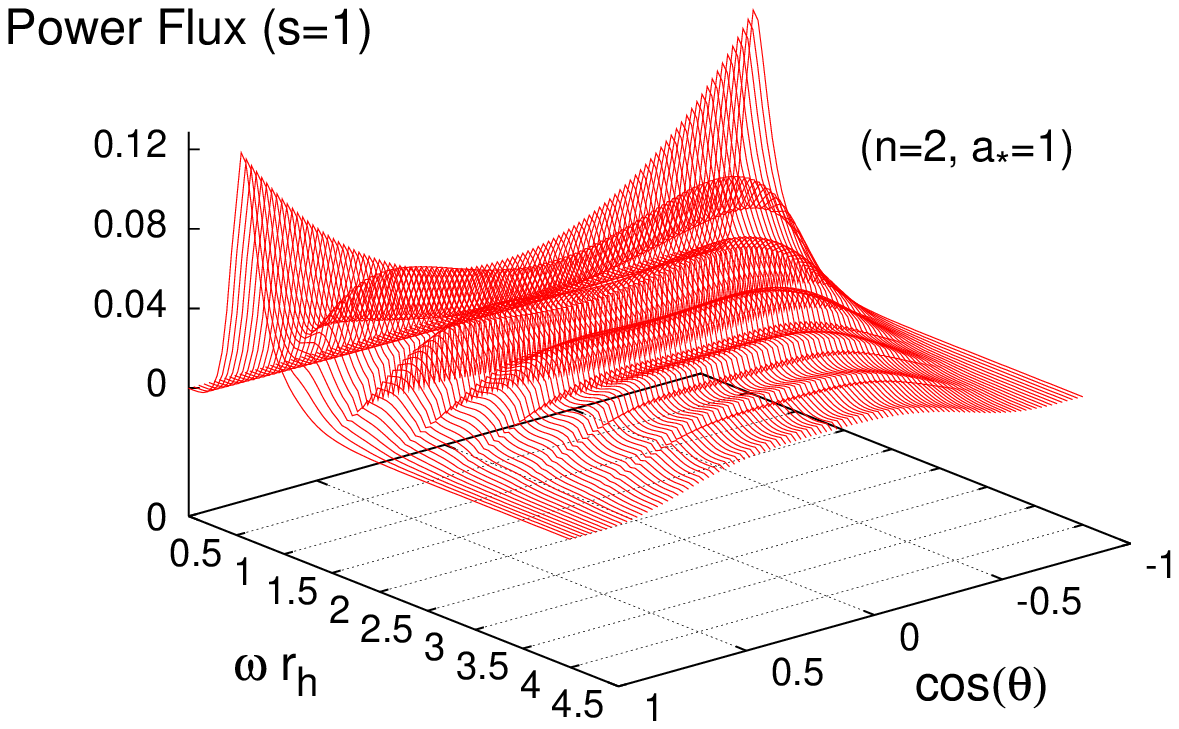}}
\caption{ Angular distribution of the power spectra, for {\bf (a)} scalar
particle emission, and {\bf (b)} gauge particle emission, from a 6-dimensional
black hole with $a_*=1$ on the brane.\label{comparison1}}
\end{center}
\end{figure}

The angular distribution of the emitted particles and energy was also
studied in detail. That became possible after the exact values of the
spin-1 weighted spheroidal harmo\-nics, valid for arbitrary values of
the angular momentum of the black hole and energy of the emitted particle,
were determined. The angular variation in the pattern of the emitted
radiation can be a distinctive signature of emission from a rotating
black hole, but also of the spin and energy of the particles emitted.
In Figs. \ref{comparison1}(a) and (b), we present for comparison the
angular distribution of the energy emission spectrum for a 6-dimensional
($n=2$) rotating black hole emitting radiation on the brane, in the
form of scalar and gauge particles, respectively. As expected, the
uniform distribution of particles seen in the case of a non-rotating
spherically-symmetric black hole is now replaced by a clearly oriented
one, however the exact pattern is strongly dependent on both the spin and
energy of the emitted particle. In the case of scalar particles \cite{DHKW},
the centrifugal force exerted by the rotating black hole forces the
scalar particles to concentrate on the equatorial plane ($\theta=0$);
nevertheless, this angular variation disappears in the low energy part
of the spectrum where the spherically-symmetric $\ell=m=0$ modes
dominate, while it strongly appears in the intermediate and high-energy
part of the spectrum where the non-spherically-symmetric modes become
important. In the case of gauge particles, as we saw in this work, the
angular distribution comes as the result of two forces: the centrifugal
one, and the spin-rotation coupling due to the non-vanishing spin of
the particle. It turns out that the latter one dominates for low-energy
particles, forcing them to align parallel to the rotation axis
($\theta=\pm \pi$), while it becomes subdominant compared to the centrifugal
force in the high-energy regime, causing the particles to concentrate again
in the equatorial plane. A similar behaviour for the emission of gauge
particles was seen also in the 4-dimensional case \cite{Casals:2005kr};
in the present case, it was found that the number of additional spacelike
dimensions work towards enhancing further the centrifugal force.

An important quantity associated to the emission of Hawking radiation
from a black hole is the total emissivity, that describes the emission
of either particles, energy or angular momentum per unit time, over the
whole frequency band. In fact, in our analysis, lower bounds for the
total emissivities were computed instead of the actual total ones,
as the differential emission rates were integrated over the energy from
zero to the chosen cut-off value of 4. This task was performed for
all three emission rates, namely for the particle, energy and angular
momentum fluxes. Their dependence on the azimuthal angle, the angular
momentum of the black hole and the number of additional dimensions was
investigated. In terms of the first quantity, both the particle and
energy emissivities were found
to have a twin-peak pattern under the influence of the centrifugal force
and the spin-rotation coupling. In terms of the other two parameters,
all emissivities were found to increase substantially as $n$ and $a_*$
increased; for instance, the energy emission per unit time increased
5-fold, for a 10-dimensional black hole, as $a_*$ increased from 0 to 1,
and 50-fold, for a black hole with $a_*=1$, when $n$ increased from
1 to 7.

\begin{figure}[ht]
\begin{center}
\mbox{
\includegraphics[height=5.4cm]{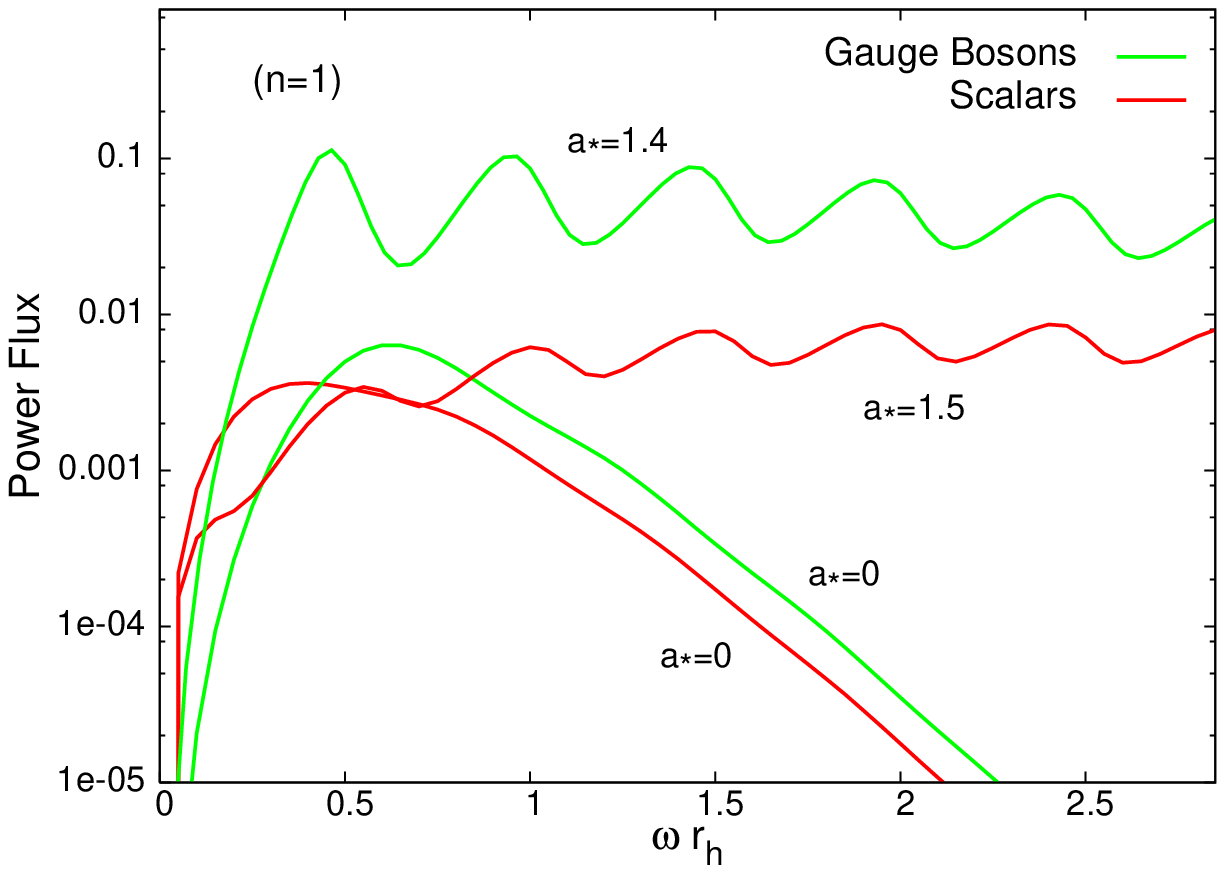}} \hspace*{-0.1cm}
{\includegraphics[height=5.4cm]{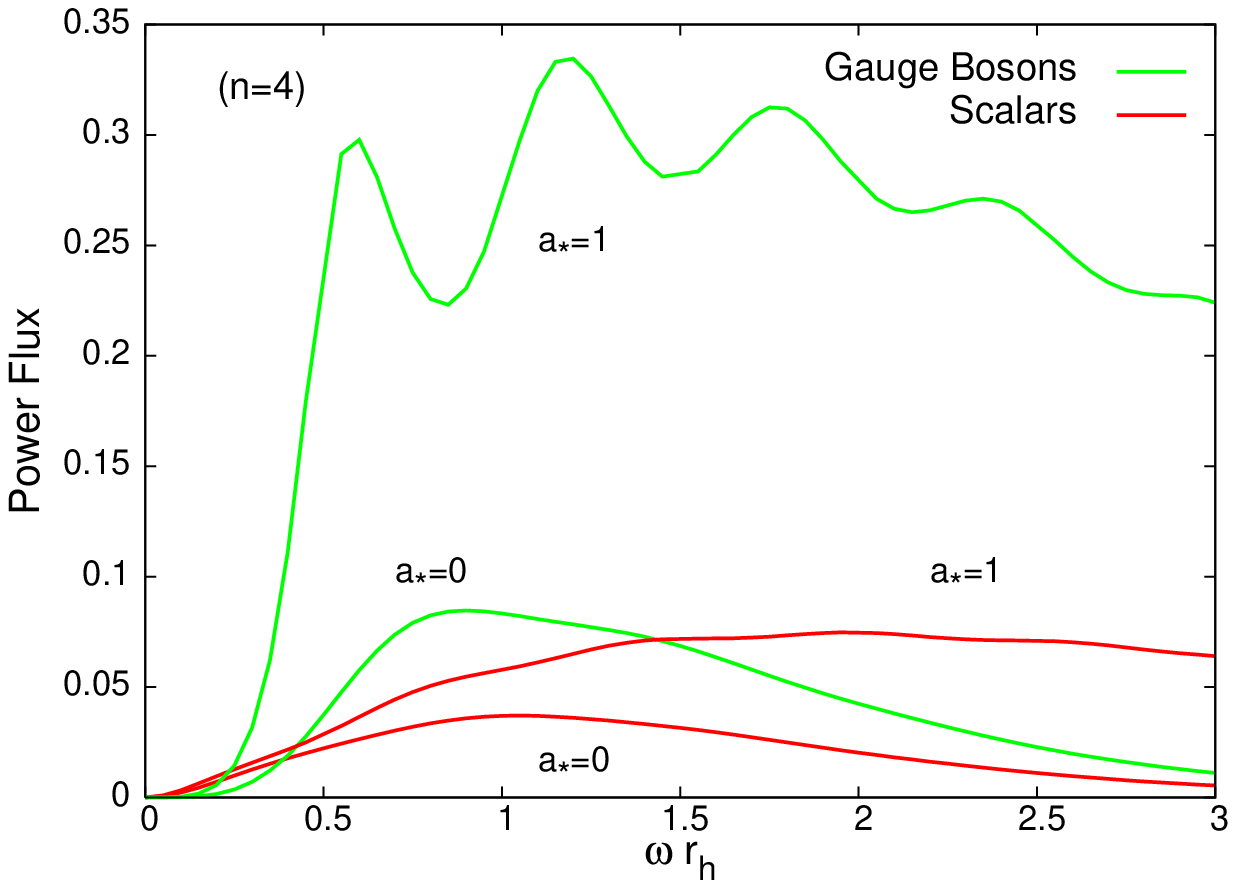}}
\caption{Energy emission rates for scalars and gauge bosons, for {\bf (a)}
$n=1$, and {\bf (b)} $n=4$, and various values of $a_*$. \label{comparison2}}
\end{center}
\end{figure}

An interesting question that arises when one combines results on the various
emission rates for different types of fields (i.e. scalars, gauge bosons and
fermions) is which type of fields is preferably emitted by the black hole
as both $n$ and $a_*$ vary. In the case of non-rotating black holes ($a_*=0$),
it was found \cite{Kanti, HK1} that a 4-dimensional black hole emits mainly
scalar particles, a 6-dimensional black hole prefers a democratic
distribution of particles, while the spectrum of a 10-dimensional
black hole is dominated by gauge bosons. A similar behaviour is expected
in the case of a rotating black hole. Looking at the vertical axes of
Figs. \ref{comparison1}(a,b), one can easily see that the emission rates
for scalars and gauge bosons differ significantly, for the same values of
$n$ and $a_*$. In order to investigate this in more detail, we have displayed
in Figs. \ref{comparison2}(a) and (b) the emission rates for scalars and
gauge bosons, for fixed $n$ ($n=1$ and $n=4$, respectively), and for
similar values of $a_*$. Figure \ref{comparison2}(a) confirms the fact
that for a non-rotating black hole living in a spacetime with a low value
of $n$, the emission rate for gauge bosons becomes comparable to that of
scalars only when both polarizations are included\footnote{In
\cite{Kanti, HK1}, the results presented refer to a single polarization
of the emitted gauge field.}. When, however, the angular momentum
of the black hole is turned on, the emission rate for gauge bosons
becomes significantly larger than the one for scalar fields over the
whole energy regime. Moving to the $n=4$ case, we see that indeed,
for $a_*=0$, the gauge bosons already clearly dominate over the scalar
fields; when the angular momentum is added on top, the increase becomes
even larger. We may therefore conclude that a rotating black hole
strongly prefers to emit gauge bosons over scalar particles for all
values of $n$.

The final issue addressed in this work was the superradiance effect
associated to the amplification of a bosonic wave incident on a rotating
black hole. We have found that, in the case of gauge fields, similarly
to the scalar case, this effect is present and greatly enhanced by the
addition of extra spacelike dimensions transverse to the brane.
The effect in this case is actually more important than in the scalar one:
the superradiance effect increases from 4.4\% to 21.7\% as the dimensionality
of spacetime goes from $D=4$ to $D=11$, even as the angular momentum of the
black hole remains the same. Vice versa, for fixed dimensionality, i.e.
$n=6$, the energy amplification varied from 0\%, for $a_*=0$, as expected,
to 36.5\%, for $a_*=1.2$. As the superradiance effect is clearly augmented by
both $n$ and $a_*$, the maximum energy amplification was sought by looking
at the extreme case of $n=7$ and $a_*=a^\text{max}_*=4.5$: the lowest
superradiant mode $\ell=m=1$ provided again the maximum amplification
that was found to reach the amazing figure of 4200\%.

The study of the emission of gauge fields on the brane by a higher-dimensional
rotating black hole, performed in this work, has revealed a number of distinct
features associated with the behaviour and magnitude of the different emission
rates, the angular distribution of particles and energy, the relative enhancement
compared to the scalar fields, and the magnitude of the superradiance effect.
Apart from the theoretical interest in studying modifications of the properties
of black holes submerged into a higher-dimensional space, we expect these features
to comprise clear signatures of the emission of Hawking radiation of a
brane-world black hole during its spin-down phase upon successful detection
of this effect during an experiment. Having starting with the study of the
emission of salar fields on the brane by such a black hole \cite{DHKW}, we have
offered in this work a similar, comprehensive analysis of the case of gauge
fields; we hope to return soon with complete results for the emission of
fermions that will eventually complete the study of the decay of a small,
higher-dimensional rotating black hole through the emission of Standard
Model fields on the brane.


\bigskip

{\bf Acknowledgments.} We would like to thank Chris M. Harris for useful
discussions.
The work of P.K. is funded by the UK PPARC Research Grant PPA/A/S/2002/00350.
The work of E.W. is supported by UK PPARC, grant reference number
PPA/G/S/2003/ 00082, the Royal Society and the London Mathematical Society.



\end{document}